\documentclass[pra,showpacs,twocolumn,aps,superscriptaddress,a4paper,longbibliography]{revtex4-1}
\usepackage{dcolumn,amssymb,amsmath,amsfonts,graphicx,latexsym,color}
\usepackage[utf8]{inputenc}
\usepackage[T1]{fontenc}
\usepackage{epstopdf}

\usepackage{ulem}
\usepackage{epstopdf}
\usepackage{subfigure}
\usepackage{float}
\usepackage{mathrsfs}
\usepackage{bbm}
\usepackage{psfrag}

\usepackage{verbatim}
\usepackage{multirow}
\usepackage{diagbox}
\usepackage[colorlinks,citecolor=blue]{hyperref}

\def\blue{\textcolor{blue}}
\def\red{\textcolor{red}}

\begin{document}

\def\qv{\vec{q}}
\def\red{\textcolor{red}}
\def\blue{\textcolor{blue}}
\def\magenta{\textcolor{magenta}}
\def\apricot{\textcolor{Apricot}}

\def\GJ{\textcolor{black}}
\def\TT{\textcolor{ForestGreen}}
\definecolor{ora}{rgb}{1,0.45,0.2}
\def\LH{\textcolor{black}}

\newcommand{\norm}[1]{\left\lVert#1\right\rVert}
\newcommand{\ad}[1]{\text{ad}_{S_{#1}(t)}}
\newcommand{\rz}{\textcolor[rgb]{0,1.0,0}}


\title{Universal competitive spectral scaling from the critical non-Hermitian skin effect}

\author{Fang Qin}
\email{qinfang@nus.edu.sg}
\affiliation{Department of Physics, National University of Singapore, Singapore 117551, Singapore}

\author{Ye Ma}
\affiliation{Department of Physics, National University of Singapore, Singapore 117551, Singapore}

\author{Ruizhe Shen}
\affiliation{Department of Physics, National University of Singapore, Singapore 117551, Singapore}

\author{Ching Hua Lee}
\email{phylch@nus.edu.sg}
\affiliation{Department of Physics, National University of Singapore, Singapore 117551, Singapore}

\date{\today}

\date{\today}
\begin{abstract}
Recently, it was discovered that certain non-Hermitian systems can exhibit qualitative different properties at different system sizes, such as being gapless at small sizes and having topological edge modes at large sizes $L$. This dramatic system size sensitivity is known as the critical non-Hermitian skin effect (cNHSE), and occurs due to the competition between two or more non-Hermitian pumping channels. In this work, we rigorously develop the notion of a size-dependent generalized Brillouin zone (GBZ) in a general multi-component cNHSE model ansatz, and found that the GBZ exhibits a universal $a+b^{1/(L+1)}$ scaling behavior. In particular, we provided analytical estimates of the scaling rate $b$ in terms of model parameters, and demonstrated their good empirical fit with two paradigmatic models, the coupled Hatano-Nelson model with offset, and the topologically coupled chain model with offset. We also provided analytic result for the critical size $L_c$, below which cNHSE scaling is frozen. The cNHSE represents the result of juxtaposing different channels for bulk-boundary correspondence breaking, and can be readily demonstrated in non-Hermitian metamaterials and circuit arrays.
\end{abstract}

\maketitle

\section{Introduction}\label{1}

Non-Hermitian systems harbor a host of interesting physics not found in equilibrium systems, such as exceptional point sensitivity and robustness~\cite{zhong2019sensing,djorwe2019exceptional,guo2021sensitivity,chen2021nanophotonic,chen2019sensitivity,nikzamir2022achieve,lee2022exceptional,chang2020entanglement,dora2022correlations,wang2022scaling,zhou2019exceptional,zhang2021tidal,wiersig2020robustness,wiersig2020prospects,wiersig2020review,zhu2022exceptional,xiao2021observation}, enlarged symmetry classes~\cite{ashida2020non,el2018non,kawabata2019symmetry,bergholtz2021exceptional}, and intrinsically non-equilibrium topological phases~\cite{nakagawa2014nonequilibrium,xiao2020non,song2019non,zhang2021universal,yu2021topological,jia2021dynamically,zhang2021quench-induced,yu2021quantum,zhang2020unified,yi2019observing,wang2019experimental,zhang2021nonequilibrium,sun2018uncover,zhang2018dynamical}. Once thought to exist almost exclusively as mathematical constructs, these novel phenomena have one by one been experimentally demonstrated in the recent years, thanks to rapid technical advances in ultracold atomic gases~\cite{li2019observation,lapp2019engineering,ren2022chiral,liang2022observation,gou2020tunable}, electrical circuits~\cite{helbig2020generalized,hofmann2019chiral,liu2020gain,liu2021non,zou2021observation,stegmaier2021topological,hohmann2022observation,lv2021realization,zhang2022anomalous,lenggenhager2022simulating,shang2022experimental,zhang2022observation,wu2022evidencing}, photonic systems~\cite{roccati2022exotic,zhou2019exceptional,feng2017non,tzuang2014non,zhou2018observation,zhen2015spawning,zeuner2015observation,miri2019exceptional,feng2011nonreciprocal,shen2016experimental,xiao2020non},  coupled acoustic cavities~\cite{ding2016emergence,tang2020exceptional,ding2018experimental,tang2021direct,tang2022experimental}, as well as other metamaterials~\cite{park2020observation,coulais2017static,zhu2018simultaneous,ghatak2020observation,brandenbourger2019non,gao2021non,qin2022light,wang2022experimental,gu2021controlling,qin2022phase,wen2022unidirectional,qin2020theory,gupta2022requisites}. 

A particularly intriguing type of non-Hermitian phenomenon is the breaking of conventional bulk-boundary correspondences (BBCs), which generically occurs whenever reciprocity is also broken. Topological BBCs relate boundary topological states with bulk topological invariants, and are cherished tenets in topological classification~\cite{chiu2013classification,chiu2016classification,potter2016classification,barkeshli2013classification,mcginley2019classification,else2016classification}. The most well-studied type of non-Hermitian BBC is the non-Hermitian skin effect (NHSE)~\cite{yao2018edge,xiong2018does,lee2019anatomy,kunst2018biorthogonal,yokomizo2019non,imura2019generalized,jin2019bulk,borgnia2020non,li2021quantized,lv2021curving,yang2022designing,jiang2022dimensional,tai2022zoology,lee2021many,lee2020unraveling,li2022non,shen2022non,qin2022non}, which is characterized by exponentially large boundary state accumulation that leads to very different energy spectra under open and periodic boundary conditions (OBCs and PBCs). To restore an effective bulk theory, the customary approach has been to define a generalized Brillouin zone (GBZ) with complexified momentum~\cite{lee2019hybrid,yokomizo2019non,kawabata2020non,yokomizo2020non,yi2020non,yokomizo2020topological,yang2020non,yokomizo2021non,deng2019non}, such that quantities computed in the GBZ correctly correspond to physical observations.
 
Most interesting is the relatively little-understood scenario of \emph{critical} NHSE (cNHSE)~\cite{li2020critical}, where even the scaling properties of the system are drastically modified by non-Hermiticity. For instance, the same metamaterial exhibiting cNHSE can behave qualitatively differently at different system sizes, such as being gapless (metallic) at small sizes but topologically insulating at large sizes~\cite{li2020critical,rafi2022critical,liu2020helical,rafi2022system}. Physically, such peculiar size-dependent transitions are due to the competition between multiple NHSE channels (non-reciprocity strengths) in the system -- at different length scales, the same physical coupling can be ``renormalized'' to very different values dependent on the dominant NHSE channel. Due to their peculiar size dependency, the cNHSE systems also harbor different entanglement scaling laws~\cite{li2020critical} from those of other Hermitian and non-Hermitian phases~\cite{calabrese2009entanglement,nishioka2009holographic,swingle2012entanglement,lee2015free,gu2016holographic,herviou2019entanglement,ortega2022polarization,okuma2021quantum,chen2022quantum,chang2020entanglement,lee2022exceptional,dora2022correlations,kawabata2022entanglement}.

In this work, we focus on addressing the following open question: How exactly can we understand cNHSE scaling behavior in terms of the GBZ, which is widely used for restoring the BBC in the thermodynamic limit? Specifically, we find that a cNHSE system of finite size can be accurately described through an ``interpolated'' GBZ that lies between the competing GBZs describing the same (but behaviorally distinct) system in the small and large size limits. Furthermore, this interpolation occurs at a rate obeying a universal exponential scaling law, with exponent inversely proportional to system size. Since the effective GBZ allows one to represent the system with a Hamiltonian with an effective Bloch description, this scaling law carries over into most physical properties of cNHSE lattices.

To motivate and substantiate our results, we consider a generic two-component ansatz for modeling a cNHSE system with two competing NHSE channels. Our ansatz encompasses the minimal model studied in Ref.~\cite{yokomizo2021scaling}, and showcases how some of its results can be generalized in the context of arbitrary NHSE channels. By subsequently specializing into two paradigmatic models, we provide detailed derivations of the universal $a+b^{1/(L+1)}$ scaling behavior governing the effective finite-size GBZ, where $L$ is the system size, and $a,b$ constants depending on the model details. We also provide detailed and empirically verified estimates of the lower critical system size $L_c$ above which such a scaling relation holds.

We pause to briefly elaborate on the experimental prospects for the cNHSE models discussed in this work. Most directly, electrical circuits, i.e., ``topolectrical circuits'' can be connected in very versatile manners, and are thus readily suited for their experimental implementation~\cite{helbig2020generalized,hofmann2019chiral,liu2020gain,liu2021non,zou2021observation,stegmaier2021topological,hohmann2022observation,lv2021realization,zhang2022anomalous,lenggenhager2022simulating,shang2022experimental,zhang2022observation,wu2022evidencing}. In general, operation amplifiers serve as almost perfectly linear components with asymmetric Laplacians~\cite{hofmann2019chiral}, and are thus ideal building blocks for models with the asymmetric couplings necessary for cNHSE. Recently the coupled Hatano-Nelson cNHSE model of this work has also been realized in an even simpler experimental circuit platform~\cite{zhang2022observation} involving only RLC circuit components, since asymmetric couplings can be rendered symmetric via a basis rotation in this model. Circuit realizations can also be largely generalized to photonic platforms~\cite{xie2023antihelical,parto2020non,el2019dawn,midya2018non,wang2019exceptional,feng2017non,wang2021topological,de2022non}. Coupled resonator arrays can be used to experimentally realize the arrays in our models, with the ring resonators~\cite{xie2023antihelical} (which are the primary resonators) representing the sites in our model chains. Experimental values of the hoppings in such photonic systems are highly tunable, ranging approximately from 5 GHz to 30 GHz~\cite{hafezi2013imaging,mittal2019photonic}. The optical gain and loss in a photonic system can be used to experimentally realize the gain and loss in our non-Hermitian models.

The paper is organized as follows: In section \ref{2}, we set up the cNHSE formalism using a general two-component ansatz.  Next, we illustrate our results through detailed calculations on two paradigmatic models, a coupled Hatano-Nelson model with energy offset (section \ref{3}) and a model with size-dependent topology (section \ref{4}). We show how their OBC spectra and effective GBZs depend greatly on the system size, and provide quantitative derivations of their exponential scale dependence, as well as the critical system size above which such scaling holds. In section \ref{5}, we demonstrate the robustness of the scaling of imaginary energy against substantial disorder. Finally, we summarize the key findings in the discussion in section \ref{6}.

\section{General two-component cNHSE ansatz}\label{2}

To understand the cNHSE phenomenon, we first review the concept of the GBZ. The GBZ formalism restores the BBC via a complex momentum deformation. For a momentum-space one-dimensional (1D) Hamiltonian $H(z)$ with $z=e^{ik}$, the GBZ corresponding to an eigenenergy $E$ can be obtained from solving for $z=e^{ik}$ in the following characteristic
Laurent polynomial~\cite{lee2019hybrid,lee2019anatomy}:
\begin{equation}
	f(z, E):=\operatorname{det}[H(z)-E\,\mathbb{I}]=0.
	\end{equation} 
For $E$ that does not coincide with any of the PBC eigenenergies, i.e., eigenvalues of $H(e^{ik})$ for real $k$, we must have complex $k=-i\ln z$. Such $E$ lies in the OBC spectrum when the latter is very different from the PBC spectrum. It can be shown that~\cite{yao2018edge,lee2019anatomy,yokomizo2019non,tai2022zoology,lee2020unraveling,yang2020non} in the thermodynamic limit, the OBC eigenenergies are given by~\footnote{This double degeneracy in $\text{Im}(k)$ is required for the state to vanish at two boundaries that are arbitrary far apart.} solutions of $k$ that are doubly degenerate in both $\text{Im}(k)$ and $E$: For such solutions, we define the GBZ as $\kappa(k)$, where the complex momentum deformation is given by $k \rightarrow k+i \kappa(k)$. In other words, we say that the conventional (Bloch) BZ is replaced by the (non-Bloch) GBZ defined by $z \rightarrow e^{i k} e^{-\kappa(k)}$. 

To understand how the GBZ formalism needs to be modified in a cNHSE system, we start from a generic two-component ansatz cNHSE Hamiltonian, written in the component basis $C_{{\bf k}}=\left(c_{{\bf k},A}^{},c_{{\bf k},B}^{}\right)^{T}$ as
\begin{align}
	{\cal H}_{g}(z)
	\!=\!\begin{pmatrix}
		\mathcal{H}^{aa}(z) &\mathcal{H}^{ab}(z) \\
		\mathcal{H}^{ba}(z) &\mathcal{H}^{bb}(z)
	\end{pmatrix}=\sum_{n=-n_{-}}^{n_{+}}\begin{pmatrix}
		h_{n}^{aa} & h_{n}^{ab} \\
		h_{n}^{ba} & h_{n}^{bb}
	\end{pmatrix}z^{n},\label{eq:Lee_Hk_general}
\end{align} where $n_{\pm}\in\mathbb{Z}$, $z=e^{ik}$. 
In principle, cNHSE exists as long as $\mathcal{H}^{aa}$ and $\mathcal{H}^{bb}$ exhibit dissimilar inverse skin localization lengths $\kappa(k)$, and couplings $\mathcal{H}^{ab},\mathcal{H}^{ba}\neq 0$. The former condition is equivalent to having asymmetric hoppings $h_{n}^{aa}\neq h_{-n}^{aa}$ and $h_{n}^{bb}\neq h_{-n}^{bb}$ for some $n$, as well as $h_{n}^{aa}/h_{-n}^{aa}\neq h_{n}^{bb}/h_{-n}^{bb}$.

To implement OBCs, we first Fourier transform to real space, where one obtains the real-space tight-binding Hamiltonian
\begin{align}
	H_{gr}\!=\!
	\sum_{i=1}^{L}\sum_{n=-n_{-}}^{n_{+}}C_{i}^{\dagger}\begin{pmatrix}
		h_{n}^{aa}  & h_{n}^{ab} \\
		h_{n}^{ba} & h_{n}^{bb}
	\end{pmatrix}C_{i+n},\label{eq:Lee_Hr_general}
\end{align} where $L$ is the system size, i.e., number of unit cells, $1\leqslant n_{\pm}\leqslant L/2$, $C_{i}=\left(c_{i,A}^{},c_{i,B}^{}\right)^{T}$ with  the annihilation (creation) operator $c_{i,\alpha}^{}$ ($c_{i,\alpha}^\dag$) on site $\alpha$ ($\alpha={\rm A},{\rm B}$) in cell $i$. For a real-space wave function $|\psi\rangle=\left(\psi_{1,{\rm A}},\psi_{1,{\rm B}},\psi_{2,{\rm A}},\psi_{2,{\rm B}},\cdots,\psi_{L,{\rm A}},\psi_{L,{\rm B}}\right)^{\rm T}$, we express the real-space Schr{\"o}dinger equation ${\cal H}_{gr}|\psi\rangle=E_{\rm OBC}|\psi\rangle$ as
\begin{eqnarray}
	\!\left\{ \begin{array}{l}
		\!\sum_{n=-n_{-}}^{n_{+}}\left(h_{n}^{aa}\psi_{i+n,{\rm A}}\!+\!h_{n}^{ab}\psi_{i+n,{\rm B}}\right)\!=\!E_{\rm OBC}\psi_{i,{\rm A}}, \vspace{7pt}\\
		\!\sum_{n=-n_{-}}^{n_{+}}\left(h_{n}^{ba}\psi_{i+n,{\rm A}}\!+\!h_{n}^{bb}\psi_{i+n,{\rm B}}\right)\!=\!E_{\rm OBC}\psi_{i,{\rm B}},
	\end{array}\right.
	\label{eq:OBC_Schrodinger_general}
\end{eqnarray} where ${\cal H}_{gr}$ is the Hamiltonian matrix of $H_{gr}$ in the basis $(C_{1},C_{2},\cdots,C_{L})^T$ and $E_{\rm OBC}$ is the eigenenergy under OBC. To relate to the complex momenta present in non-Hermitian skin modes, we solve the real-space Schr{\"o}dinger equation via the ansatz
\begin{eqnarray}
	\left( \psi_{n,{\rm A}},~\psi_{n,{\rm B}}\right)^{T}=\sum_{j}(\beta_{j})^n\left(\phi_{\rm A}^{\left(j\right)},~\phi_{\rm B}^{\left(j\right)}\right)^{T}, \label{eq:eigenstates_general}
\end{eqnarray}
where ${\rm A}$, ${\rm B}$ are the site indices in the cell, and $n$ represents the position of the cell (A,B) in the real space. Here, $\beta_{j}$ are specific solutions to $z=e^{ik}$, and characterizes the spatial localization of the boundary skin-localized wave function. By substituting Eq.~(\ref{eq:eigenstates_general}) into (\ref{eq:OBC_Schrodinger_general}), we can write the bulk eigenequation as
\begin{eqnarray}
	\!\left\{ \begin{array}{l}
		\!\left[\sum_{n=-n_{-}}^{n_{+}}\!h_{n}^{aa}(\beta_{j})^n\!-\!E_{\rm OBC}\right]\phi_{{\rm A}}^{(j)}\!+\!\sum_{n=-n_{-}}^{n_{+}}\!h_{n}^{ab}(\beta_{j})^n\phi_{{\rm B}}^{(j)} \\
\!=\!0, \vspace{7pt}\\
		\!\sum_{n=-n_{-}}^{n_{+}}\!h_{n}^{ba}(\beta_{j})^n\phi_{{\rm A}}^{(j)}\!+\!\left[\sum_{n=-n_{-}}^{n_{+}}\!h_{n}^{bb}(\beta_{j})^n\!-\!E_{\rm OBC}\right]\phi_{{\rm B}}^{(j)} \\
\!=\!0.
	\end{array}\right.
	\label{eq:bulk_eigenequation_general}
\end{eqnarray} 
Equation~\eqref{eq:bulk_eigenequation_general} can be recast into the energy dispersion characteristic equation 
\begin{eqnarray}
	E_{\rm OBC}^{2}&\!-\!&\sum_{n=-n_{-}}^{n_{+}}\left(h_{n}^{aa} \!+\! h_{n}^{bb}\right)(\beta_{j})^{n}E_{\rm OBC} \nonumber\\
	&\!+\!&\left[ \sum_{n=-n_{-}}^{n_{+}}h_{n}^{aa}(\beta_{j})^{n} \right]\left[ \sum_{n=-n_{-}}^{n_{+}}h_{n}^{bb}(\beta_{j})^{n} \right]\nonumber\\
	&\!-\!&\left[ \sum_{n=-n_{-}}^{n_{+}}h_{n}^{ab}(\beta_{j})^{n} \right]\left[ \sum_{n=-n_{-}}^{n_{+}}h_{n}^{ba}(\beta_{j})^{n} \right]\!=\!0,\label{eq:characteristic_general} 
\end{eqnarray} 
where we have labeled the solutions $\beta_j$ with increasing magnitude $\left|\beta_1\right|\leqslant\left|\beta_2\right|\leqslant\dots\leqslant\left|\beta_{2M}\right|$. Here $M=n_{-}+n_{+}$. 

Importantly, the key property required for restoring BBCs -- the complex momentum deformation (effective GBZ) -- does not require intimate knowledge of most of these $\beta$ solutions. This is because fundamentally, the required complex deformation depends on the decay rate of the eigenstates, which turns out to depend only on two dominant $\beta$ solutions. Below, we derive the precise conditions from the bulk eigenequations (\ref{eq:bulk_eigenequation_general}) as well as constraints from the OBCs $\psi_{-n_{-},\alpha}=\cdots=\psi_{-1,\alpha}=\psi_{0,\alpha}=\psi_{L+1,\alpha}=\psi_{L+2,\alpha}=\cdots=\psi_{L+n_{+},\alpha}=0~(\alpha={\rm A},{\rm B};~1\leqslant n_{\pm}\leqslant L/2)$. 
\begin{widetext}
By eliminating $\phi_{\rm B}^{\left(j\right)}$ in terms of $\phi_{\rm A}^{\left(j\right)}$, we obtain $2M$ simultaneous linear equations in $\phi_{\rm A}^{\left(j\right)}$, $(j=1,2,\dots,2M)$, which yield nonvanishing solutions only if the determinant 
\begin{equation}
\begin{vmatrix}
F_{1}^{(a,1)}\beta_{1} & F_{2}^{(a,1)}\beta_{2} & \cdots & F_{2M}^{(a,1)}\beta_{2M} \\
F_{1}^{(b,1)}\beta_{1} & F_{2}^{(b,1)}\beta_{2} & \cdots & F_{2M}^{(b,1)}\beta_{2M} \\
\vdots & \vdots & \vdots & \vdots \\
F_{1}^{(a,n_{+})}\left(\beta_{1}\right)^{n_{+}} & F_{2}^{(a,n_{+})}\left(\beta_{2}\right)^{n_{+}} & \cdots & F_{2M}^{(a,n_{+})}\left(\beta_{2M}\right)^{n_{+}} \\
F_{1}^{(b,n_{+})}\left(\beta_{1}\right)^{n_{+}} & F_{2}^{(b,n_{+})}\left(\beta_{2}\right)^{n_{+}} & \cdots & F_{2M}^{(b,n_{+})}\left(\beta_{2M}\right)^{n_{+}} \\
G_{1}^{(a,1)}\left(\beta_{1}\right)^{L\!-\!(n_{-}\!-\!1)} & G_{2}^{(a,1)}\left(\beta_{2}\right)^{L\!-\!(n_{-}\!-\!1)} & \cdots & G_{2M}^{(a,1)}\left(\beta_{2M}\right)^{L\!-\!(n_{-}\!-\!1)} \\
G_{1}^{(b,1)}\left(\beta_{1}\right)^{L\!-\!(n_{-}\!-\!1)} & G_{2}^{(b,1)}\left(\beta_{2}\right)^{L\!-\!(n_{-}\!-\!1)} & \cdots & G_{2M}^{(b,1)}\left(\beta_{2M}\right)^{L\!-\!(n_{-}\!-\!1)} \\
\vdots & \vdots & \vdots & \vdots \\
G_{1}^{(a,n_{-})}\left(\beta_{1}\right)^{L} & G_{2}^{(a,n_{-})}\left(\beta_{2}\right)^{L} & \cdots & G_{2M}^{(a,n_{-})}\left(\beta_{2M}\right)^{L} \\
G_{1}^{(b,n_{-})}\left(\beta_{1}\right)^{L} & G_{2}^{(b,n_{-})}\left(\beta_{2}\right)^{L} & \cdots & G_{2M}^{(b,n_{-})}\left(\beta_{2M}\right)^{L} 
\end{vmatrix}=0,
	\label{eqapp2a_general}
\end{equation} 
\end{widetext} 
as derived in more detail in Appendix \ref{Appendix_I}. This determinant expression captures the constraints from OBCs at both boundaries. In general, it is a complicated expression, but can still be written explicitly in terms of $\beta_j$ and $E_{\rm OBC}$ for the two-band ansatz:
\begin{align}
F_{j}^{(a,i)}&= \sum_{n=-(i-1)}^{n_{+}}(h_{n}^{aa}+f_{j}h_{n}^{ab})\left(\beta_{j}\right)^{n} - E_{\rm OBC},\\
F_{j}^{(b,i)}&= \sum_{n=-(i-1)}^{n_{+}}(h_{n}^{ba}+f_{j}h_{n}^{bb})\left(\beta_{j}\right)^{n} - f_{j}E_{\rm OBC},\\
G_{j}^{(a,i)}&= \sum_{n=-n_{-}}^{n_{-}-i}(h_{n}^{aa}+f_{j}h_{n}^{ab})\left(\beta_{j}\right)^{n} - E_{\rm OBC},\\
G_{j}^{(b,i)}&= \sum_{n=-n_{-}}^{n_{-}-i}(h_{n}^{ba}+f_{j}h_{n}^{bb})\left(\beta_{j}\right)^{n} - f_{j}E_{\rm OBC},
\end{align}
where
\begin{align}
f_{j}=\frac{\phi^{(j)}_{\rm B}}{\phi_{\rm A}^{(j)}} &= \frac{E_{\rm OBC}-\sum_{n=-n_{-}}^{n_{+}}h_{n}^{aa}\left(\beta_{j}\right)^{n}}{\sum_{n=-n_{-}}^{n_{+}}h_{n}^{ab}\left(\beta_{j}\right)^{n}} \nonumber\\
&= \frac{\sum_{n=-n_{-}}^{n_{+}}h_{n}^{ba}\left(\beta_{j}\right)^{n}}{E_{\rm OBC}-\sum_{n=-n_{-}}^{n_{+}}h_{n}^{bb}\left(\beta_{j}\right)^{n}}.
\end{align} 

Equation~(\ref{eqapp2a_general}) can be rearranged in a compact multivariate polynomial form
\begin{align}
\sum_{P,Q}J(\beta_{i\in P},\beta_{j\in Q},E_{\rm OBC})\!\left[\prod_{i\in P}\left(\beta_{i}\right)^{k}\right]\! \left[\prod_{j\in Q}\left(\beta_{j}\right)^{k'}\right]\!=\!0,\label{eqapp3a_general} \nonumber\\
\end{align} where $k=1,\cdots,n_{+}$, $k'=L\!-\!(n_{-}\!-\!1),\cdots,L$, sets $P$ and $Q$ are two disjoint subsets of the set $\{1,2,\cdots,2M\}$ with $M$ elements, respectively, and $J(\beta_{i\in P},\beta_{j\in Q},E_{\rm OBC})$ is the $E_{\rm OBC}$-dependent coefficient corresponding to a particular permutation of $P$ and $Q$. By separating the product contributions of the $\beta$s which are exponentiated by $L$, we can extract out contributions that scale differently with $L$.

Furthermore, in the case $n_{+}=n_{-}$ where the maximal left and right hopping distances are the same, Eq.~(\ref{eqapp3a_general}) simplifies to
\begin{align}
\sum_{P,Q}J(\beta_{i\in P},\beta_{j\in Q},E_{\rm OBC})\!\left[\prod_{i\in P}\left(\beta_{i}\right)^{L+1}\right]\!=\!0.\label{eqapp4a_general}
\end{align} 
In the thermodynamic limit, the large $L$ in the exponents picks up the slowest decaying terms, and these would be the physically dominant contributions amidst the complicated jungle of terms. Specifically, in Eq.~(\ref{eqapp4a_general}), we find that there are two leading terms proportional to $(\beta_{M}\beta_{M+2}\beta_{M+3}\cdots\beta_{2M})^{L+1}$ and $(\beta_{M+1}\beta_{M+2}\beta_{M+3}\cdots\beta_{2M})^{L+1}$, which yield in the limit of large system size $L$
\begin{eqnarray}
	\left|\frac{\beta_{M}}{\beta_{M+1}}\right|
	\simeq\left|-\frac{J(\beta_{i\in P_1},\beta_{j\in Q_1},E_{\rm OBC})}{J(\beta_{i\in P_2},\beta_{j\in Q_2},E_{\rm OBC})}\right|_{E_{\rm OBC}=E_{\infty}}^{\frac{1}{L+1}},
	\label{eq:boundary_equation1_general}
\end{eqnarray} where $P_1=\{M+1,M+2,M+3\cdots,2M\}$, $Q_1=\{1,2,3,\cdots,M\}$, $P_2=\{M,M+2,M+3,\cdots,2M\}$, $Q_2=\{1,2,\cdots,M-2,M-1,M+1\}$, $M=n_++n_-$, and $L$ is the system size with $L\to\infty$. 
We emphasize that the form of this result Eq.~\eqref{eq:boundary_equation1_general} with $L\to\infty$ still holds for general higher-component or multi-band models [see Eq.~\eqref{eq:boundary_equation1_multi}, albeit with more complicated $J$ functions], as derived in Appendix \ref{Appendix_II}. The details are complicated, but physically, we expect qualitatively similar behavior because the critical NHSE essentially arises from the competition between the NHSE and the couplings, and with greater number of bands, we will have more avenues for the competition. But unless the model is fine tuned, we will generically still see the direct competition between pairs of bands, which thus reduces qualitatively to two-band behavior.

We comment on a few key takeways from Eq.~\eqref{eq:boundary_equation1_general}. Without any assumption on the detailed hoppings in the 1D tight-binding model, we showed how the requirement of satisfying OBCs at both ends generically lead to Eq.~\eqref{eq:boundary_equation1_general}, which relates $|\beta_M/\beta_{M+1}|^{L+1}$ with a combination of $L$-independent model parameters. It picks out the solutions $\beta_M$ and $\beta_{M+1}$ of Eq.~\eqref{eq:characteristic_general} as the dominant ones at large $L$, although in this regime, the $L$ dependence is also generally weak since the exponent $1/(L+1)$ changes slowly. Below, we discuss further on the large $L$ and moderate $L$ regimes separately.

In the thermodynamic limit of $L\to\infty$, the right hand side of Eq.~\eqref{eq:boundary_equation1_general} tends to unity, giving rise to the standard GBZ result $|\beta_{M}|=|\beta_{M+1}|$ discussed in \cite{lee2019anatomy,yokomizo2019non,yokomizo2021scaling,yang2020non,li2020critical,zhang2020correspondence,yao2018edge}. Hence, to draw the GBZ for $L\to\infty$, we uniformly vary the relative phase between $\beta_{M}$ and $\beta_{M+1}$, and trace out the trajectory ${\cal C}_{\beta}$ satisfying $|\beta_{M}|=|\beta_{M+1}|$. Since $L$ is large, each point in the GBZ curve is separated by a $2\pi/L$ phase interval that converges to a continuum, resulting in continuum complex energy bands.

For finite $L$ away from the thermodynamic limit, we emphasize that this standard GBZ construction for $E_\infty=\text{lim}_{L\to\infty}E_{\rm OBC}$ may no longer be valid. While in many cases, $E_{\rm OBC}$ does not change significantly as $L$ is extrapolated down to moderate [i.e., $L\sim \mathcal{O}(10)$], in cNHSE cases, the spectra and hence other physical properties vary strongly with system size. To characterize such cNHSE scenarios at finite $L$, we note from Eq.~\eqref{eq:boundary_equation1_general} that the magnitudes $|\beta_M|$ and $|\beta_{M+1}|$ can no longer be treated as equal. Physically, this implies that the OBC eigenstates are superpositions of different modes with inverse spatial decay lengths of either $-\ln|\beta_M|$ or $-\ln|\beta_{M+1}|$. As such, the effective cNHSE GBZ is described by \emph{both} $|\beta_M|$ and $|\beta_{M+1}|$, which are no longer equal. Contributions from other $\beta_{j}$ solutions affect the eigenstate decay rates negligibly even in the presence of cNHSE, as numerically verified for our illustrative coupled Hatano-Nelson model in Appendix \ref{Appendix_III}.

In the following two sections, we shall elaborate on how the pair of GBZ solutions $|\beta_M|$ and $|\beta_{M+1}|$ scale with system size $L$. Since the exact scaling dependencies can be highly complicated, we shall illustrate our results concretely through two paradigmatic cNHSE models, the minimal coupled Hatano-Nelson model with energy offset in section \ref{3} and a model with size-dependent topological states in section \ref{4}.

\section{Coupled Hatano-Nelson model with energy offset}\label{3}

\begin{figure}
    \includegraphics[width=0.45\linewidth]{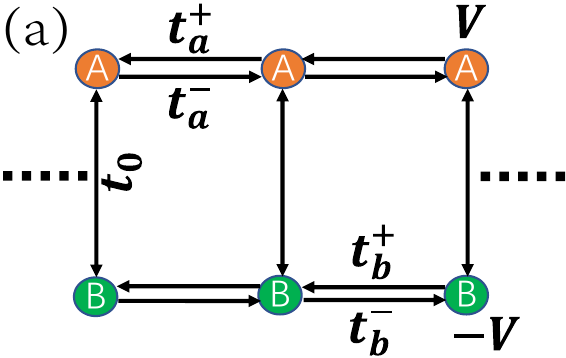}
	\includegraphics[width=1.1\linewidth]{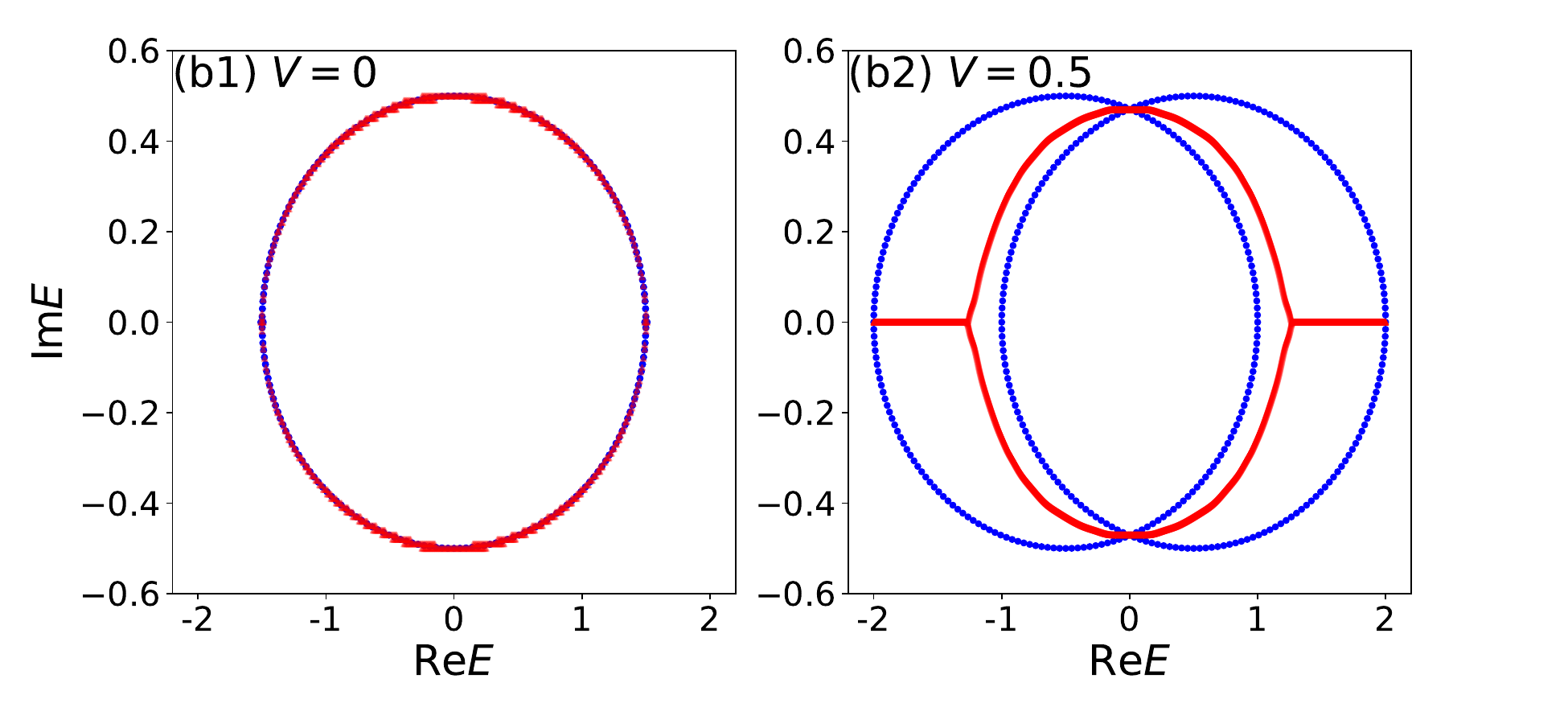}
	\caption{(a) Coupled Hatano-Nelson chain model [Eq.~(\ref{eq:Lee_Hr})] with inter-chain hopping $t_0$, intra-chain hopping asymmetries $t_{a}^{\pm}=t_1\pm \delta_a$ and $t_{b}^{\pm}=t_1\pm \delta_b$, and chain energy offsets $\pm V$. 
	(b) Energy spectra of Eq.~(\ref{eq:Lee_Hr}) under PBCs (blue) and OBC (red) in the $L\rightarrow \infty$ limit for (b1) $V=0$ and (b2) $V=0.5$. Parameters are $t_0=0.01$, $t_1=0.75$, and $\delta_a=-\delta_b=0.25$. While the PBC and OBC spectra coincide in the $V=0$ case studied in \cite{yokomizo2021scaling}, they deviate when $V\neq 0$, leading to broken bulk-boundary correspondence. }\label{fig:E_V}
\end{figure}

In this section, we elaborate on a cNHSE model formed by coupling the simplest possible NHSE chains -- two equal and oppositely oriented Hatano-Nelson chains. Going beyond the minimal model introduced in Ref.~\cite{yokomizo2021scaling}, which provided elegant analytic results, we additionally introduce on-site energy offsets $\pm V$ on the two chains, respectively, such that the inter-chain coupling now also faces nontrivial competition from the energetic separation of $2|V|$. The coupled chains are illustrated in Fig.~\ref{fig:E_V}(a), with each chain constituting one of the sublattices $A$ and $B$. In the basis $C_{{\bf k}}=\left(c_{{\bf k},A}^{},c_{{\bf k},B}^{}\right)^{T}$, its momentum-space Hamiltonian is
\begin{align}
{\cal H}(z)=\begin{pmatrix}
t_a^{+}z+t_a^{-}/z+V & t_0\\
t_0 & t_b^{+}z+t_b^{-}/z-V
\end{pmatrix},\label{eq:Lee_Hk} 
\end{align} where $t_a^{\pm}=t_1\pm\delta_a$, $t_b^{\pm}=t_1\pm\delta_b$, $t_0$ is the inter-chain hopping, and $\pm V$ is the on-site potential energy. We denote $z=e^{ik}$ as before, where $k$ is the momentum. It is related via Fourier transformation to the corresponding real-space tight-binding Hamiltonian 
\begin{eqnarray}
	H_{r}&=&\sum_{n}\left(t_{a}^{+}c_{n,{\rm A}}^\dag c_{n+1,{\rm A}}+t_{a}^{-}c_{n+1,{\rm A}}^\dag c_{n,{\rm A}}+ t_{0}c_{n,{\rm A}}^\dag c_{n,{\rm B}} \right. \nonumber\\
	&&\left.+t_{b}^{+}c_{n,{\rm B}}^\dag c_{n+1,{\rm B}}+t_{b}^{-}c_{n+1,{\rm B}}^\dag c_{n,{\rm B}}+ t_{0}c_{n,{\rm B}}^\dag c_{n,{\rm A}} \right. \nonumber\\
	&&\left.+ Vc_{n,{\rm A}}^\dag c_{n,{\rm A}} - Vc_{n,{\rm B}}^\dag c_{n,{\rm B}} \right),
	\label{eq:Lee_Hr}
\end{eqnarray} where $c_{n,\alpha}^{}$ ($c_{n,\alpha}^\dag$) is the annihilation (creation) operator on site $\alpha$ ($\alpha={\rm A},{\rm B}$) in unit cell $n$. Evidently, $t^+_a/t^-_a$ and $t^+_b/t^-_b$ are the hopping asymmetries of chains A and B. 

The energy eigenvalues of the Hamiltonian (\ref{eq:Lee_Hk}) under PBCs are given by
\begin{eqnarray}
	E_{\rm PBC}^{(\pm)}(k)&=&2t_{1}\cos k + i(\delta_a+\delta_b)\sin k \nonumber\\
	&\pm& \sqrt{[i(\delta_a-\delta_b)\sin k + V]^2+t_0^2},
\end{eqnarray}
where $k\in\mathbb{R}$ and $t_a^{\pm}=t_1\pm\delta_a$, $t_b^{\pm}=t_1\pm\delta_b$. In Fig.~\ref{fig:E_V}(b1), we see that in the large-$L$ limit, the PBC spectrum (blue) agrees well with the OBC spectrum (red) only in the $V=0$ case which Ref.~\cite{yokomizo2021scaling} has considered. When $V\neq 0$ [Fig.~\ref{fig:E_V}(b2)], the OBC spectrum lies in the interior of the PBC loops and can only agree with $E_{\rm PBC}^{(\pm)}(k)$ if we perform an appropriate complex momentum deformation $k\rightarrow k+ i \kappa(k)$~\cite{li2020critical,lee2019anatomy,yokomizo2019non,tai2022zoology,lee2020unraveling,yang2020non}. While it may appear here that the $V=0$ case does not experience BBC breaking (i.e., the NHSE), that is actually untrue once we consider finite system sizes~\cite{yokomizo2021scaling}.  Below, we show that this model exhibits cNHSE at finite system sizes for \emph{all} values of $V$, and compare some analytic approximations with numerical results.

\subsection{Finite-size scaling from the cNHSE}\label{3.1}

To understand how the PBC and OBC spectra differ beyond the thermodynamic limit shown in Figs.~\ref{fig:E_V}(b1) and \ref{fig:E_V}(b2), we examine the real-space Schr{\"o}dinger's equation ${\cal H}_{r}|\psi\rangle=E_{\rm OBC}|\psi\rangle$, where $|\psi\rangle=\left(\psi_{1,{\rm A}},\psi_{1,{\rm B}},\psi_{2,{\rm A}},\psi_{2,{\rm B}},\cdots,\psi_{n,{\rm A}},\psi_{n,{\rm B}},\cdots\right)^{\rm T}$:
\begin{eqnarray}
	\!\left\{ \begin{array}{l}
		\!t_{a}^{-}\psi_{n-1,{\rm A}}\!+\!t_{0}\psi_{n,{\rm B}}\!+\!t_{a}^{+}\psi_{n+1,{\rm A}}\!+\!V\psi_{n,{\rm A}}\!=\!E_{\rm OBC}\psi_{n,{\rm A}}, \vspace{7pt}\\
		\!t_{b}^{-}\psi_{n-1,{\rm B}}\!+\!t_{0}\psi_{n,{\rm A}}\!+\!t_{b}^{+}\psi_{n+1,{\rm B}}\!-\!V\psi_{n,{\rm B}}\!=\!E_{\rm OBC}\psi_{n,{\rm B}},
	\end{array}\right.
	\label{eq:OBC_Schrodinger}
\end{eqnarray}  where ${\cal H}_{r}$ is the Hamiltonian matrix of $H_{r}$ in the basis $(C_{1},C_{2},\cdots,C_{j},\cdots)^T$.
Based on the approach developed in the section \ref{2}, we can use as an eigenstate ansatz which is a linear combination of $\beta$ solutions, such as to solve the real-space Schr{\"o}dinger equation ~\cite{yokomizo2021scaling,yao2018edge,deng2019non}:
\begin{eqnarray}
	\left( \begin{array}{c}
		\psi_{n,{\rm A}} \vspace{5pt}\\
		\psi_{n,{\rm B}}
	\end{array}\right)=\sum_{j=1}^4(\beta_j)^n\left( \begin{array}{c}
		\phi_{\rm A}^{\left(j\right)} \vspace{5pt}\\
		\phi_{\rm B}^{\left(j\right)}
	\end{array}\right).
\end{eqnarray}
This allows us to rewrite Eq.~(\ref{eq:OBC_Schrodinger}) as
\begin{eqnarray}
	\!\left(\! \begin{array}{cc}
		\!t_{a}^{+}\beta\!+\!t_{a}^{-}\beta^{-1}\!+\!V\! & t_{0}                 \\
		t_{0}                 & \!t_{b}^{+}\beta\!+\!t_{b}^{-}\beta^{-1}\!-\!V\!
	\end{array}\!\right)\!\!\left(\! \begin{array}{c}
		\!\phi_{\rm A}\! \\
		\!\phi_{\rm B}\!
	\end{array}\!\right)\!
	\!=\!E_{\rm OBC}\!\left(\! \begin{array}{c}
		\!\phi_{\rm A}\! \\
		\!\phi_{\rm B}\!
	\end{array}\!\right)\!, \label{eq:bulk_eigenequation} \nonumber\\
\end{eqnarray}
where we have written $\beta_j=\beta$ and $\phi_\alpha^{\left(j\right)}=\phi_\alpha~(\alpha={\rm A},{\rm B})$ for notational simplicity, since Eq.~\eqref{eq:bulk_eigenequation} applies separately to different $j$. Essentially, this ansatz has allowed us to replace $z=e^{ik}$ by $\beta$. Nontrivial solutions to Eq.~\eqref{eq:bulk_eigenequation} satisfy the bulk characteristic dispersion equation 
\begin{eqnarray}
	&&t_{a}^{+}t_{b}^{+}\beta^2-[(t_{a}^{+}+t_{b}^{+})E_{\rm OBC}+(t_{a}^{+}-t_{b}^{+})V ]\beta\nonumber\\ 
	&\!+\!& \left(t_{a}^{+}t_{b}^{-}+t_{a}^{-}t_{b}^{+}+E_{\rm OBC}^2-t_{0}^2-V^2\right) \nonumber\\
	&\!-\!&[(t_{a}^{-}+t_{b}^{-})E_{\rm OBC}+(t_{a}^{-}-t_{b}^{-})V ]\beta^{-1} 
	\!+\!t_{a}^{-}t_{b}^{-}\beta^{-2}\!=\!0. \label{eq:characteristic}\nonumber\\
\end{eqnarray}
For each value of $E_{\rm OBC}$, there are four solutions $\beta=\beta_j$, $j=1,2,3,4$, since the maximal and minimal powers of $\beta$ are $n_+=n_-=1$. 

\subsubsection{Finite-size scaling of the OBC spectra}

\begin{figure}[h]
	\includegraphics[width=1\linewidth]{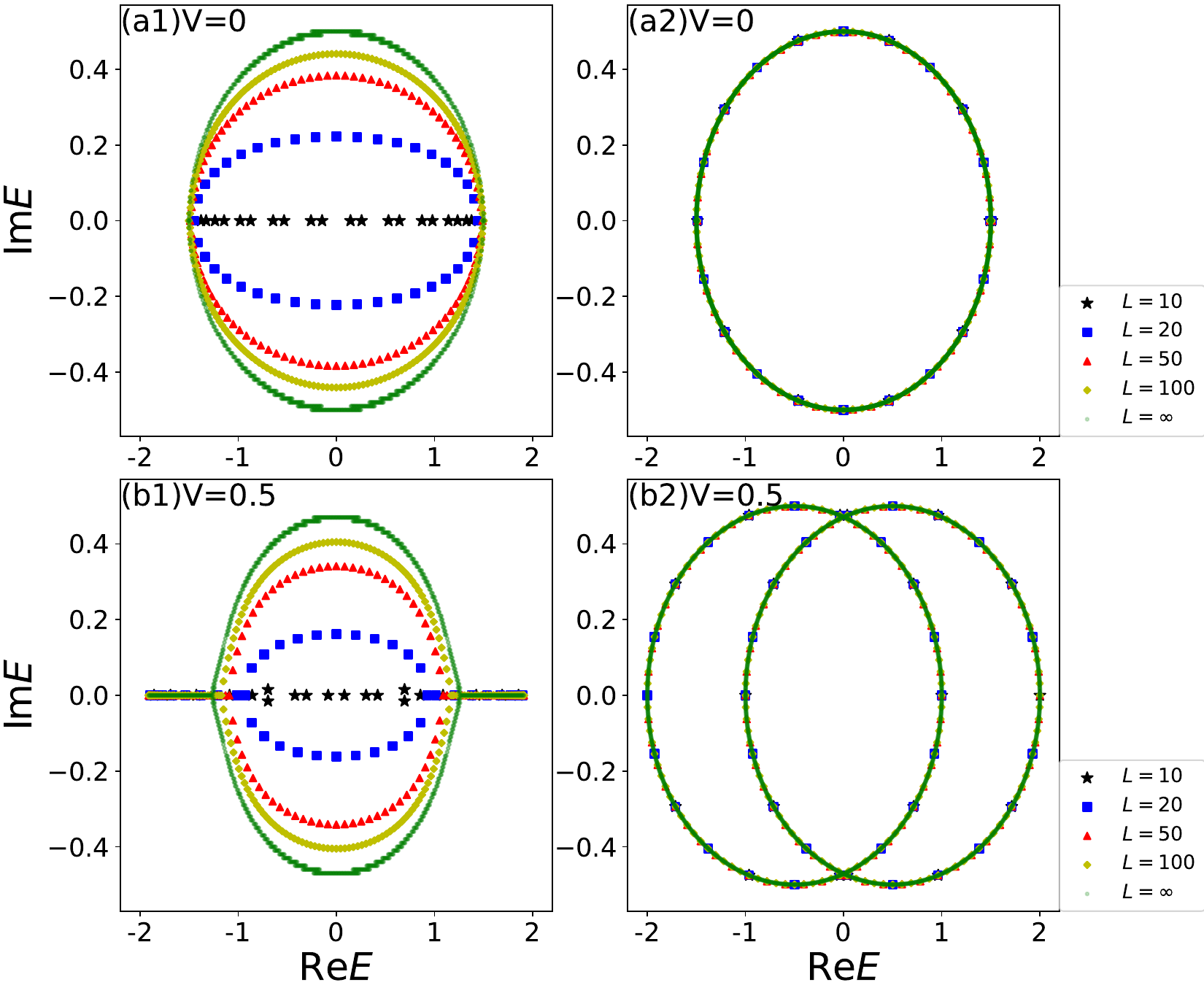}
	\caption{OBC energy spectra of the coupled Hatano-Nelson model Hamiltonian (\ref{eq:Lee_Hr}) with (a1) $V=0$ and (b1) $V=0.5$ at different finite system sizes $L=10$ (black), $20$ (blue), $50$ (red), $100$ (yellow), $\infty$ (green). When $V=0$, the spectrum is real for short chains, but complex for long chains due to the strong effective couplings from large $L$. But interestingly for $V\neq 0$, short chains can possess some complex energies, and long chains possess some real energies. PBC energy spectra of the coupled Hatano-Nelson model Hamiltonian (\ref{eq:Lee_Hr}) with (a2) $V=0$ and (b2) $V=0.5$ at different finite system sizes $L=10$ (black), $20$ (blue), $50$ (red), $100$ (yellow), $\infty$ (green). Parameters are $t_0=0.01$, $t_1=0.75$, and $\delta_a=-\delta_b=0.25$, the same as those in Fig.~\ref{fig:E_V}.}\label{fig:E_OBC_V05_L_together1}
\end{figure}
To understand the OBC spectrum $E_{\rm OBC}$ in terms of non-Bloch theory, we need to obtain its effective GBZ. For finite $L$, the GBZ comprises the two dominant $\beta$ solutions such that $E_\text{PBC}(-i\log \beta)$ numerically coincides with $E_{\rm OBC}$. The numerically computed $E_{\rm OBC}$ is shown in Fig.~\ref{fig:E_OBC_V05_L_together1} for both (a) $V=0$ and (b) $V=0.5$. Evidently, the OBC spectra in both cases depends strongly on $L$, being real for small $L$ (i.e., $L=10$), and gradually morphing into the large $L$ spectrum previously shown in Fig.~\ref{fig:E_V}. Physically, the spectrum remains real when the couplings (here with small bare values $t_0=0.01$) are strong enough for the directed amplifications from both chains to cancel~\footnote{Thus directed amplification provides an alternative mechanism for achieving real non-Hermitian spectra~\cite{yang2022designing,zhang2022real}, unrelated to PT symmetry.}; as the system gets larger, the cNHSE becomes exponentially stronger and the couplings serve to ``close up''~\cite{lee2020ultrafast,liu2020helical,zhang2022observation} the amplification loops, causing unchecked amplification that corresponds to complex energies. For $V\neq 0$, some eigenenergies can remain real even at arbitrarily large system sizes presumably because the potential offsets obstruct unchecked amplification.

\subsubsection{From OBC spectra to size-dependent cNHSE GBZs}

\begin{figure}[h]
\includegraphics[width=1\linewidth]{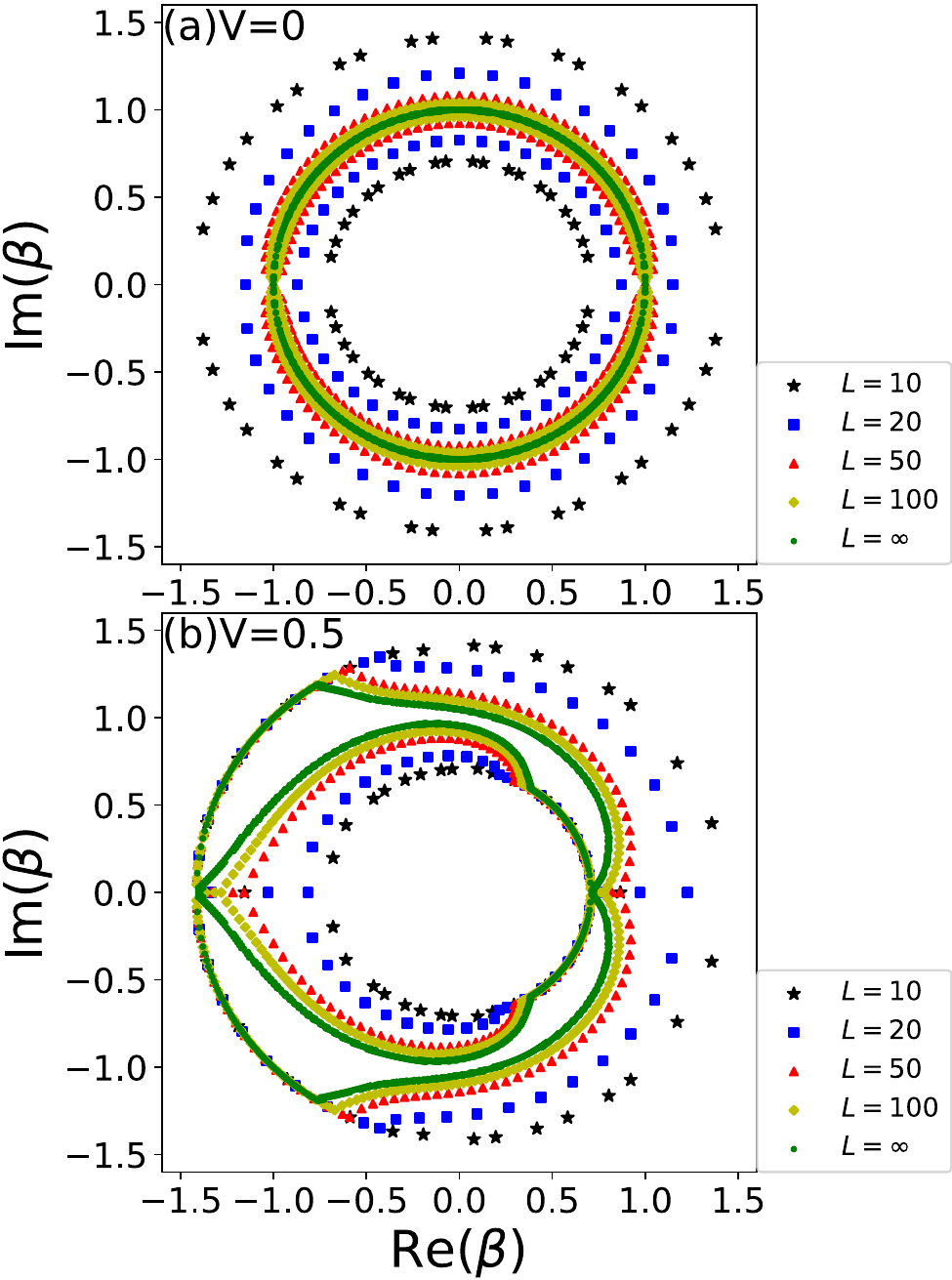}
\caption{GBZ of the coupled Hatano-Nelson model Hamiltonian (\ref{eq:Lee_Hr}) at different finite system sizes $L=10$ (black), $20$ (blue), $50$ (red), $100$ (yellow), $\infty$ (green) for (a) $V=0$ and (b) $V=0.5$. At finite $L$, the GBZ is given by solutions $\beta_M=\beta_2$ and $\beta_{M+1}=\beta_3$; as $L\rightarrow \infty$, the $\beta_2$ and $\beta_3$ loops converge towards the standard GBZ solution $|\beta_2|=|\beta_3|$. Note that this standard GBZ can consist of two loops (as in (b)), since this is a two-band model. Parameters are $t_0=0.01$, $t_1=0.75$, and $\delta_a=-\delta_b=0.25$, the same as those in Fig.~\ref{fig:E_V}. 
}
\label{fig:GBZ_1}
\end{figure}

While the size-dependent spectra in Fig.~\ref{fig:E_OBC_V05_L_together1} unambiguously signify the presence of cNHSE, size-dependencies in the spectra are model-specific. Key to more fundamental understanding of cNHSE scaling is the scaling behavior of the GBZ~\footnote{Although the GBZ also depends on the model, at least it remains invariant across models related by conformal transforms in the complex $E_{\rm OBC}$ plane~\cite{lee2020unraveling,tai2022zoology}.}. To compute the GBZ, we substitute the OBC energies into the characteristic equation (\ref{eq:characteristic}) and obtain the $\beta$ solutions. Here, for each $E_{\rm OBC}$ point, we have four solutions $\left|\beta_1\right|\leqslant\left|\beta_2\right|\leqslant\left|\beta_3\right|\leqslant\left|\beta_4\right|$ and the GBZ is given by the two solutions $\beta_M=\beta_2$ and $\beta_{M+1}=\beta_3$. Figure~\ref{fig:GBZ_1} shows the GBZ computed at various finite system sizes $L=10$, $20$, $50$, $100$; the $L=\infty$ case (green) is plotted by solving Eq.~\eqref{eq:characteristic} with the standard condition $|\beta_2|=|\beta_3|$ (i.e., intersecting $\beta_2$ and $\beta_3$ solution curves)~\cite{yokomizo2019non,kawabata2020non,yokomizo2020non,yi2020non,yokomizo2020topological,yang2020non,yokomizo2021non,deng2019non} valid in the thermodynamic limit. 

For the finite-size cases under $V=0$ shown in Fig.~\ref{fig:GBZ_1}(a), there are two loops in the Re($\beta$)-Im($\beta$) plane for each value of $L$, corresponding to the $\beta_2$ and $\beta_3$ solutions. As the system size $L$ increases to infinity, they converge towards each other, as expected from the condition $|\beta_2|=|\beta_3|$. 
Similarly, for the $V\neq0$ case in Fig.~\ref{fig:GBZ_1}(b), the two loops in the Re($\beta$)-Im($\beta$) plane get closer and closer to each other as the system size $L$ increases. However, in this case, they do not converge into one single loop because the GBZ solution $|\beta_2|=|\beta_3|$ itself consists of two loops [green in Fig.~\ref{fig:GBZ_1}(b)]. Here the GBZ solutions are also highly anisotropic in the wave number $\text{arg}(\beta)$, exhibiting cusps at $\beta$ corresponding to branch points in the spectrum~\cite{lee2020unraveling,tai2022zoology}.

\subsubsection{Finite scaling behavior of the GBZ}
Having numerically seen how the GBZ varies with system size, we now rigorously derive the scaling rules governing it. To do so, we examine the OBC constraints in detail. As elaborated in Appendix \ref{Appendix_IV}, imposing open boundaries at $x=1$ and $x=L$, i.e., $\psi_{0,\alpha}=\psi_{L+1,\alpha}=0$ gives rise to the condition
\begin{eqnarray}
	&&~~X_{1,4}X_{2,3}\left[\left(\beta_1\beta_4\right)^{L+1}+\left(\beta_2\beta_3\right)^{L+1}\right] \nonumber\\
	&&-X_{1,3}X_{2,4}\left[\left(\beta_1\beta_3\right)^{L+1}+\left(\beta_2\beta_4\right)^{L+1}\right] \nonumber\\
	&&+X_{1,2}X_{3,4}\left[\left(\beta_1\beta_2\right)^{L+1}+\left(\beta_3\beta_4\right)^{L+1}\right]=0,
	\label{eq:boundary_equation0}
\end{eqnarray}
where $X_{i,j}$ are defined as
\begin{equation}
	X_{i,j}\equiv t_{a}^{+}(\beta_j-\beta_i)+t_{a}^{-}(\beta_j^{-1}-\beta_i^{-1})
	\label{eq:Xij}
\end{equation}
with $i,j=1,2,3,4$. This result is equivalent to Eq.~\eqref{eqapp2a_general}, but specialized to our coupled Hatano-Nelson model Hamiltonian. Interestingly, it is independent of $V$ and $E_\text{OBC}$, even though they both definitely affect the values of $\beta_j$, since the individual $\beta_j$ solutions are determined by the characteristic dispersion equation~\eqref{eq:characteristic}. When $L$ is varied, the $\beta_j$ solutions of Eq.~\eqref{eq:characteristic} vary since $E_{\rm OBC}$ changes with $L$. How exactly $E_\text{\rm OBC}$ can change is indirectly constrained by Eq.~\eqref{eq:boundary_equation0}, which imposes a $L$-dependent relation between the $\beta_j$ solutions corresponding to the value of $E_{\rm OBC}$. 

\begin{figure}[h!]
	\includegraphics[width=1\linewidth]{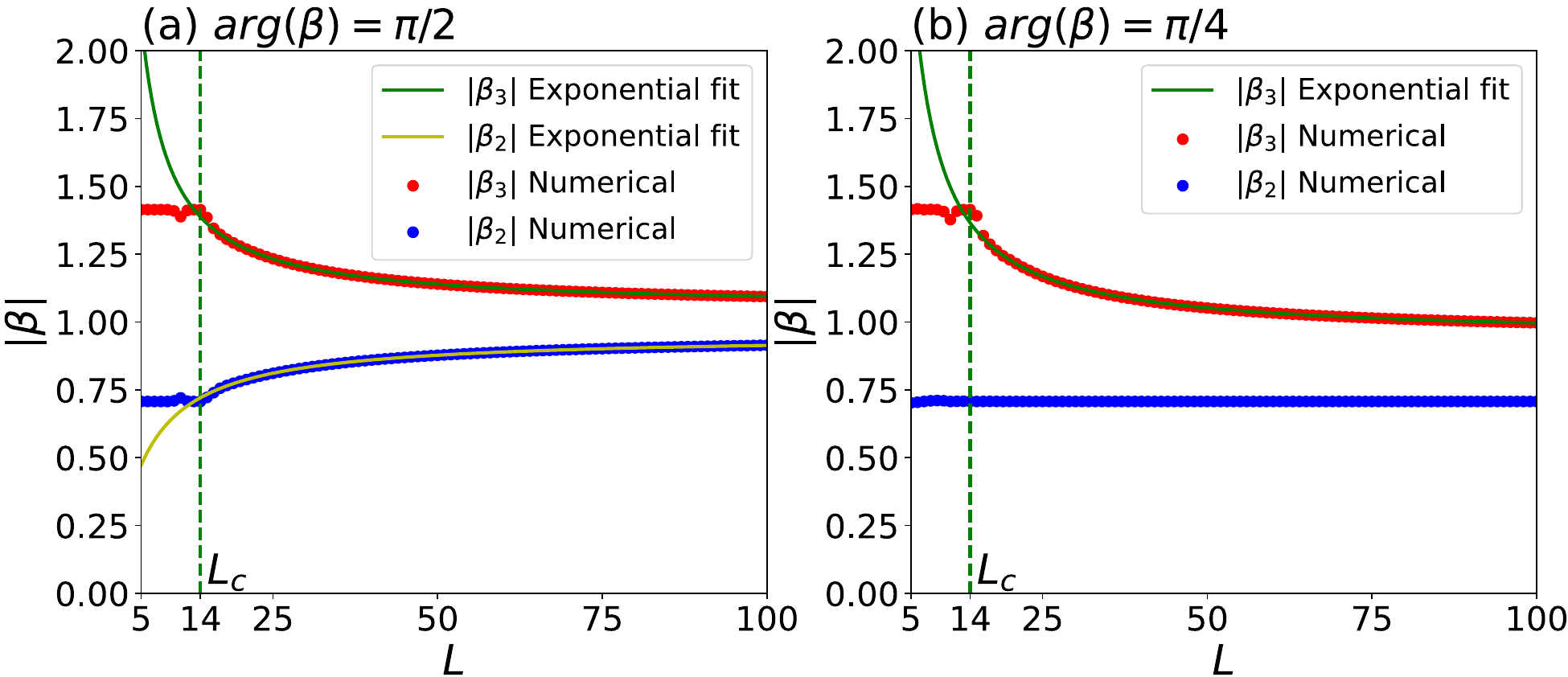}
	\caption{The GBZ radii $|\beta_2|$ and $|\beta_3|$ of our coupled Hatano-Nelson model Hamiltonian (\ref{eq:Lee_Hr}) with $V=0.5$, plotted against the system size $L$ for (a) $\arg(\beta)=\frac{\pi}{2}$ and (b) $\arg(\beta)=\frac{\pi}{4}$. Results obtained from the numerical OBC spectra exhibit excellent fitting with the exponential scaling of Eq.~(\ref{eq:beta3}), with fitted parameters $a(\arg(\beta)=\frac{\pi}{2})\approx0.050$, $b(\arg(\beta)=\frac{\pi}{2})\approx78.65$, $a(\arg(\beta)=\frac{\pi}{4})\approx-0.057$, and $b(\arg(\beta)=\frac{\pi}{4})\approx198.11$. The scaling is frozen below the lower critical length $L_{c}\approx14$, limited by the bare asymmetric couplings $t_a^\pm$ and $t_b^\pm$. Parameters are $t_0=0.01$, $t_1=0.75$, and $\delta_a=-\delta_b=0.25$, the same as those in Fig.~\ref{fig:E_V}. }
	\label{fig:beta_L_1}
\end{figure}

To make progress in deriving the finite-size scaling properties of the $\beta$s, our strategy is to consider the large-$L$ limit and obtain the leading-order scaling behavior. In this limit, we can approximate the boundary equation~(\ref{eq:boundary_equation0}) by retaining only the two dominant terms $-X_{1,3}X_{2,4}\left(\beta_2\beta_4\right)^{L+1}$ and $X_{1,2}X_{3,4}\left(\beta_3\beta_4\right)^{L+1}$. To make further headway, we note that the cNHSE is already well manifested when the bare value of the coupling $t_0$ is very small, i.e., $t_0=0.01$ as in Fig.~\ref{fig:E_V}. (In fact, if $t_0$ is of the same order as the two Hatano-Nelson chains, it would be difficult to see the weak coupling/small-$L$ limit with real spectra.) As such, we can expand up to the second order of the coupling parameter $t_{0}$ (see Appendix \ref{Appendix_V}) to obtain
\begin{eqnarray}
\!\left|\frac{\beta_2}{\beta_3}\right|
&\!\simeq\!&\left|\frac{X_{1,2}X_{3,4}}{X_{1,3}X_{2,4}}\right|^{\frac{1}{L+1}} \nonumber\\
&\!\approx\!&\left|(t_{a}^{+}t_{b}^{-} \!-\! t_{a}^{-}t_{b}^{+})^{2}\!f_{\infty}(E_{\infty})t_{0}^{2}\right|^{\frac{1}{L+1}},\!
\label{eq:boundary_equation1}
\end{eqnarray} where $E_{\infty}\equiv\lim_{L\to\infty}E_{\rm OBC}$, and
\begin{eqnarray}
f_{\infty}(E_{\infty})=
\frac{\sqrt{(E_{\infty}-V)^{2} - 4t_{a}^{+}t_{a}^{-}}  
\sqrt{(E_{\infty}+V)^{2} - 4t_{b}^{+}t_{b}^{-} }}{h^{2}(E_{\infty}) },\notag\\
\end{eqnarray} 
$h(E_{\infty})=E_{\infty}^{2}(t_{a}^{-}-t_{b}^{-})(t_{a}^{+}-t_{b}^{+}) + (t_{a}^{+}t_{b}^{-} - t_{a}^{-}t_{b}^{+})^{2} + 2E_{\infty}(t_{a}^{+}t_{a}^{-} - t_{b}^{+}t_{b}^{-})V + (t_{a}^{+}+t_{b}^{+})(t_{a}^{-}+t_{b}^{-})V^{2}$. Notice that $E_{\infty}$ in Eq.~\eqref{eq:boundary_equation1} depends on $t_{0}$. Equation~\eqref{eq:boundary_equation1} is Eq.~\eqref{eq:boundary_equation1_general} specialized to our coupled Hatano-Nelson model Hamiltonian [Eq.~\eqref{eq:Lee_Hr}]. It expresses the ratio of the GBZ quantities $|\beta_2|$ and $|\beta_3|$ as a constant exponentiated by $1/(L+1)$, which is a scaling behavior that is universal across cNHSE models.

\begin{figure}
\includegraphics[width=1\linewidth]{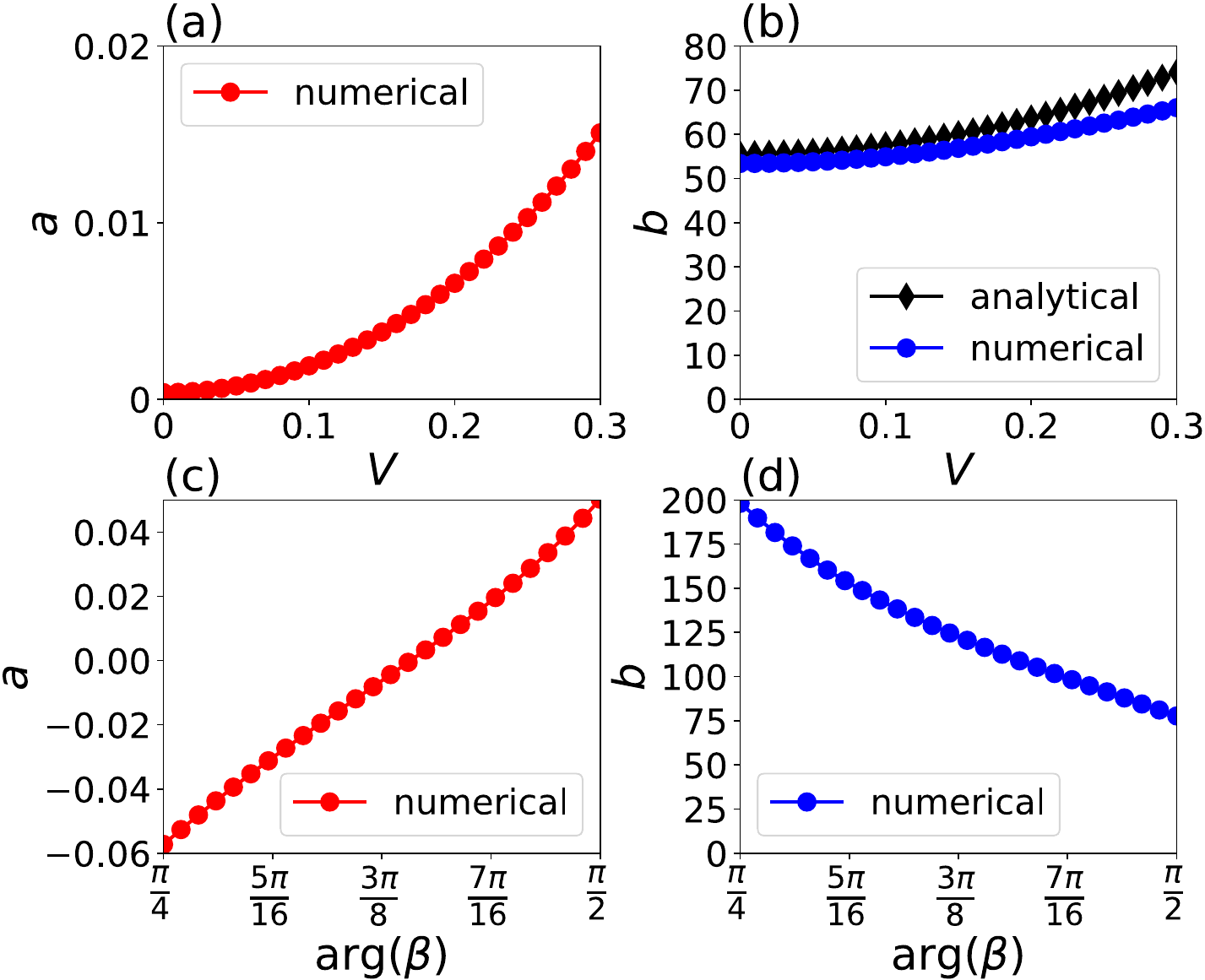}
\caption{(a), (b) Exponential scaling parameters $a$ and $b$ [Eq.~(\ref{eq:beta3})] of $|\beta_3|$ as a function of $V$ at $\arg(\beta)=\frac{\pi}{2}$. Their numerical values are extracted from the plot of $|\beta_3|$ against $L$, which is computed from the numerical $E_{\rm OBC}$ data. In (b), this numerically obtained $b$ is shown to be well predicted from the model parameters through the analytic result Eq.~\eqref{eq:b}, which is derived under the small $V$ approximation. (c), (d) Show the numerically obtained $a$ and $b$ as a function of $\arg(\beta)$, at fixed $V=0.5$. Parameters are $t_0=0.01$, $t_1=0.75$, and $\delta_a=-\delta_b=0.25$, the same as those in Fig.~\ref{fig:E_V}.}
\label{fig:a_b_V_arg}
\end{figure}

While the $1/(L+1)$ exponential scaling behavior holds generally for the ratio $ |\beta_M/\beta_{M+1}|$, it can apply to $|\beta_M|$ or $|\beta_{M+1}|$ individually if they are related in special ways. In Fig.~\ref{fig:beta_L_1}, we show the numerically extracted $|\beta_{2}|$ and $|\beta_{3}|$ at two special values of $\text{arg}(\beta)$, where $|\beta_2|\approx 1/|\beta_3|$ in Fig.~\ref{fig:beta_L_1}(a) and $|\beta_2|$ is constant in Fig.~\ref{fig:beta_L_1}(b). As such, $|\beta_2/\beta_3|\approx|\beta_2|^2\approx|\beta_3|^{-2}$ in Fig.~\ref{fig:beta_L_1}(a) and $|\beta_2/\beta_3|\propto |\beta_3|^{-1}$ in Fig.~\ref{fig:beta_L_1}(b), hence allowing for $|\beta_{3}|$ to be fitted to an exponential form
\begin{eqnarray}\label{eq:beta3}
|\beta_{3}|=a+b^{\frac{1}{L+1}},
\end{eqnarray} 
where the parameters $a,b\in\mathbb{R}$, $b>0$, and $|a|\ll1\ll|b|$. In general, this exponential relation fits the numerically obtained $|\beta|$s very well for sufficiently large $L$, as demonstrated in Fig.~\ref{fig:beta_L_1}. The actual values of fitting parameters $a$ and $b$ are shown in Fig.~\ref{fig:a_b_V_arg} as functions of the on-site energy $V$ [Figs.~\ref{fig:a_b_V_arg}(a) and \ref{fig:a_b_V_arg}(b)] and $\arg(\beta)$ [Figs.~\ref{fig:a_b_V_arg}(c) and \ref{fig:a_b_V_arg}(d)]. It is found that both $a$ and $b$ are monotonically increasing functions of the on-site energy $V$ at $\arg(\beta)=\frac{\pi}{2}$ [Figs.~\ref{fig:a_b_V_arg}(a) and \ref{fig:a_b_V_arg}(b)]. Also, in the range of $\arg(\beta)\in[\frac{\pi}{4},~\frac{\pi}{2}]$ for $V=0.5$, $a$ is a monotonically increasing function of $\arg(\beta)$, but $b$ is a monotonically decreasing function of $\arg(\beta)$. We see that the condition $|a|\ll|b|$ is always satisfied with different on-site energy $V$ and $\arg(\beta)$.

The correctness of our exponential fit can be checked by comparing against analytic results involving the model parameters. From Eq.~\eqref{eq:boundary_equation1}, we see that in the case of $|\beta_2|\approx 1/|\beta_3|$, the parameter $b$ in the exponential scaling relation is approximately given by 
\begin{eqnarray}\label{eq:b}
b\approx\left|(t_{a}^{+}t_{b}^{-} \!-\! t_{a}^{-}t_{b}^{+})^{2}\!f_{\infty}(E_{\infty})t_{0}^{2}\right|^{-1/2}.
\end{eqnarray}
As shown in Fig.~\ref{fig:a_b_V_arg}(b), both the analytical and numerical results agree well with each other when the on-site energy $V$ is smaller than $0.2$, where the $|\beta_2|\approx 1/|\beta_3|$ approximation accurately holds. For different fixed $\text{arg}(\beta)$, $E_\text{OBC}$ would be different, leading to different values of $b$. Indeed, as evident in Fig.~\ref{fig:GBZ_1}(b), the convergence behavior of $|\beta|$ and hence $b$ varies significantly with $\text{arg}(\beta)$ [Fig.~\ref{fig:a_b_V_arg}(d)].

\subsection{Lower critical system size for the cNHSE}\label{3.2}

\begin{figure}
\includegraphics[width=1\linewidth]{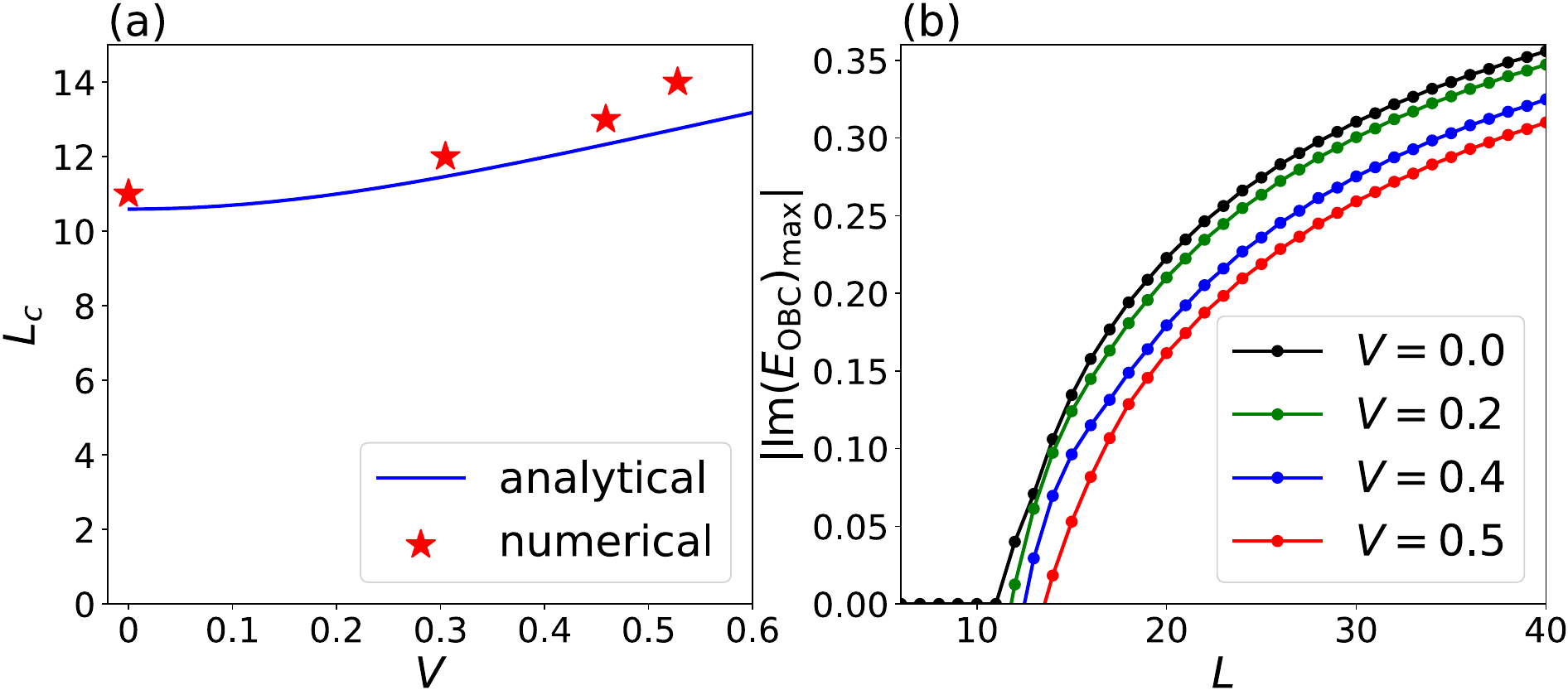}
\caption{(a) Critical system size $L_c$ versus $V$ at $\arg(E_{\rm OBC})=\frac{\pi}{2}$. The analytical result given by Eq.~(\ref{eq:Lc}) (blue) agrees reasonably well with numerical results (red stars) estimated by the threshold system size $L=L_c$, below which the spectrum is unaffected by $L$. We observe that $L_c$ increases with $V$, confirming the intuition that the inter-chain energy offset $V$ obstructs critical NHSE hybridization. (b) Absolute value of the maximal imaginary part of the eigenvalues $|{\rm Im}(E_{\rm OBC})_{\rm max}|$ as a function of the system size $L$, also at $\arg(E_{\rm OBC})=\frac{\pi}{2}$. The onset of complex $E_\text{OBC}$ typically occurs at $L_c$, except for small systems ($L=10$), where the nonzero $V$ offset can give rise to complex energies [see Fig.~\ref{fig:E_OBC_V05_L_together1}(b)]. Parameters are $t_0=0.01$, $t_1=0.75$, and $\delta_a=-\delta_b=0.25$, the same as those in Fig.~\ref{fig:E_V}.}\label{fig:Lc}
\end{figure}

As seen in Fig.~\ref{fig:beta_L_1}, the scaling of the GBZ parameters $|\beta_{M}|$ and $|\beta_{M+1}|$ ($M=2$ here) is only exponential and described by Eq.~(\ref{eq:beta3}) above a certain lower critical system size $L_c$. Below that, they remain effectively constant, indicative of the absence of the cNHSE. The reason is that the spatial skin decay lengths $-1/\ln|\beta_{M}|$ and $-1/\ln|\beta_{M+1}|$ cannot be faster than that of the physical NHSE chains in the cNHSE model. In our model, noting that $|\beta_{3}|>|\beta_{2}|$, we must have $|\beta_{3c}|=\sqrt{t_{a}^{+}/t_{a}^{-}}$ and $|\beta_{2c}|=\sqrt{t_{b}^{+}/t_{b}^{-}}$, corresponding to the $|\beta|$s of the individual Hatano-Nelson chains. 

Substituting $|\beta_2/\beta_3|$ with $|\beta_{2c}/\beta_{3c}|$ in Eq.~(\ref{eq:boundary_equation1}), we obtain the critical system size $L_c$ of our coupled Hatano-Nelson model as 
\begin{equation}\label{eq:Lc}
L_c\!\approx\!\frac{2\ln\left|(t_{a}^{+}t_{b}^{-} - t_{a}^{-}t_{b}^{+})^{2}f_{\infty}(E_{\infty})t_0^2\right|}{\ln\left|t^+_bt_a^-/(t^-_bt_a^+)\right|}-1.
\end{equation}  
As shown in Fig.~\ref{fig:Lc}(a) for $\arg(E_{\rm OBC})=\frac{\pi}{2}$, this analytic expression [Eq.~\eqref{eq:Lc}] for $L_c$ (blue curve) agrees rather well with its numerical determination (red stars), i.e., from plots such as Fig.~\ref{fig:beta_L_1}. Not surprisingly, it increases monotonically with the inter-chain energy offset $V$, since the offset impedes energy matching and acts as an obstruction to the critical coupling between the Hatano-Nelson chains. 

$L_c$ can also be thought of as the lower critical length above which the inter-chain coupling $t_0$ is ``switched on'' to cause the cNHSE. As seen in Fig.~\ref{fig:Lc}(b), the energy spectrum becomes complex precisely above $L_c$. Since our OBC Hatano-Nelson chains have real spectra when uncoupled, it means that they become effective coupled only when $L\geq L_c$. Naively, we would expect small $t_0$ to continuously give rise to small imaginary energies; yet, in reality, there exists a sharp real-to-complex spectral transition~\cite{yokomizo2021scaling,li2020critical} controlled by $L_c$. We note that $L_c\rightarrow \infty$ as $t_0\rightarrow 0$, consistent with the expectation that uncoupled chains will never experience the cNHSE.

\section{Topologically coupled chain model}\label{4}

To complement the exposition of our coupled Hatano-Nelson cNHSE model above, we next consider more sophisticated inter-chain couplings which lead to size-controlled topological states, as first designed in~\cite{li2020critical}. In the basis $C_{{\bf k}}=\left(c_{{\bf k},A}^{},c_{{\bf k},B}^{}\right)^{T}$, it is given by
\begin{align}
{\cal H}_\text{top}(z)=\left(\begin{matrix}
t_{a}^{+}z+t_{a}^{-}/z+V & \delta_{ab}(z+1/z) \\
-\delta_{ab}(z+1/z) & t_{b}^{+}z+t_{b}^{-}/z-V
\end{matrix}\right), \label{eq:Lee_Hk_topo}
\end{align} where $t_a^{\pm}=t_1\pm\delta_a$, $t_b^{\pm}=t_1\pm\delta_b$, and $z=e^{ik}$. Here, the simple inter-chain couplings $t_0$ of the coupled Hatano-Nelson model are replaced by criss-crossing inter-chain couplings $\pm \delta_{ab}$ which can potentially introduce topological flux~\cite{lee2014lattice}.
\begin{figure}[h]
	\includegraphics[width=0.5\linewidth]{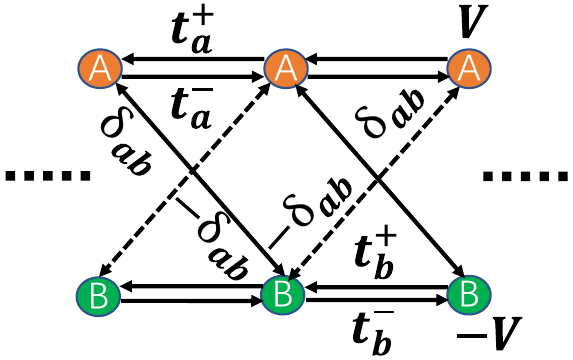}
	\caption{Topologically coupled two-chain model [Eq.~(\ref{eq:Lee_Hr_topo})] with criss-crossing inter-chain non-reciprocal couplings $\pm\delta_{ab}$ and asymmetric hoppings $t_{a}^{\pm}=t_1\pm \delta_a$ and $t_{b}^{\pm}=t_1\pm\delta_b$ in chains A and B. The chains are given energy offsets of $\pm V$. 
	}
	\label{fig:model_topo}
\end{figure}
\begin{figure}[h]
	\includegraphics[width=.8\linewidth]{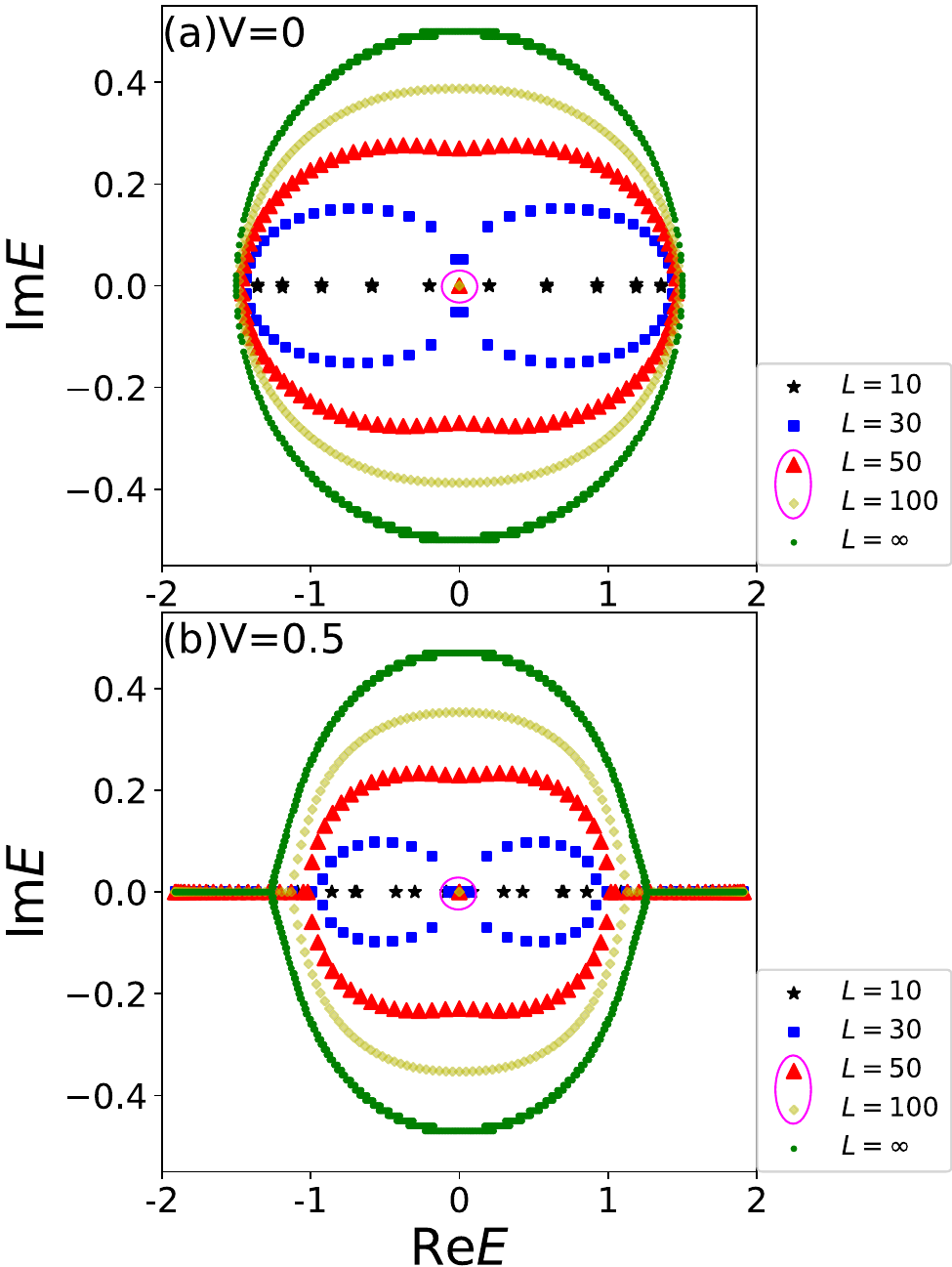}
	\caption{OBC energy spectra of the topologically coupled chain model Hamiltonian (\ref{eq:Lee_Hr_topo}) with (a) $V=0$ and (b) $V=0.5$ at different system sizes $L=10$ (black), $30$ (blue), $50$ (red), $100$ (yellow), $\infty$ (green). Notably, topological zero modes (circled) appear at $E=0$ in the point gap only at sufficiently large system sizes of $L=50,100$. The other parameters are $\delta_{ab}=0.5\times10^{-3}$, $t_1=0.75$, and $\delta_a=-\delta_b=0.25$. }
	\label{fig:E_OBC_V05_L_together2}
\end{figure}
Under PBCs, the energy eigenvalues can be simply obtained from the Hamiltonian (\ref{eq:Lee_Hk_topo}) as
\begin{eqnarray}
E_{\rm PBC}^{(\pm)}(k)&\!=\!&2t_{1}\cos k + i(\delta_a+\delta_b)\sin k \nonumber\\
&\!\pm\!& \sqrt{[i(\delta_a-\delta_b)\sin k \!+\! V]^2 \!-\! 4\delta_{ab}^{2}\cos^{2} k}. \label{eq:E_PBC_topo}
\end{eqnarray}
with $k=-i\ln z\in\mathbb{R}$ and $t_a^{\pm}=t_1\pm\delta_a$, $t_b^{\pm}=t_1\pm\delta_b$. By Fourier transformation, one obtains the real-space tight-binding Hamiltonian (Fig.~\ref{fig:model_topo}):
\begin{align}
H_\text{t}&=\sum_{n}\left(t_{a}^{+}c_{n,{\rm A}}^\dag c_{n+1,{\rm A}}+t_{a}^{-}c_{n+1,{\rm A}}^\dag c_{n,{\rm A}} + \delta_{ab}c_{n,{\rm A}}^\dag c_{n+1,{\rm B}} \right. \nonumber\\
&\left.- \delta_{ab}c_{n+1,{\rm B}}^\dag c_{n,{\rm A}} + t_{b}^{+}c_{n,{\rm B}}^\dag c_{n+1,{\rm B}}+t_{b}^{-}c_{n+1,{\rm B}}^\dag c_{n,{\rm B}} \right. \nonumber\\
&\left.+ \delta_{ab}c_{n+1,{\rm A}}^\dag c_{n,{\rm B}} - \delta_{ab}c_{n,{\rm B}}^\dag c_{n+1,{\rm A}} + Vc_{n,{\rm A}}^\dag c_{n,{\rm A}} \right. \nonumber\\
&\left.- Vc_{n,{\rm B}}^\dag c_{n,{\rm B}} \right), \label{eq:Lee_Hr_topo}
\end{align} where $c_{n,\alpha}^{}$ ($c_{n,\alpha}^\dag$) is the annihilation (creation) operator on site $\alpha$ ($\alpha={\rm A},{\rm B}$) in cell $n$.

Following the similar derivations as Eq.~(\ref{eq:characteristic}), we can obtain the characteristic energy dispersion equation
\begin{eqnarray}
(t_{a}^{+}t_{b}^{+}+\delta_{ab}^{2})\beta^2&+&[-(t_{a}^{+}+t_{b}^{+})E_{\rm OBC} - (t_{a}^{+}-t_{b}^{+})V ]\beta \nonumber\\
&+& \left(t_{a}^{+}t_{b}^{-}+t_{a}^{-}t_{b}^{+}+2\delta_{ab}^{2}+E_{\rm OBC}^2-V^2\right) \nonumber\\
&+&[-(t_{a}^{-}+t_{b}^{-})E_{\rm OBC} - (t_{a}^{-}-t_{b}^{-})V ]\beta^{-1} \nonumber\\
&+&(t_{a}^{-}t_{b}^{-}+\delta_{ab}^{2})\beta^{-2}=0.
\label{eq:characteristic_topo}
\end{eqnarray}

\begin{figure}[h!]
	\centering
	\includegraphics[width=1\linewidth]{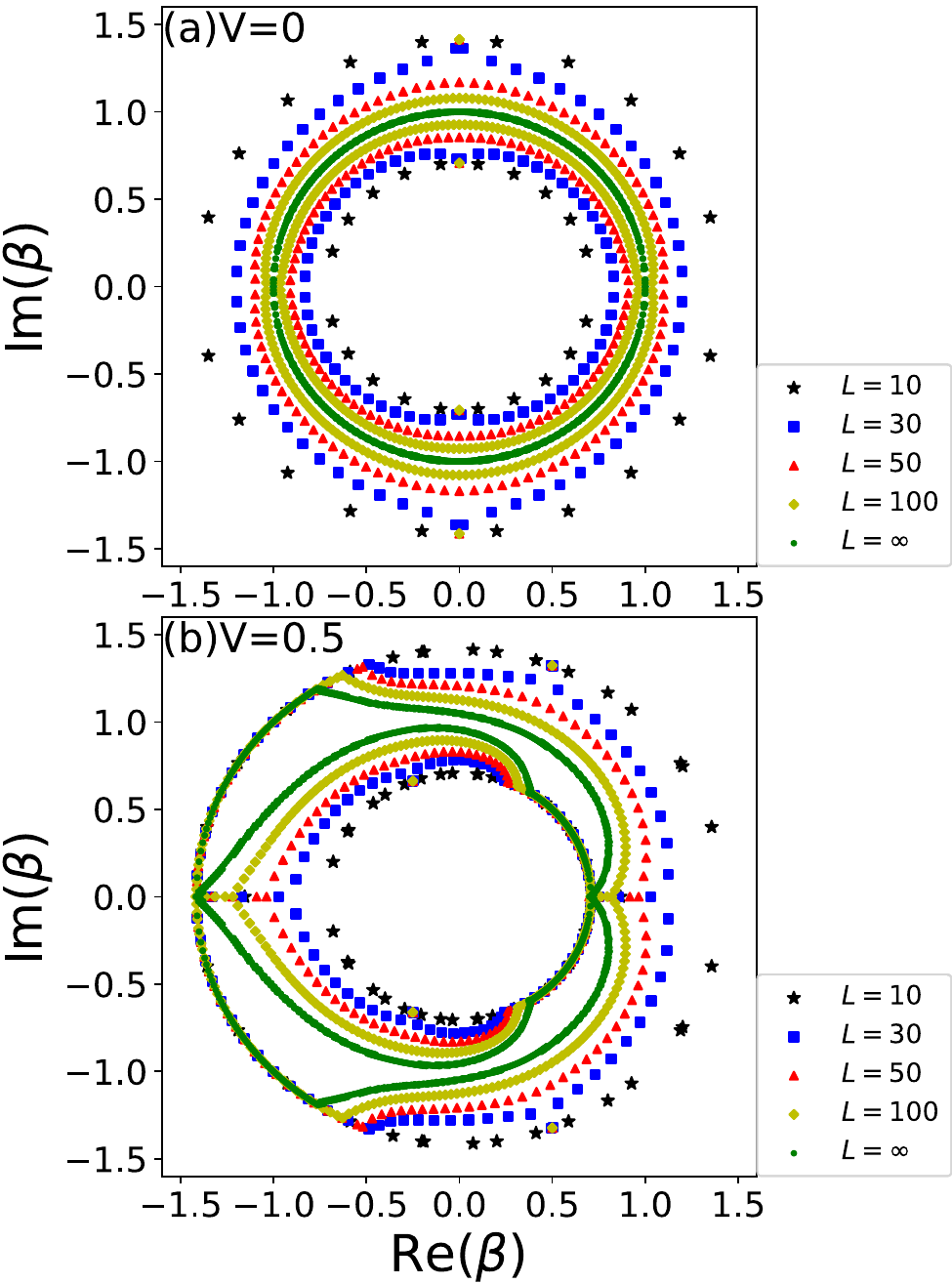}
	\caption{GBZ of the topologically coupled chain model Hamiltonian (\ref{eq:Lee_Hr_topo}) at different finite system sizes $L=10$ (black), $20$ (blue), $50$ (red), $100$ (yellow), $\infty$ (green) with (a) $V=0$ and (b) $V=0.5$. Parameters are $\delta_{ab}=0.5\times10^{-3}$, $t_1=0.75$, and $\delta_a=-\delta_b=0.25$, the same as those in Fig.~\ref{fig:E_OBC_V05_L_together2}. The GBZ is qualitatively similar to that in Fig.~\ref{fig:GBZ_1}, apart from the isolated topological modes (dirty red and yellow) which by definition do not belong to any continuum of states.}
	\label{fig:GBZ_2}
\end{figure}

\begin{figure}[h!]
	\includegraphics[width=1\linewidth]{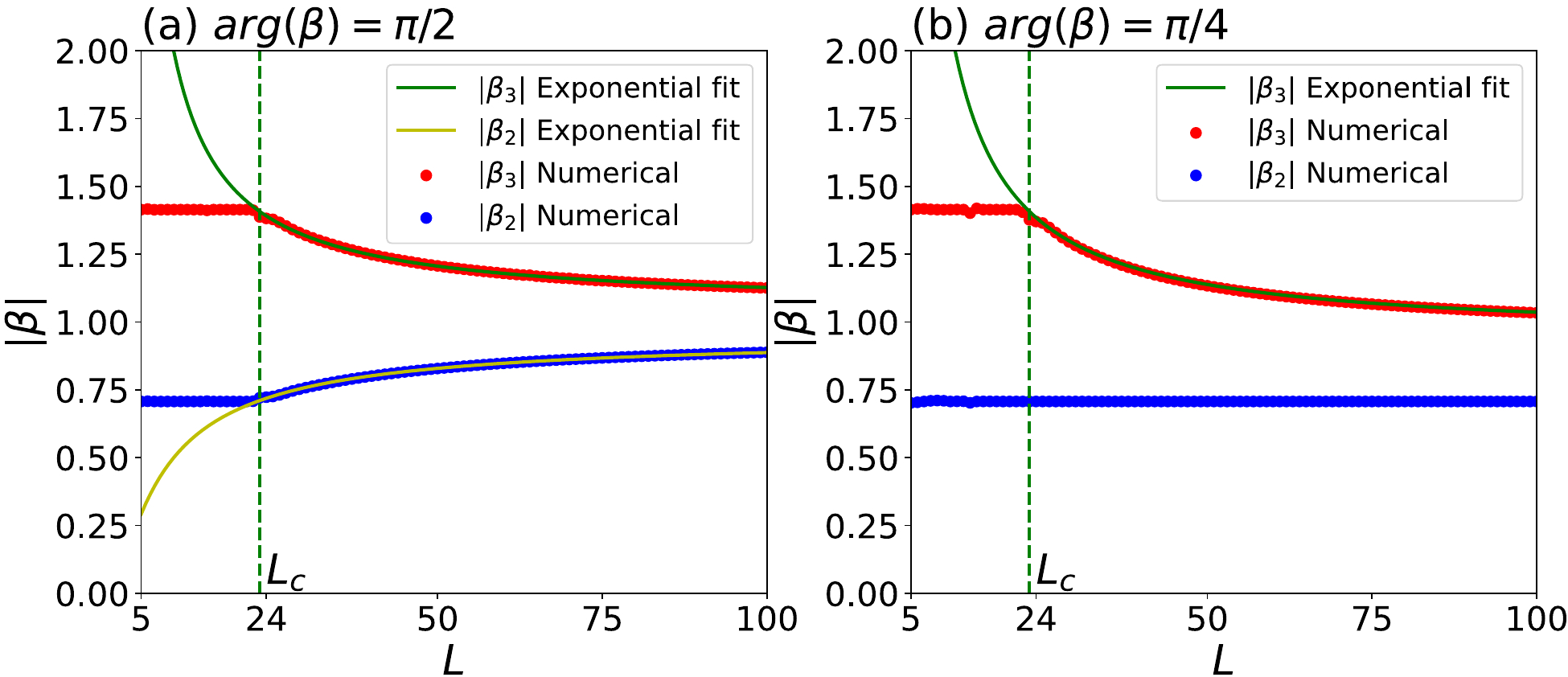}
	\caption{The GBZ radii $|\beta_2|$ and $|\beta_3|$ of the topologically coupled chain model Hamiltonian (\ref{eq:Lee_Hr_topo}) versus the system size $L$ at (a) $\arg(\beta)=\frac{\pi}{2}$ and (b) $\arg(\beta)=\frac{\pi}{4}$. The exponential fits of $|\beta_3|$ correspond to scaling parameters $a(\arg(\beta)=\frac{\pi}{2})\approx0.0520$, $b(\arg(\beta)=\frac{\pi}{2})\approx1476.563$, $a(\arg(\beta)=\frac{\pi}{4})\approx-0.0607$, and $b(\arg(\beta)=\frac{\pi}{4})\approx8791.616$. The cNHSE scaling is frozen below $L=L_{c}\approx23$, but at $\text{arg}(\beta)=\frac{\pi}{4}$, $|\beta_2|$ remains constant across all $L$.  
Here, $V=0.5$ and the other parameters $\delta_{ab}=0.5\times10^{-3}$, $t_1=0.75$, and $\delta_a=-\delta_b=0.25$ are the same as those in Fig.~\ref{fig:GBZ_2}. }
	\label{fig:beta_L_2}
\end{figure}
Similarly as before, we can compute the OBC energy spectra and the GBZ of the topological coupled chain model Hamiltonian (\ref{eq:Lee_Hr_topo}) at different finite system sizes $L=10$, $30$, $50$, $100$, $\infty$ as shown in Figs.~\ref{fig:E_OBC_V05_L_together2} and \ref{fig:GBZ_2}, respectively. We find that they are qualitatively similar to those of the coupled Hatano-Nelson model, except that there is a topological zero mode at $E=0$ (dirty red and yellow). These topological modes also correspond to isolated solutions in the GBZ plot (Fig.~\ref{fig:GBZ_2}), although they are exempted from the finite-size scaling behavior.
It is found that the topological zero modes appear at $E=0$ in the point gap only at sufficiently large system sizes as shown in Fig.~\ref{fig:E_OBC_V05_L_together2}. The reason is that the GBZ depends strongly on the system size as shown in Fig.~\ref{fig:GBZ_2}, and so does the OBC spectrum as shown in Fig.~\ref{fig:E_OBC_V05_L_together2}. When we tune the system size $L$ (regarding $L$ as a parameter), the OBC spectrum changes. At a critical $L$, the OBC spectrum's gap closes and after that, topological zero modes appear, as shown in Fig.~\ref{fig:E_OBC_V0_L_together2_S} in  Appendix \ref{Appendix_VI}. 
Different from the famous single-chain Su-Schrieffer-Heeger model~\cite{su1979solitons,atala2013direct,wang2013topological,lohse2016thouless,nakajima2016topological,leder2016real}, our topologically coupled chain model has two coupled chains, i.e., the coupling between these two chains plays an important role here. In the topologically coupled chain model, the competition between the coupling of the two chains and the finite system size determines the existence or absence of the topological zero modes. This conclusion can be found by calculating the topological phase diagram of the topologically coupled chain model as shown in Fig. 4(d) in a previous work~\cite{li2020critical}. Therefore, the strength of the coupling of the two chains determines the threshold system size which is required for the topological modes.

By enforcing OBCs in the real-space Hamiltonian, we arrive at
\begin{eqnarray}
&&~\left[Z_{1,4}^{(b)}Z_{2,3}^{(a)}\left(\beta_1\beta_4\right)^{L+1}\!+\!Z_{1,4}^{(a)}Z_{2,3}^{(b)}\left(\beta_2\beta_3\right)^{L+1}\right] \nonumber\\
&&\!-\!\left[Z_{1,3}^{(b)}Z_{2,4}^{(a)}\left(\beta_1\beta_3\right)^{L+1}\!+\!Z_{1,3}^{(a)}Z_{2,4}^{(b)}\left(\beta_2\beta_4\right)^{L+1}\right] \nonumber\\
&&\!+\!\left[Z_{1,2}^{(b)}Z_{3,4}^{(a)}\left(\beta_1\beta_2\right)^{L+1}\!+\!Z_{1,2}^{(a)}Z_{3,4}^{(b)}\left(\beta_3\beta_4\right)^{L+1}\right]\!=\!0,
\label{eq:boundary_equation0_topo}
\end{eqnarray}
where dispersion relation solutions $\beta_j~(j=1,2,3,4)$ are arranged so that $\left|\beta_1\right|\leqslant\left|\beta_2\right|\leqslant\left|\beta_3\right|\leqslant\left|\beta_4\right|$, and $Z_{i,j}^{(c)}~(i,j=1,2,3,4;~c=a,b)$ are defined as
\begin{eqnarray}
Z_{i,j}^{(c)}=X_{i}^{(c)}Y_{j}^{(c)}-X_{j}^{(c)}Y_{i}^{(c)}.
\end{eqnarray}
Here, $X_j^{(c)}$ and $Y_j^{(c)}$ are defined as
\begin{eqnarray}
X_{j}^{(a)}&=&E_{\rm OBC}\!-\!(t_{a}^{+}\!-\!t_{a}^{-})\beta_{j}-V,\\
Y_{j}^{(a)}&=&E_{\rm OBC}\!-\!(t_{b}^{+}\!-\!t_{b}^{-})\beta_{j}+V,\\
X_{j}^{(b)}&=&E_{\rm OBC} \!+\! (t_{a}^{+} \!-\! t_{a}^{-})\beta_{j}^{-1}-V,\\
Y_{j}^{(b)}&=&E_{\rm OBC} \!+\! (t_{b}^{+} \!-\! t_{b}^{-})\beta_{j}^{-1}+V.
\end{eqnarray} The corresponding derivation of Eq.~(\ref{eq:boundary_equation0_topo}) is given in Appendix \ref{Appendix_VII}.

To deal with Eq.~(\ref{eq:boundary_equation0_topo}), we only consider the two dominant terms $-Z_{1,3}^{(a)}Z_{2,4}^{(b)}\left(\beta_2\beta_4\right)^{L+1}$ and $Z_{1,2}^{(a)}Z_{3,4}^{(b)}\left(\beta_3\beta_4\right)^{L+1}$ on the left-hand side. In this case, by substituting the solutions of the characteristic equation (\ref{eq:characteristic_topo}) into this approximated boundary equation, we can approximate Eq.~(\ref{eq:boundary_equation0_topo}) as
\begin{eqnarray}
&&\left|\frac{\beta_2}{\beta_3}\right|
\!\simeq\!\left|\frac{Z_{1,2}^{(a)}Z_{3,4}^{(b)}}{Z_{1,3}^{(a)}Z_{2,4}^{(b)}}\right|_{E_{\rm OBC}=E_{\infty}}^{\frac{1}{L+1}} \nonumber\\
&\!\approx\!&\left|\!\frac{\Delta_{a}(E_{\infty}\!-\!V\!+\!\Delta_{a})\Delta_{b}(E_{\infty}\!+\!V\!+\!\Delta_{b})}{2t_{a}^{+}t_{b}^{-}\left[V^{2} \!-\! E_{\infty}^{2} \!+\! 2(t_{a}^{+}t_{b}^{-}\!+\!t_{b}^{+}t_{a}^{-}) \!+\! \Delta_{a}\Delta_{b} \right]}\!\right|^{\frac{1}{L\!+\!1}}\!, \nonumber\\
\label{eq:boundary_equation1_topo}
\end{eqnarray} where $\Delta_a=\sqrt{(E_{\infty}-V)^{2} - 4t_{a}^{+}t_{a}^{-} }$,
$\Delta_b=\sqrt{(E_{\infty}+V)^{2} - 4t_{b}^{+}t_{b}^{-} }$, and we have used the approximation $\delta_{ab}\to 0$ under the condition of weak inter-chain couplings. Notice that $E_{\infty}$ in Eq.~\eqref{eq:boundary_equation1_topo} depends on $\delta_{ab}$. Therefore, we can also follow Eq.~(\ref{eq:beta3}) and postulate an exponential fitting ansatz of $|\beta_{3}|$ as
\begin{eqnarray}\label{eq:beta3_topo}
|\beta_{3}|=a+b^{\frac{1}{L+1}},
\end{eqnarray}
for cases where $|\beta_{2}|\approx1/|\beta_{3}|$. The scaling behavior of $a$ and $b$ in Eq.~(\ref{eq:beta3_topo}) can be extracted or estimated from the asymptotic result Eq.~(\ref{eq:boundary_equation1_topo}) with the model parameters.

In Figs.~\ref{fig:beta_L_2}(a) and \ref{fig:beta_L_2}(b), we show $|\beta_{2}|$ and $|\beta_{3}|$ for the topologically coupled chain model Hamiltonian (\ref{eq:Lee_Hr_topo}) as a function of the system size $L$ both from the exponential formula in Eq.~(\ref{eq:beta3_topo}) and from numerical diagonalization. We observe an exponential scaling behavior qualitatively similar to that of the coupled Hatano-Nelson model, which should also universally holds for other cNHSE models.

\section{Robust spectral scaling behavior under disorder}\label{5} 

\begin{figure}[h]
	\centering
	\includegraphics[width=1\linewidth]{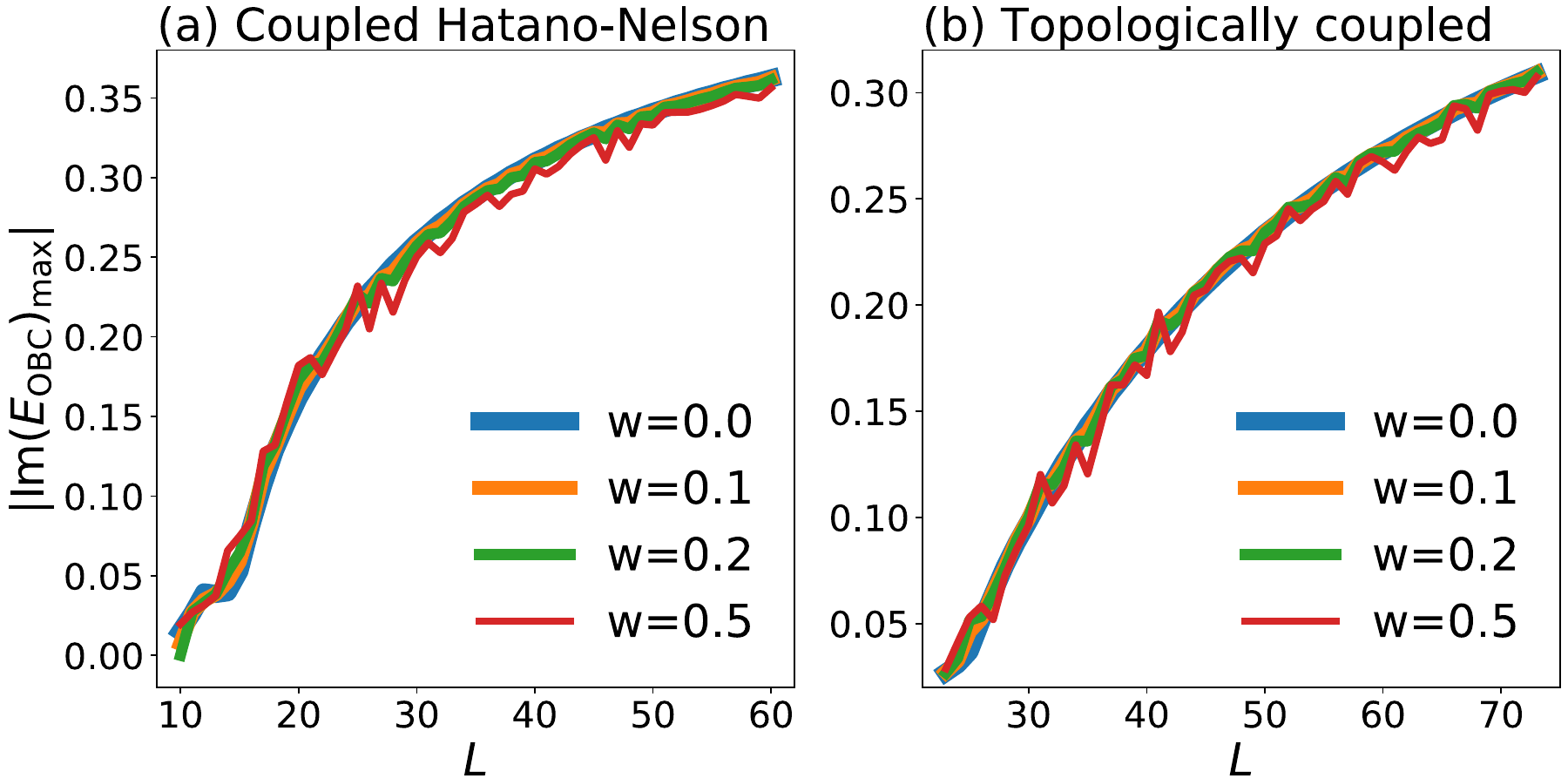}
	\caption{Absolute value of the maximal imaginary part of the eigenvalues $|{\rm Im}(E_{\rm OBC})_{\rm max}|$ for different system sizes $L$, which is satisfied for the eigenenergy with $\arg(E_{\rm OBC})=\frac{\pi}{2}$. In both  (a) the coupled Hatano-Nelson model [Eq.~\eqref{eq:Lee_Hr}] and (b) the topologically coupled chain model [Eq.~\eqref{eq:Lee_Hr_topo}], the spectral scaling behavior is very robust up to disorder strength $w=0.2$, as defined in Eq.~\eqref{dis}. Even at much larger disorder $w=0.5$, the same qualitative spectral scaling prevails. Here, $V=0.5$ and the other parameters are $t_0=0.01$, $\delta_{ab}=0.5\times10^{-3}$, $t_1=0.75$, and $\delta_a=-\delta_b=0.25$, the same as those in previous figures on these respective models. }\label{fig:disorder}
\end{figure}

In this section, we check the robustness of the scaling behavior of the OBC spectra in the presence of uniformly distributed on-site disorder
\begin{equation}\label{dis}
	H_{\rm dis}=\sum_{n,\alpha} \tilde{w}(n,\alpha)c^{\dagger}_{n,\alpha}c_{n,\alpha}
\end{equation}
with random number $\tilde{w}\in[-w/2,w/2]$ and $\alpha={\rm A},{\rm B}$ are the site indices in the cell $n$. Since the GBZ is directly determined through the OBC spectrum, robustness in the scaling behavior in the spectrum would also imply similar robustness in the GBZ.

In Fig.~\ref{fig:disorder}, we plot the absolute value of the maximal imaginary part of the eigenenergies $|\text{Im}(E_{\rm OBC})_\text{max}|$ as a function of the system size $L$ under different disorder strengths from $w=0$ to $0.5$. $|\text{Im}(E_{\rm OBC})_\text{max}|$ determines the ``width'' of the spectrum in the imaginary direction and can be used as a measure of how the shape of the spectrum is deformed under disorder. For our models, $|\text{Im}(E_{\rm OBC})_\text{max}|$ usually occurs when $\arg(E_{\rm OBC})=\frac{\pi}{2}$, but that is not necessarily universal.
From Fig.~\ref{fig:disorder}, we find that relatively weak disorder ($w<0.2$) affects the spectrum negligibly, but moderately large disorder ($w=0.5$) gives rise to visible spectral perturbations. However, the qualitative spectral scaling behavior remains very robust, which indicates that the cNHSE is strongly robust against on-site disorder. This is not surprising given that the cNHSE arises from the competition between different NHSE channels, and should not be affected too much by the on-site energy landscape. It has to be noted that \emph{hopping} disorder, however, can affect the long-time state dynamics and hence significantly modify the overall energy spectrum~\cite{li2021impurity,ezawa2022dynamical,jiang2022filling,guo2021exact,kawabata2022entanglement}.

\section{Discussion}\label{6}

Systems experiencing the critical non-Hermitian skin effect (cNHSE) are particularly sensitive to the system size, exhibiting qualitatively different spectra and spatial eigenstate behavior at different sizes $L$. How the cNHSE scaling is exactly described by the GBZ, particularly for a system of finite size, is an open question. As we already know, the GBZ can be used to restore the BBC in the thermodynamic limit. But for a system of finite size, can GBZ still be a valid theoretical framework? Using the GBZ as a tool to investigate the cNHSE scaling behavior provides an effective way to understand the physical picture of the finite-size effect on the competing NHSE tendencies between small and large size limits. 

In this work, we considered a generic two-component cNHSE ansatz model with two competing NHSE channels, and provided detailed studies of two paradigmatic models, of which the minimal model studied by Ref.~\cite{yokomizo2021scaling} is a special case. We find that our effective finite-size GBZ obeys a universal exponential scaling law, with exponent inversely proportional to the system size, and scaling rate $b$ expressible in term of the model parameters in certain cases. Based on this, we provide detailed and empirically verified estimates of the critical system size $L_c$ where such a scaling relation begins to hold both analytically and numerically.

Such cNHSE phenomena can be readily experimentally demonstrated in non-Hermitian metamaterials with well-controlled gain/loss and effective couplings, such as photonic crystal arrays and electrical circuits. Since the non-reciprocity from different NHSE can cancel, the setup may not even require physical asymmetric couplings, such as in the recent experiment~\cite{zhang2022observation}. Moving forward, it would be immensely interesting to explore the interplay of cNHSE and many-body interactions in emerging and rapidly progressing platforms such as ultracold atomic arrays and quantum circuits.

\begin{acknowledgments}
We thank Zhesen Yang, Kai Zhang, and Chen Fang for helpful discussions.
F.Q. is supported by the Singapore National Research Foundation (Grant No.~NRF2021-QEP2-02-P09). 
\end{acknowledgments}
 
\bibliography{references_cNHSE}



\clearpage
\appendix
\onecolumngrid



\section{Derivation of the determinant form of the OBC constraints for a two-band cNHSE model [Eq.~\eqref{eqapp2a_general}]}\label{Appendix_I}

From the bulk eigenequation in Eq.~(\ref{eq:bulk_eigenequation_general}), we obtain:
\begin{eqnarray}
\!\left\{ \begin{array}{l}
\!\left[\sum_{n=-n_{-}}^{n_{+}}h_{n}^{aa}\left(\beta_{j}\right)^{n}\!-\!E_{\rm OBC}\right]\phi_{{\rm A}}^{(j)}\!+\!\sum_{n=-n_{-}}^{n_{+}}h_{n}^{ab}\left(\beta_{j}\right)^{n}\phi_{{\rm B}}^{(j)}\!=\!0,~\frac{\phi_{{\rm B}}^{(j)}}{\phi_{{\rm A}}^{(j)}}=\frac{E_{\rm OBC}\!-\!\sum_{n=-n_{-}}^{n_{+}}h_{n}^{aa}\left(\beta_{j}\right)^{n}}{\sum_{n=-n_{-}}^{n_{+}}h_{n}^{ab}\left(\beta_{j}\right)^{n}},  \\
\!\sum_{n=-n_{-}}^{n_{+}}h_{n}^{ba}\left(\beta_{j}\right)^{n}\phi_{{\rm A}}^{(j)}\!+\!\left[\sum_{n=-n_{-}}^{n_{+}}h_{n}^{bb}\left(\beta_{j}\right)^{n}\!-\!E_{\rm OBC}\right]\phi_{{\rm B}}^{(j)}\!=\!0,~\frac{\phi_{{\rm B}}^{(j)}}{\phi_{{\rm A}}^{(j)}}=\frac{\sum_{n=-n_{-}}^{n_{+}}h_{n}^{ba}\left(\beta_{j}\right)^{n}}{E_{\rm OBC}-\sum_{n=-n_{-}}^{n_{+}}h_{n}^{bb}\left(\beta_{j}\right)^{n}},
\end{array}\right.
\end{eqnarray} i.e.,
\begin{align}\label{eq:phi}
&\frac{\phi^{(j)}_{\rm B}}{\phi_{\rm A}^{(j)}} = \frac{E_{\rm OBC}-\sum_{n=-n_{-}}^{n_{+}}h_{n}^{aa}\left(\beta_{j}\right)^{n}}{\sum_{n=-n_{-}}^{n_{+}}h_{n}^{ab}\left(\beta_{j}\right)^{n}} = \frac{\sum_{n=-n_{-}}^{n_{+}}h_{n}^{ba}\left(\beta_{j}\right)^{n}}{E_{\rm OBC}-\sum_{n=-n_{-}}^{n_{+}}h_{n}^{bb}\left(\beta_{j}\right)^{n}} = f_{j},\\
&\left[ \sum_{n=-n_{-}}^{n_{+}}h_{n}^{ab}\left(\beta_{j}\right)^{n} \right]\left[ \sum_{n=-n_{-}}^{n_{+}}h_{n}^{ba}\left(\beta_{j}\right)^{n} \right]=\left[E_{\rm OBC}-\sum_{n=-n_{-}}^{n_{+}}h_{n}^{aa}\left(\beta_{j}\right)^{n}\right]\left[E_{\rm OBC}-\sum_{n=-n_{-}}^{n_{+}}h_{n}^{bb}\left(\beta_{j}\right)^{n}\right] \\
&\phi^{(j)}_{\rm B}=f_{j}\phi^{(j)}_{\rm A}.\label{eq:ratio_general}
\end{align} 
which also relates $E_{\rm OBC}$ with $\beta_j$ solutions.

Substituting these real-space eigenequations under the OBC constraints $\psi_{-n_{-},\alpha}=\cdots=\psi_{-1,\alpha}=\psi_{0,\alpha}=\psi_{L+1,\alpha}=\psi_{L+2,\alpha}=\cdots=\psi_{L+n_{+},\alpha}=0~(\alpha={\rm A},{\rm B};~1\leqslant n_{\pm}\leqslant L/2)$ into the real-space Schr{\"o}dinger equation ${\cal H}_{gr}|\psi\rangle=E_{\rm OBC}|\psi\rangle$ [where ${\cal H}_{gr}$ is the Hamiltonian matrix of $H_{gr}$ in the basis $(C_{1},C_{2},\cdots,C_{L})^T$], we can get
\begin{eqnarray}
\left\{ \begin{array}{l}
\sum_{n=0}^{n_{+}}h_{n}^{aa}\psi_{1+n,{\rm A}}+\sum_{n=0}^{n_{+}}h_{n}^{ab}\psi_{1+n,{\rm B}}\!=\!E_{\rm OBC}\psi_{1,{\rm A}},   \vspace{5pt}\\
\sum_{n=0}^{n_{+}}h_{n}^{ba}\psi_{1+n,{\rm A}}+\sum_{n=0}^{n_{+}}h_{n}^{bb}\psi_{1+n,{\rm B}}\!=\!E_{\rm OBC}\psi_{1,{\rm B}},   \vspace{5pt}\\
\sum_{n=-1}^{n_{+}}h_{n}^{aa}\psi_{2+n,{\rm A}}+\sum_{n=-1}^{n_{+}}h_{n}^{ab}\psi_{2+n,{\rm B}}\!=\!E_{\rm OBC}\psi_{2,{\rm A}},   \vspace{5pt}\\
\sum_{n=-1}^{n_{+}}h_{n}^{ba}\psi_{2+n,{\rm A}}+\sum_{n=-1}^{n_{+}}h_{n}^{bb}\psi_{2+n,{\rm B}}\!=\!E_{\rm OBC}\psi_{2,{\rm B}},   \vspace{5pt}\\
~~~~~~~~~~\vdots \\
\sum_{n=-(n_{+}-1)}^{n_{+}}h_{n}^{aa}\psi_{n_{+}+n,{\rm A}}+\sum_{n=-(n_{+}-1)}^{n_{+}}h_{n}^{ab}\psi_{n_{+}+n,{\rm B}}\!=\!E_{\rm OBC}\psi_{n_{+},{\rm A}},~~1\leqslant 2n_{+}\leqslant L,~~1\leqslant n_{+}\leqslant L/2,   \vspace{5pt}\\
\sum_{n=-(n_{+}-1)}^{n_{+}}h_{n}^{ba}\psi_{n_{+}+n,{\rm A}}+\sum_{n=-(n_{+}-1)}^{n_{+}}h_{n}^{bb}\psi_{n_{+}+n,{\rm B}}\!=\!E_{\rm OBC}\psi_{n_{+},{\rm B}},~~1\leqslant 2n_{+}\leqslant L,~~1\leqslant n_{+}\leqslant L/2,   \vspace{10pt}\\
\sum_{n=-n_{-}}^{n_{-}\!-\!1}h_{n}^{aa}\psi_{L\!-\!(n_{-}\!-\!1)+n,{\rm A}}+\sum_{n=-n_{-}}^{n_{-}\!-\!1}h_{n}^{ab}\psi_{L\!-\!(n_{-}\!-\!1)+n,{\rm B}}\!=\!E_{\rm OBC}\psi_{L\!-\!(n_{-}\!-\!1),{\rm A}},~~1\leqslant L\!-\!2n_{-}\!+\!1\leqslant L, \vspace{5pt}\\
\sum_{n=-n_{-}}^{n_{-}\!-\!1}h_{n}^{ba}\psi_{L\!-\!(n_{-}\!-\!1)+n,{\rm A}}+\sum_{n=-n_{-}}^{n_{-}\!-\!1}h_{n}^{bb}\psi_{L\!-\!(n_{-}\!-\!1)+n,{\rm B}}\!=\!E_{\rm OBC}\psi_{L\!-\!(n_{-}\!-\!1),{\rm B}},~~1\leqslant n_{-}\leqslant L/2,   \vspace{5pt}\\
~~~~~~~~~~\vdots \\
\sum_{n=-n_{-}}^{1}h_{n}^{aa}\psi_{L-1+n,{\rm A}}+\sum_{n=-n_{-}}^{1}h_{n}^{ab}\psi_{L-1+n,{\rm B}}\!=\!E_{\rm OBC}\psi_{L-1,{\rm A}}, \vspace{5pt}\\
\sum_{n=-n_{-}}^{1}h_{n}^{ba}\psi_{L-1+n,{\rm A}}+\sum_{n=-n_{-}}^{1}h_{n}^{bb}\psi_{L-1+n,{\rm B}}\!=\!E_{\rm OBC}\psi_{L-1,{\rm B}},\vspace{5pt}\\
\sum_{n=-n_{-}}^{0}h_{n}^{aa}\psi_{L+n,{\rm A}}+\sum_{n=-n_{-}}^{0}h_{n}^{ab}\psi_{L+n,{\rm B}}\!=\!E_{\rm OBC}\psi_{L,{\rm A}},~~1\leqslant L-n_{-}\leqslant L,~~0\leqslant n_{-}\leqslant L-1, \vspace{5pt}\\
\sum_{n=-n_{-}}^{0}h_{n}^{ba}\psi_{L+n,{\rm A}}+\sum_{n=-n_{-}}^{0}h_{n}^{bb}\psi_{L+n,{\rm B}}\!=\!E_{\rm OBC}\psi_{L,{\rm B}},~~1\leqslant L-n_{-}\leqslant L,~~0\leqslant n_{-}\leqslant L-1.
\end{array}\right.
\label{eqapp1a_genera1}
\end{eqnarray} 

Invoking the non-Bloch ansatz $\left( \psi_{n,{\rm A}},~\psi_{n,{\rm B}}\right)^{T}=\sum_{j=1}^{2M}\left(\beta_{j}\right)^{n}\left(\phi_{\rm A}^{\left(j\right)},~\phi_{\rm B}^{\left(j\right)}\right)^{T}$ ($M=n_{-}+n_{+}$) into the above equations (\ref{eqapp1a_genera1}), we have, generalizing~\cite{lee2019anatomy},
\begin{eqnarray}
\left\{ \begin{array}{l}
\sum_{j=1}^{2M}\sum_{n=0}^{n_{+}}h_{n}^{aa}\left(\beta_{j}\right)^{1+n}\phi_{{\rm A}}^{(j)}+\sum_{j=1}^{2M}\sum_{n=0}^{n_{+}}h_{n}^{ab}\left(\beta_{j}\right)^{1+n}\phi_{{\rm B}}^{(j)}\!=\!E_{\rm OBC}\sum_{j=1}^{2M}\left(\beta_{j}\right)\phi_{{\rm A}}^{(j)},   \vspace{5pt}\\
\sum_{j=1}^{2M}\sum_{n=0}^{n_{+}}h_{n}^{ba}\left(\beta_{j}\right)^{1+n}\phi_{{\rm A}}^{(j)}+\sum_{j=1}^{2M}\sum_{n=0}^{n_{+}}h_{n}^{bb}\left(\beta_{j}\right)^{1+n}\phi_{{\rm B}}^{(j)}\!=\!E_{\rm OBC}\sum_{j=1}^{2M}\left(\beta_{j}\right)\phi_{{\rm B}}^{(j)},   \vspace{5pt}\\
\sum_{j=1}^{2M}\sum_{n=-1}^{n_{+}}h_{n}^{aa}\left(\beta_{j}\right)^{2+n}\phi_{{\rm A}}^{(j)}+\sum_{j=1}^{2M}\sum_{n=-1}^{n_{+}}h_{n}^{ab}\left(\beta_{j}\right)^{2+n}\psi_{{\rm B}}^{(j)}\!=\!E_{\rm OBC}\sum_{j=1}^{2M}\left(\beta_{j}\right)^{2}\phi_{{\rm A}}^{(j)},   \vspace{5pt}\\
\sum_{j=1}^{2M}\sum_{n=-1}^{n_{+}}h_{n}^{ba}\left(\beta_{j}\right)^{2+n}\phi_{{\rm A}}^{(j)}+\sum_{j=1}^{2M}\sum_{n=-1}^{n_{+}}h_{n}^{bb}\left(\beta_{j}\right)^{2+n}\phi_{{\rm B}}^{(j)}\!=\!E_{\rm OBC}\sum_{j=1}^{2M}\left(\beta_{j}\right)^{2}\phi_{{\rm B}}^{(j)},   \vspace{5pt}\\
~~~~~~~~~~\vdots \\
\sum_{j=1}^{2M}\sum_{n=-(n_{+}-1)}^{n_{+}}h_{n}^{aa}\left(\beta_{j}\right)^{n_{+}+n}\phi_{{\rm A}}^{(j)}+\sum_{j=1}^{2M}\sum_{n=-(n_{+}-1)}^{n_{+}}h_{n}^{ab}\left(\beta_{j}\right)^{n_{+}+n}\phi_{{\rm B}}^{(j)}\!=\!E_{\rm OBC}\sum_{j=1}^{2M}\left(\beta_{j}\right)^{n_{+}}\phi_{{\rm A}}^{(j)}, \vspace{5pt}\\
\sum_{j=1}^{2M}\sum_{n=-(n_{+}-1)}^{n_{+}}h_{n}^{ba}\left(\beta_{j}\right)^{n_{+}+n}\phi_{{\rm A}}^{(j)}+\sum_{j=1}^{2M}\sum_{n=-(n_{+}-1)}^{n_{+}}h_{n}^{bb}\left(\beta_{j}\right)^{n_{+}+n}\phi_{{\rm B}}^{(j)}\!=\!E_{\rm OBC}\sum_{j=1}^{2M}\left(\beta_{j}\right)^{n_{+}}\phi_{{\rm B}}^{(j)}, \vspace{10pt}\\
\sum_{j=1}^{2M}\sum_{n=-n_{-}}^{n_{-}\!-\!1}h_{n}^{aa}\left(\beta_{j}\right)^{L\!-\!(n_{-}\!-\!1)+n}\phi_{{\rm A}}^{(j)}+\sum_{j=1}^{2M}\sum_{n=-n_{-}}^{n_{-}\!-\!1}h_{n}^{ab}\left(\beta_{j}\right)^{L\!-\!(n_{-}\!-\!1)+n}\phi_{{\rm B}}^{(j)}\!=\!E_{\rm OBC}\sum_{j=1}^{2M}\left(\beta_{j}\right)^{L\!-\!(n_{-}\!-\!1)}\phi_{{\rm A}}^{(j)}, \vspace{5pt}\\
\sum_{j=1}^{2M}\sum_{n=-n_{-}}^{n_{-}\!-\!1}h_{n}^{ba}\left(\beta_{j}\right)^{L\!-\!(n_{-}\!-\!1)+n}\phi_{{\rm A}}^{(j)}+\sum_{j=1}^{2M}\sum_{n=-n_{-}}^{n_{-}\!-\!1}h_{n}^{bb}\left(\beta_{j}\right)^{L\!-\!(n_{-}\!-\!1)+n}\phi_{{\rm B}}^{(j)}\!=\!E_{\rm OBC}\sum_{j=1}^{2M}\left(\beta_{j}\right)^{L\!-\!(n_{-}\!-\!1)}\phi_{{\rm B}}^{(j)},  \vspace{5pt}\\
~~~~~~~~~~\vdots \\
\sum_{j=1}^{2M}\sum_{n=-n_{-}}^{1}h_{n}^{aa}\left(\beta_{j}\right)^{L-1+n}\phi_{{\rm A}}^{(j)}+\sum_{j=1}^{2M}\sum_{n=-n_{-}}^{1}h_{n}^{ab}\left(\beta_{j}\right)^{L-1+n}\phi_{{\rm B}}^{(j)}\!=\!E_{\rm OBC}\sum_{j=1}^{2M}\left(\beta_{j}\right)^{L-1}\phi_{{\rm A}}^{(j)}, \vspace{5pt}\\
\sum_{j=1}^{2M}\sum_{n=-n_{-}}^{1}h_{n}^{ba}\left(\beta_{j}\right)^{L-1+n}\phi_{{\rm A}}^{(j)}+\sum_{j=1}^{2M}\sum_{n=-n_{-}}^{1}h_{n}^{bb}\left(\beta_{j}\right)^{L-1+n}\phi_{{\rm B}}^{(j)}\!=\!E_{\rm OBC}\sum_{j=1}^{2M}\left(\beta_{j}\right)^{L-1}\phi_{{\rm B}}^{(j)},\vspace{5pt}\\
\sum_{j=1}^{2M}\sum_{n=-n_{-}}^{0}h_{n}^{aa}\left(\beta_{j}\right)^{L+n}\phi_{{\rm A}}^{(j)}+\sum_{j=1}^{2M}\sum_{n=-n_{-}}^{0}h_{n}^{ab}\left(\beta_{j}\right)^{L+n}\phi_{{\rm B}}^{(j)}\!=\!E_{\rm OBC}\sum_{j=1}^{2M}\left(\beta_{j}\right)^{L}\phi_{{\rm A}}^{(j)}, \vspace{5pt}\\
\sum_{j=1}^{2M}\sum_{n=-n_{-}}^{0}h_{n}^{ba}\left(\beta_{j}\right)^{L+n}\phi_{{\rm A}}^{(j)}+\sum_{j=1}^{2M}\sum_{n=-n_{-}}^{0}h_{n}^{bb}\left(\beta_{j}\right)^{L+n}\phi_{{\rm B}}^{(j)}\!=\!E_{\rm OBC}\sum_{j=1}^{2M}\left(\beta_{j}\right)^{L}\phi_{{\rm B}}^{(j)}.
\end{array}\right.
\label{eqapp1a_genera2}
\end{eqnarray}

Substituting Eq.~(\ref{eq:ratio_general}), i.e., $\phi^{(j)}_{\rm B}=f_{j}\phi^{(j)}_{\rm A}$ into the above equations (\ref{eqapp1a_genera2}) such as to eliminate the $\phi^{(j)}_{\rm B}$, we have
\begin{eqnarray}
\left\{ \begin{array}{l}
\sum_{j=1}^{2M}\left[\sum_{n=0}^{n_{+}}(h_{n}^{aa}+f_{j}h_{n}^{ab})\left(\beta_{j}\right)^{n} - E_{\rm OBC} \right]\left(\beta_{j}\right)\phi_{{\rm A}}^{(j)}\!=\!0,   \vspace{5pt}\\
\sum_{j=1}^{2M}\left[\sum_{n=0}^{n_{+}}(h_{n}^{ba}+f_{j}h_{n}^{bb})\left(\beta_{j}\right)^{n} - f_{j}E_{\rm OBC} \right]\left(\beta_{j}\right)\phi_{{\rm A}}^{(j)}\!=\!0,   \vspace{5pt}\\
\sum_{j=1}^{2M}\left[\sum_{n=-1}^{n_{+}}(h_{n}^{aa}+f_{j}h_{n}^{ab})\left(\beta_{j}\right)^{n} - E_{\rm OBC} \right]\left(\beta_{j}\right)^{2}\phi_{{\rm A}}^{(j)}\!=\!0,   \vspace{5pt}\\
\sum_{j=1}^{2M}\left[\sum_{n=-1}^{n_{+}}(h_{n}^{ba}+f_{j}h_{n}^{bb})\left(\beta_{j}\right)^{n} - f_{j}E_{\rm OBC} \right]\left(\beta_{j}\right)^{2}\phi_{{\rm A}}^{(j)}\!=\!0,   \vspace{5pt}\\
~~~~~~~~~~\vdots \\
\sum_{j=1}^{2M}\left[\sum_{n=-(n_{+}-1)}^{n_{+}}(h_{n}^{aa}+f_{j}h_{n}^{ab})\left(\beta_{j}\right)^{n} - E_{\rm OBC} \right]\left(\beta_{j}\right)^{n_{+}}\phi_{{\rm A}}^{(j)}\!=\!0, \vspace{5pt}\\
\sum_{j=1}^{2M}\left[\sum_{n=-(n_{+}-1)}^{n_{+}}(h_{n}^{ba}+f_{j}h_{n}^{bb})\left(\beta_{j}\right)^{n} - f_{j}E_{\rm OBC} \right]\left(\beta_{j}\right)^{n_{+}}\phi_{{\rm A}}^{(j)}\!=\!0, \vspace{10pt}\\
\sum_{j=1}^{2M}\left[\sum_{n=-n_{-}}^{n_{-}\!-\!1}(h_{n}^{aa}+f_{j}h_{n}^{ab})\left(\beta_{j}\right)^{n} - E_{\rm OBC}\right]\left(\beta_{j}\right)^{L\!-\!(n_{-}\!-\!1)}\phi_{{\rm A}}^{(j)}\!=\!0, \vspace{5pt}\\
\sum_{j=1}^{2M}\left[\sum_{n=-n_{-}}^{n_{-}\!-\!1}(h_{n}^{ba}+f_{j}h_{n}^{bb})\left(\beta_{j}\right)^{n} - f_{j}E_{\rm OBC}\right]\left(\beta_{j}\right)^{L\!-\!(n_{-}\!-\!1)}\phi_{{\rm A}}^{(j)}\!=\!0,  \vspace{5pt}\\
~~~~~~~~~~\vdots \\
\sum_{j=1}^{2M}\left[\sum_{n=-n_{-}}^{1}(h_{n}^{aa}+f_{j}h_{n}^{ab})\left(\beta_{j}\right)^{n} - E_{\rm OBC}\right]\left(\beta_{j}\right)^{L-1}\phi_{{\rm A}}^{(j)}\!=\!0, \vspace{5pt}\\
\sum_{j=1}^{2M}\left[\sum_{n=-n_{-}}^{1}(h_{n}^{ba}+f_{j}h_{n}^{bb})\left(\beta_{j}\right)^{n} - f_{j}E_{\rm OBC}\right]\left(\beta_{j}\right)^{L-1}\phi_{{\rm A}}^{(j)}\!=\!0,\vspace{5pt}\\
\sum_{j=1}^{2M}\left[\sum_{n=-n_{-}}^{0}(h_{n}^{aa}+f_{j}h_{n}^{ab})\left(\beta_{j}\right)^{n} - E_{\rm OBC}\right]\left(\beta_{j}\right)^{L}\phi_{{\rm A}}^{(j)}\!=\!0, \vspace{5pt}\\
\sum_{j=1}^{2M}\left[\sum_{n=-n_{-}}^{0}(h_{n}^{ba}+f_{j}h_{n}^{bb})\left(\beta_{j}\right)^{n} - f_{j}E_{\rm OBC}\right]\left(\beta_{j}\right)^{L}\phi_{{\rm A}}^{(j)}\!=\!0.
\end{array}\right.
\label{eqapp1a_genera3}
\end{eqnarray} 
We can express Eq.~(\ref{eqapp1a_genera3}) in more compact notation
\begin{eqnarray}
\left\{ \begin{array}{l}
\sum_{j=1}^{2M}F_{j}^{(a,1)}\beta_{j}\phi_{\rm A}^{(j)}=0, \\
\sum_{j=1}^{2M}F_{j}^{(b,1)}\beta_{j}\phi_{\rm A}^{(j)}=0, \\
~~~~~~~~~~\vdots \\
\sum_{j=1}^{2M}F_{j}^{(a,n_{+})}\left(\beta_{j}\right)^{n_{+}}\phi_{\rm A}^{(j)}=0, \\
\sum_{j=1}^{2M}F_{j}^{(b,n_{+})}\left(\beta_{j}\right)^{n_{+}}\phi_{\rm A}^{(j)}=0, \\
\sum_{j=1}^{2M}G_{j}^{(a,1)}\left(\beta_{j}\right)^{L\!-\!(n_{-}\!-\!1)}\phi_{\rm A}^{(j)}=0, \\
\sum_{j=1}^{2M}G_{j}^{(b,1)}\left(\beta_{j}\right)^{L\!-\!(n_{-}\!-\!1)}\phi_{\rm A}^{(j)}=0, \\
~~~~~~~~~~\vdots \\
\sum_{j=1}^{2M}G_{j}^{(a,n_{-})}\left(\beta_{j}\right)^{L}\phi_{\rm A}^{(j)}=0,\\
\sum_{j=1}^{2M}G_{j}^{(b,n_{-})}\left(\beta_{j}\right)^{L}\phi_{\rm A}^{(j)}=0,
\end{array}\right.
\end{eqnarray} where 
\begin{align}
F_{j}^{(a,i)}&= \sum_{n=-(i-1)}^{n_{+}}(h_{n}^{aa}+f_{j}h_{n}^{ab})\left(\beta_{j}\right)^{n} - E_{\rm OBC},\\
F_{j}^{(b,i)}&= \sum_{n=-(i-1)}^{n_{+}}(h_{n}^{ba}+f_{j}h_{n}^{bb})\left(\beta_{j}\right)^{n} - f_{j}E_{\rm OBC},\\
G_{j}^{(a,i)}&= \sum_{n=-n_{-}}^{n_{-}-i}(h_{n}^{aa}+f_{j}h_{n}^{ab})\left(\beta_{j}\right)^{n} - E_{\rm OBC},\\
G_{j}^{(b,i)}&= \sum_{n=-n_{-}}^{n_{-}-i}(h_{n}^{ba}+f_{j}h_{n}^{bb})\left(\beta_{j}\right)^{n} - f_{j}E_{\rm OBC}.
\end{align}

For a nontrivial state $\phi_{A}^{\left(j\right)}~(j=1,2,\dots,2M)$ that does not vanish, we hence require the vanishing determinant
\begin{equation}
\begin{vmatrix}
F_{1}^{(a,1)}\beta_{1} & F_{2}^{(a,1)}\beta_{2} & \cdots & F_{2M}^{(a,1)}\beta_{2M} \\
F_{1}^{(b,1)}\beta_{1} & F_{2}^{(b,1)}\beta_{2} & \cdots & F_{2M}^{(b,1)}\beta_{2M} \\
\vdots & \vdots & \vdots & \vdots \\
F_{1}^{(a,n_{+})}\left(\beta_{1}\right)^{n_{+}} & F_{2}^{(a,n_{+})}\left(\beta_{2}\right)^{n_{+}} & \cdots & F_{2M}^{(a,n_{+})}\left(\beta_{2M}\right)^{n_{+}} \\
F_{1}^{(b,n_{+})}\left(\beta_{1}\right)^{n_{+}} & F_{2}^{(b,n_{+})}\left(\beta_{2}\right)^{n_{+}} & \cdots & F_{2M}^{(b,n_{+})}\left(\beta_{2M}\right)^{n_{+}} \\
G_{1}^{(a,1)}\left(\beta_{1}\right)^{L\!-\!(n_{-}\!-\!1)} & G_{2}^{(a,1)}\left(\beta_{2}\right)^{L\!-\!(n_{-}\!-\!1)} & \cdots & G_{2M}^{(a,1)}\left(\beta_{2M}\right)^{L\!-\!(n_{-}\!-\!1)} \\
G_{1}^{(b,1)}\left(\beta_{1}\right)^{L\!-\!(n_{-}\!-\!1)} & G_{2}^{(b,1)}\left(\beta_{2}\right)^{L\!-\!(n_{-}\!-\!1)} & \cdots & G_{2M}^{(b,1)}\left(\beta_{2M}\right)^{L\!-\!(n_{-}\!-\!1)} \\
\vdots & \vdots & \vdots & \vdots \\
G_{1}^{(a,n_{-})}\left(\beta_{1}\right)^{L} & G_{2}^{(a,n_{-})}\left(\beta_{2}\right)^{L} & \cdots & G_{2M}^{(a,n_{-})}\left(\beta_{2M}\right)^{L} \\
G_{1}^{(b,n_{-})}\left(\beta_{1}\right)^{L} & G_{2}^{(b,n_{-})}\left(\beta_{2}\right)^{L} & \cdots & G_{2M}^{(b,n_{-})}\left(\beta_{2M}\right)^{L} 
\end{vmatrix}=0.
\end{equation}

\section{Derivation of the determinant form of the OBC constraints for a general multi-band model}\label{Appendix_II}

Here, we generalize the above derivation to a general multi-band model, and show that the OBC constraints result in an analogous vanishing determinant expression. In momentum space, an $N$-band model Hamiltonian in the basis $C_{{\bf k}}=\left(c_{{\bf k},1}^{},c_{{\bf k},2}^{},\cdots,c_{{\bf k},N}^{}\right)^{T}$ is given by
\begin{align}
	{\cal H}_{mb}(z)
	\!=\!\sum_{n=-n_{-}}^{n_{+}}\begin{pmatrix}
		h_{n}^{11} & h_{n}^{12} & \cdots & h_{n}^{1N} \\
		h_{n}^{21} & h_{n}^{22} & \cdots & h_{n}^{2N} \\
		\vdots & \vdots & \vdots & \vdots \\
		h_{n}^{N1} & h_{n}^{N2} & \cdots & h_{n}^{NN}
	\end{pmatrix}z^{n}, \label{eq:Lee_Hk_multi}
\end{align} where $N$ is the number of bands, which we set to be an even number.

By Fourier transformation, one obtains the real-space tight-binding Hamiltonian of this system as
\begin{align}
	H_{mbr}\!=\!
	\sum_{j=1}^{L}\sum_{n=-n_{-}}^{n_{+}}C_{j}^{\dagger}\begin{pmatrix}
		h_{n}^{11} & h_{n}^{12} & \cdots & h_{n}^{1N} \\
		h_{n}^{21} & h_{n}^{22} & \cdots & h_{n}^{2N} \\
		\vdots & \vdots & \vdots & \vdots \\
		h_{n}^{N1} & h_{n}^{N2} & \cdots & h_{n}^{NN}
	\end{pmatrix}C_{j+n},\label{eq:Lee_Hr_multi}
\end{align} where $C_{j}=\left(c_{j,1},c_{j,2},\cdots,c_{j,N}\right)^{T}$.

With $|\psi\rangle=\left(\psi_{1,1},\psi_{1,2},\cdots,\psi_{1,N},\psi_{2,1},\psi_{2,2},\cdots,\psi_{2,N},\cdots,\psi_{L,1},\psi_{L,2},\cdots,\psi_{L,N}\right)^{\rm T}$, the solutions of the real-space Schr{\"o}dinger equation ${\cal H}_{mbr}|\psi\rangle=E_{\rm OBC}|\psi\rangle$ [where ${\cal H}_{mbr}$ is the Hamiltonian matrix of $H_{mbr}$ in the basis $(C_{1},C_{2},\cdots,C_{L})^T$] can be given by
\begin{eqnarray}
	\left( \begin{array}{c}
		\psi_{n,1} \\
		\psi_{n,2} \\
		\vdots \\
		\psi_{n,N}
	\end{array}\right)=\sum_{j=1}^{2M}\left(\beta_j\right)^n\left( \begin{array}{c}
		\phi_{1}^{\left(j\right)} \\
		\phi_{2}^{\left(j\right)} \\
		\vdots \\
		\phi_{N}^{\left(j\right)}
	\end{array}\right), \label{eq:eigenstates_multi}
\end{eqnarray}
where $2M=N\times(n_{-}+n_{+})$ and $\beta=\beta_j$ are the solutions of the characteristic equation
\begin{eqnarray}
	{\rm Det}\left[{\cal H}_{mb}(\beta)-E_{\rm OBC} \right]=0,\label{eq:characteristic_multi}
\end{eqnarray} where ${\cal H}_{mb}(\beta)$ is the non-Bloch matrix~\cite{yokomizo2021scaling} as
\begin{align}
	{\cal H}_{mb}(\beta)
	\!=\!\sum_{n=-n_{-}}^{n_{+}}\begin{pmatrix}
		h_{n}^{11} & h_{n}^{12} & \cdots & h_{n}^{1N} \\
		h_{n}^{21} & h_{n}^{22} & \cdots & h_{n}^{2N} \\
		\vdots & \vdots & \vdots & \vdots \\
		h_{n}^{N1} & h_{n}^{N2} & \cdots & h_{n}^{NN}
	\end{pmatrix}\beta^{n}.
\end{align}
In general, the characteristic equation~(\ref{eq:characteristic_multi}) has $2M$ solutions for $\beta$, where $M=N\times(n_{-}+n_{+})/2$ is an integer and $N$ is an even number. We label these solutions such that $\left|\beta_1\right|\leqslant\left|\beta_2\right|\leqslant\dots\leqslant\left|\beta_{2M}\right|$. 

From the eigenequations, we obtain
\begin{eqnarray}
\!\left\{ \begin{array}{l}
\!\sum_{n=-n_{-}}^{n_{+}}h_{n}^{11}\left(\beta_{j}\right)^{n}\phi_{1}^{(j)}\!+\!\sum_{n=-n_{-}}^{n_{+}}h_{n}^{12}\left(\beta_{j}\right)^{n}\phi_{2}^{(j)}\!+\!\cdots\!+\!\sum_{n=-n_{-}}^{n_{+}}h_{n}^{1N}\left(\beta_{j}\right)^{n}\phi_{N}^{(j)}\!=\!E_{\rm OBC}\phi_{1}^{(j)}, \\
\!\sum_{n=-n_{-}}^{n_{+}}h_{n}^{21}\left(\beta_{j}\right)^{n}\phi_{1}^{(j)}\!+\!\sum_{n=-n_{-}}^{n_{+}}h_{n}^{22}\left(\beta_{j}\right)^{n}\phi_{2}^{(j)}\!+\!\cdots\!+\!\sum_{n=-n_{-}}^{n_{+}}h_{n}^{2N}\left(\beta_{j}\right)^{n}\phi_{N}^{(j)}\!=\!E_{\rm OBC}\phi_{2}^{(j)}, \\
\vdots\\
\!\sum_{n=-n_{-}}^{n_{+}}h_{n}^{N1}\left(\beta_{j}\right)^{n}\phi_{1}^{(j)}\!+\!\sum_{n=-n_{-}}^{n_{+}}h_{n}^{N2}\left(\beta_{j}\right)^{n}\phi_{2}^{(j)}\!+\!\cdots\!+\!\sum_{n=-n_{-}}^{n_{+}}h_{n}^{NN}\left(\beta_{j}\right)^{n}\phi_{N}^{(j)}\!=\!E_{\rm OBC}\phi_{N}^{(j)}, \\
\end{array}\right.
\end{eqnarray} i.e.,
\begin{eqnarray}
\!\left\{ \begin{array}{l}
\!\left[\sum_{n=-n_{-}}^{n_{+}}h_{n}^{11}\left(\beta_{j}\right)^{n}\!-\!E_{\rm OBC}\right]\phi_{1}^{(j)}\!+\!\sum_{n=-n_{-}}^{n_{+}}h_{n}^{12}\left(\beta_{j}\right)^{n}\phi_{2}^{(j)}\!+\!\cdots\!+\!\sum_{n=-n_{-}}^{n_{+}}h_{n}^{1N}\left(\beta_{j}\right)^{n}\phi_{N}^{(j)}\!=\!0, \\
\!\sum_{n=-n_{-}}^{n_{+}}h_{n}^{21}\left(\beta_{j}\right)^{n}\phi_{1}^{(j)}\!+\!\left[\sum_{n=-n_{-}}^{n_{+}}h_{n}^{22}\left(\beta_{j}\right)^{n}\!-\!E_{\rm OBC}\right]\phi_{2}^{(j)}\!+\!\cdots\!+\!\sum_{n=-n_{-}}^{n_{+}}h_{n}^{2N}\left(\beta_{j}\right)^{n}\phi_{N}^{(j)}\!=\!0, \\
\vdots\\
\!\sum_{n=-n_{-}}^{n_{+}}h_{n}^{N1}\left(\beta_{j}\right)^{n}\phi_{1}^{(j)}\!+\!\sum_{n=-n_{-}}^{n_{+}}h_{n}^{N2}\left(\beta_{j}\right)^{n}\phi_{2}^{(j)}\!+\!\cdots\!+\!\left[\sum_{n=-n_{-}}^{n_{+}}h_{n}^{NN}\left(\beta_{j}\right)^{n}\!-\!E_{\rm OBC}\right]\phi_{N}^{(j)}\!=\!0, \\
\end{array}\right.
\end{eqnarray} i.e.,
\begin{eqnarray}
\!\left\{ \begin{array}{l}
\!\left[\sum_{n=-n_{-}}^{n_{+}}h_{n}^{11}\left(\beta_{j}\right)^{n}\!-\!E_{\rm OBC}\right]\phi_{1}^{(j)}\!+\!\sum_{n=-n_{-}}^{n_{+}}h_{n}^{12}\left(\beta_{j}\right)^{n}f_{j}^{(2)}\phi_{1}^{(j)}\!+\!\cdots\!+\!\sum_{n=-n_{-}}^{n_{+}}h_{n}^{1N}\left(\beta_{j}\right)^{n}f_{j}^{(N)}\phi_{1}^{(j)}\!=\!0, \\
\!\sum_{n=-n_{-}}^{n_{+}}h_{n}^{21}\left(\beta_{j}\right)^{n}\phi_{1}^{(j)}\!+\!\left[\sum_{n=-n_{-}}^{n_{+}}h_{n}^{22}\left(\beta_{j}\right)^{n}\!-\!E_{\rm OBC}\right]f_{j}^{(2)}\phi_{1}^{(j)}\!+\!\cdots\!+\!\sum_{n=-n_{-}}^{n_{+}}h_{n}^{2N}\left(\beta_{j}\right)^{n}f_{j}^{(N)}\phi_{1}^{(j)}\!=\!0, \\
\vdots\\
\!\sum_{n=-n_{-}}^{n_{+}}h_{n}^{N1}\left(\beta_{j}\right)^{n}\phi_{1}^{(j)}\!+\!\sum_{n=-n_{-}}^{n_{+}}h_{n}^{N2}\left(\beta_{j}\right)^{n}f_{j}^{(2)}\phi_{1}^{(j)}\!+\!\cdots\!+\!\left[\sum_{n=-n_{-}}^{n_{+}}h_{n}^{NN}\left(\beta_{j}\right)^{n}\!-\!E_{\rm OBC}\right]f_{j}^{(N)}\phi_{1}^{(j)}\!=\!0, \\
\end{array}\right.
\end{eqnarray} where $f_{j}^{(\alpha)}=\phi_{\alpha}^{(j)}/\phi_{1}^{(j)}$ with $\alpha=1,2,\cdots,N$, i.e.,
\begin{align}
&\phi^{(j)}_{\alpha}=f_{j}^{(\alpha)}\phi^{(j)}_{1},
\end{align} 
where  $\alpha=1,2,\cdots,N$.

As we know, Eq.~(\ref{eq:eigenstates_multi}) has $2M\times N$ unknown coefficients,
but with the real-space Schr{\"o}dinger equation ${\cal H}_{mbr}|\psi\rangle=E_{\rm OBC}|\psi\rangle$ and an additional $2M$ boundary conditions, the $2M\times N$ coefficients can be reduced to $2M$-independent coefficients.
By rewriting the coupling constraints in terms of $\phi_{1}^{\left(j\right)}~(j=1,2,\dots,2M)$, which should have nonzero values, we have, analogously as before,
\begin{equation}
	\begin{vmatrix}
\tilde{F}_{1}^{(1,1)}\beta_{1} & \tilde{F}_{2}^{(1,1)}\beta_{2} & \cdots & \tilde{F}_{2M}^{(1,1)}\beta_{2M} \\
\tilde{F}_{1}^{(2,1)}\beta_{1} & \tilde{F}_{2}^{(2,1)}\beta_{2} & \cdots & \tilde{F}_{2M}^{(2,1)}\beta_{2M} \\
\vdots & \vdots & \vdots & \vdots \\
\tilde{F}_{1}^{(N,1)}\beta_{1} & \tilde{F}_{2}^{(N,1)}\beta_{2} & \cdots & \tilde{F}_{2M}^{(N,1)}\beta_{2M} \\
\vdots & \vdots & \vdots & \vdots \\
\tilde{F}_{1}^{(1,n_{+})}\left(\beta_{1}\right)^{n_{+}} & \tilde{F}_{2}^{(1,n_{+})}\left(\beta_{2}\right)^{n_{+}} & \cdots & \tilde{F}_{2M}^{(1,n_{+})}\left(\beta_{2M}\right)^{n_{+}} \\
\tilde{F}_{1}^{(2,n_{+})}\left(\beta_{1}\right)^{n_{+}} & \tilde{F}_{2}^{(2,n_{+})}\left(\beta_{2}\right)^{n_{+}} & \cdots & \tilde{F}_{2M}^{(2,n_{+})}\left(\beta_{2M}\right)^{n_{+}} \\
\vdots & \vdots & \vdots & \vdots \\
\tilde{F}_{1}^{(N,n_{+})}\left(\beta_{1}\right)^{n_{+}} & \tilde{F}_{2}^{(N,n_{+})}\left(\beta_{2}\right)^{n_{+}} & \cdots & \tilde{F}_{2M}^{(N,n_{+})}\left(\beta_{2M}\right)^{n_{+}} \\
\tilde{G}_{1}^{(1,1)}\left(\beta_{1}\right)^{L\!-\!(n_{-}\!-\!1)} & \tilde{G}_{2}^{(1,1)}\left(\beta_{2}\right)^{L\!-\!(n_{-}\!-\!1)} & \cdots & \tilde{G}_{2M}^{(1,1)}\left(\beta_{2M}\right)^{L\!-\!(n_{-}\!-\!1)} \\
\tilde{G}_{1}^{(2,1)}\left(\beta_{1}\right)^{L\!-\!(n_{-}\!-\!1)} & \tilde{G}_{2}^{(2,1)}\left(\beta_{2}\right)^{L\!-\!(n_{-}\!-\!1)} & \cdots & \tilde{G}_{2M}^{(2,1)}\left(\beta_{2M}\right)^{L\!-\!(n_{-}\!-\!1)} \\
\vdots & \vdots & \vdots & \vdots \\
\tilde{G}_{1}^{(N,1)}\left(\beta_{1}\right)^{L\!-\!(n_{-}\!-\!1)} & \tilde{G}_{2}^{(N,1)}\left(\beta_{2}\right)^{L\!-\!(n_{-}\!-\!1)} & \cdots & \tilde{G}_{2M}^{(N,1)}\left(\beta_{2M}\right)^{L\!-\!(n_{-}\!-\!1)} \\
\vdots & \vdots & \vdots & \vdots \\
\tilde{G}_{1}^{(1,n_{-})}\left(\beta_{1}\right)^{L} & \tilde{G}_{2}^{(1,n_{-})}\left(\beta_{2}\right)^{L} & \cdots & \tilde{G}_{2M}^{(1,n_{-})}\left(\beta_{2M}\right)^{L} \\
\tilde{G}_{1}^{(2,n_{-})}\left(\beta_{1}\right)^{L} & \tilde{G}_{2}^{(2,n_{-})}\left(\beta_{2}\right)^{L} & \cdots & \tilde{G}_{2M}^{(2,n_{-})}\left(\beta_{2M}\right)^{L} \\
\vdots & \vdots & \vdots & \vdots \\
\tilde{G}_{1}^{(N,n_{-})}\left(\beta_{1}\right)^{L} & \tilde{G}_{2}^{(N,n_{-})}\left(\beta_{2}\right)^{L} & \cdots & \tilde{G}_{2M}^{(N,n_{-})}\left(\beta_{2M}\right)^{L}
	\end{vmatrix}=0,
	\label{eqapp2a_multi}
\end{equation}
where
\begin{align}
\tilde{F}_{j}^{(1,i)}&= \sum_{n=-(i-1)}^{n_{+}}(h_{n}^{11}+f_{j}^{(2)}h_{n}^{12}+\cdots+f_{j}^{(N)}h_{n}^{1N})\left(\beta_{j}\right)^{n} - E_{\rm OBC},\\
\tilde{F}_{j}^{(2,i)}&= \sum_{n=-(i-1)}^{n_{+}}(h_{n}^{21}+f_{j}^{(2)}h_{n}^{22}+\cdots+f_{j}^{(N)}h_{n}^{2N})\left(\beta_{j}\right)^{n} - f_{j}^{(2)}E_{\rm OBC},\\
& \vdots \nonumber\\
\tilde{F}_{j}^{(N,i)}&= \sum_{n=-(i-1)}^{n_{+}}(h_{n}^{N1}+f_{j}^{(2)}h_{n}^{N2}+\cdots+f_{j}^{(N)}h_{n}^{NN})\left(\beta_{j}\right)^{n} - f_{j}^{(N)}E_{\rm OBC},\\
\tilde{G}_{j}^{(1,i)}&= \sum_{n=-n_{-}}^{n_{-}-i}(h_{n}^{11}+f_{j}^{(2)}h_{n}^{12}+\cdots+f_{j}^{(N)}h_{n}^{1N})\left(\beta_{j}\right)^{n} - E_{\rm OBC},\\
\tilde{G}_{j}^{(2,i)}&= \sum_{n=-n_{-}}^{n_{-}-i}(h_{n}^{12}+f_{j}^{(2)}h_{n}^{22}+\cdots+f_{j}^{(N)}h_{n}^{2N})\left(\beta_{j}\right)^{n} - f_{j}^{(2)}E_{\rm OBC}, \\
& \vdots \nonumber\\
\tilde{G}_{j}^{(N,i)}&= \sum_{n=-n_{-}}^{n_{-}-i}(h_{n}^{N1}+f_{j}^{(2)}h_{n}^{N2}+\cdots+f_{j}^{(N)}h_{n}^{NN})\left(\beta_{j}\right)^{n} - f_{j}^{(N)}E_{\rm OBC}.
\end{align} 
We can collect the terms and express Eq.~(\ref{eqapp2a_multi}) as a multivariate polynomial of the form
\begin{align}
	\sum_{P,Q}\tilde{J}(\beta_{i\in P},\beta_{j\in Q},E_{\rm OBC})\!\left[\prod_{i\in P}\left(\beta_{i}\right)^{k}\right]\! \left[\prod_{j\in Q}\left(\beta_{j}\right)^{k'}\right]\!=\!0,\label{eqapp3a_multi}
\end{align} where $k=1,\cdots,n_{+}$, $k'=L\!-\!(n_{-}\!-\!1),\cdots,L$, the sets $P$ and $Q$ are two disjoint subsets of the set $\{1,2,\cdots,2M\}$ with $M$ elements, respectively. 

By setting $n_{+}=n_{-}$, Eq.~(\ref{eqapp3a_multi}) can be reduced to
\begin{align}
\sum_{P,Q}\tilde{J}(\beta_{i\in P},\beta_{j\in Q},E_{\rm OBC})\!\left[\prod_{i\in P}\left(\beta_{i}\right)^{L+1}\right]\!=\!0.\label{eqapp4a_multi}
\end{align} 

In Eq.~(\ref{eqapp4a_multi}), there are two leading terms proportional to $(\beta_{M}\beta_{M+2}\beta_{M+3}\cdots\beta_{2M})^{L+1}$ and $(\beta_{M+1}\beta_{M+2}\beta_{M+3}\cdots\beta_{2M})^{L+1}$.
Therefore, in the limit of large system size $L$, we can reduce (\ref{eqapp4a_multi}), which solves the characteristic dispersion equation (\ref{eq:characteristic_multi}) and open boundary conditions, to the familiar form
\begin{eqnarray}
	\left|\frac{\beta_{M}}{\beta_{M+1}}\right|
	\simeq\left|-\frac{\tilde{J}(\beta_{i\in P_1},\beta_{j\in Q_1},E_{\rm OBC})}{\tilde{J}(\beta_{i\in P_2},\beta_{j\in Q_2},E_{\rm OBC})}\right|_{E_{\rm OBC}=E_{\infty}}^{\frac{1}{L+1}},
	\label{eq:boundary_equation1_multi}
\end{eqnarray} where $P_1=\{M+1,M+2,M+3,\cdots,2M\}$, $Q_1=\{1,2,3,\cdots,M\}$, $P_2=\{M,M+2,M+3,\cdots,2M\}$, $Q_2=\{1,2,\cdots,M-2,M-1,M+1\}$, and $L$ is the system size with $L\to\infty$. For large $L$, the right hand side (RHS) tends towards unity, and hence $|\beta_M|\approx |\beta_{M+1}|$ for the OBC eigenfunctions in the thermodynamic limit (in practice, $L\simeq 20$ is usually sufficient large when the cNHSE is \emph{absent}).

\section{Numerical confirmation of the validity of the GBZ upon extrapolating to finite-size systems}\label{Appendix_III}

\begin{figure}[H]
\includegraphics[width=1\linewidth]{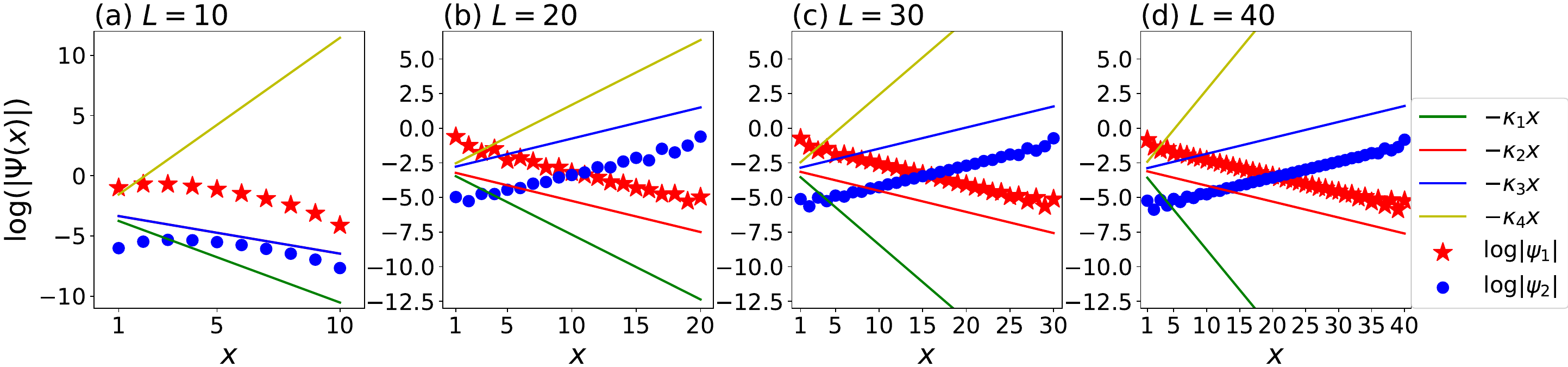}
\caption{Spatial decay of eigenstates and how they are determined by $\beta_M$ and $\beta_{M+1}$ ($M=2$). Plotted are the $\log(|\psi(x)|)$ ($\log(|\psi(x)|)=\ln(|\psi(x)|)$) of two representative eigenstates with different left/right localizations (red stars and blue disks), at different finite system sizes $L=10,20,30,40$ (a)--(d). Compared against them are the decay profiles corresponding to the four $\kappa=-\log|\beta|=-\ln|\beta|$ solutions. We see that $|\beta_2|=e^{-\kappa_2}$ and $|\beta_3|=e^{-\kappa_3}$ controls the eigenstate decay rate very well down to $L=20$, even though in principle, they rigorously determine the decay rate only in the thermodynamic limit. (a) The eigenstates correspond to $\arg(E_{\rm OBC})=0$ and Max(Re($E_{\rm OBC}$)). (b)--(d) The eigenstates correspond to $\arg(E_{\rm OBC})=\frac{\pi}{2}$. Here, $V=0.5$ and the other parameters are $t_0=0.01$, $t_1=0.75$, and $\delta_a=-\delta_b=0.25$, which are the same as those in Fig.~\ref{fig:E_V} of the main text.}
\label{fig:kappa_L}
\end{figure}

Here in Fig.~\ref{fig:kappa_L}, we numerically confirm that for our coupled Hatano-Nelson model, the GBZ solutions $\beta_M=\beta_2$ and $\beta_{M+1}=\beta_3$ still largely determine the eigensolution decay rates down to small system sizes.

\section{Derivation of Eq.~(\ref{eq:boundary_equation0}) for the OBC constraints of the coupled Hatano-Nelson model}\label{Appendix_IV}

From the real-space eigenequations (\ref{eq:OBC_Schrodinger}) subjected to OBCs $\psi_{0,\alpha}=\psi_{L+1,\alpha}=0~(\alpha={\rm A},{\rm B})$, we have
\begin{eqnarray}
\left\{ \begin{array}{l}
t_{0}\psi_{1,{\rm B}}+t_{a}^{+}\psi_{2,{\rm A}}+V\psi_{1,{\rm A}}=E_{\rm OBC}\psi_{1,{\rm A}},   \vspace{5pt}\\
t_{0}\psi_{1,{\rm A}}+t_{b}^{+}\psi_{2,{\rm B}}-V\psi_{1,{\rm B}}=E_{\rm OBC}\psi_{1,{\rm B}},   \vspace{5pt}\\
t_{a}^{-}\psi_{L-1,{\rm A}}+t_{0}\psi_{L,{\rm B}}+V\psi_{L,{\rm A}}=E_{\rm OBC}\psi_{L,{\rm A}}, \vspace{5pt}\\
t_{b}^{-}\psi_{L-1,{\rm B}}+t_{0}\psi_{L,{\rm A}}-V\psi_{L,{\rm B}}=E_{\rm OBC}\psi_{L,{\rm B}}.
\end{array}\right.
\label{eqapp1a}
\end{eqnarray}
By substituting the ansatz $\left( \psi_{n,{\rm A}},\psi_{n,{\rm B}}\right)^{\rm T}=\sum_{j=1}^4\beta_j^n\left(\phi_{\rm A}^{(j)},\phi_{\rm B}^{(j)}\right)^{\rm T}$ into Eq.~(\ref{eqapp1a}), we can get
\begin{eqnarray}
\left\{ \begin{array}{l}
t_{0}\sum_{j=1}^{4}\beta_{j}\phi_{\rm B}^{(j)}+t_{a}^{+}\sum_{j=1}^{4}\beta_{j}^{2}\phi_{\rm A}^{(j)}+V\sum_{j=1}^{4}\beta_{j}\phi_{\rm A}^{(j)}=E_{\rm OBC}\sum_{j=1}^{4}\beta_{j}\phi_{\rm A}^{(j)},   \vspace{5pt}\\
t_{0}\sum_{j=1}^{4}\beta_{j}\phi_{\rm A}^{(j)}+t_{b}^{+}\sum_{j=1}^{4}\beta_{j}^{2}\phi_{\rm B}^{(j)}-V\sum_{j=1}^{4}\beta_{j}\phi_{\rm B}^{(j)}=E_{\rm OBC}\sum_{j=1}^{4}\beta_{j}\phi_{\rm B}^{(j)},   \vspace{5pt}\\
t_{a}^{-}\sum_{j=1}^{4}\beta_{j}^{L-1}\phi_{\rm A}^{(j)}+t_{0}\sum_{j=1}^{4}\beta_{j}^{L}\phi_{\rm B}^{(j)}+V\sum_{j=1}^{4}\beta_{j}^{L}\phi_{\rm A}^{(j)}=E_{\rm OBC}\sum_{j=1}^{4}\beta_{j}^{L}\phi_{\rm A}^{(j)}, \vspace{5pt}\\
t_{b}^{-}\sum_{j=1}^{4}\beta_{j}^{L-1}\phi_{\rm B}^{(j)}+t_{0}\sum_{j=1}^{4}\beta_{j}^{L}\phi_{\rm A}^{(j)}-V\sum_{j=1}^{4}\beta_{j}^{L}\phi_{\rm B}^{(j)}=E_{\rm OBC}\sum_{j=1}^{4}\beta_{j}^{L}\phi_{\rm B}^{(j)},
\end{array}\right.
\end{eqnarray} Furthermore, by using the bulk eigenequation in Eq.~(\ref{eq:bulk_eigenequation}):
\begin{eqnarray}
\left\{ \begin{array}{l}
(t_{a}^{+}\beta+t_{a}^{-}\beta^{-1}+V-E_{\rm OBC})\phi_{{\rm A}}+t_{0}\phi_{{\rm B}}=0,~\frac{\phi_{{\rm B}}}{\phi_{{\rm A}}}=-\frac{(t_{a}^{+}\beta+t_{a}^{-}\beta^{-1}+V-E_{\rm OBC})}{t_{0}},   \\
t_{0}\phi_{{\rm A}}+(t_{b}^{+}\beta+t_{b}^{-}\beta^{-1}-V-E_{\rm OBC})\phi_{{\rm B}}=0,~\frac{\phi_{{\rm B}}}{\phi_{{\rm A}}}=-\frac{t_{0}}{(t_{b}^{+}\beta+t_{b}^{-}\beta^{-1}-V-E_{\rm OBC})},
\end{array}\right.
\end{eqnarray}  i.e.,
\begin{align}\label{eq:phi}
&\frac{\phi^{(j)}_{\rm B}}{\phi_{\rm A}^{(j)}} = \frac{(E_{\rm OBC}-t_{a}^{+}\beta_{j}-t_{a}^{-}\beta_{j}^{-1}-V)}{t_{0}} = \frac{t_{0}}{(E_{\rm OBC}-t_{b}^{+}\beta_{j}-t_{b}^{-}\beta_{j}^{-1}+V)} = f_{j},\\
&t_{0}^{2}=(E_{\rm OBC}-t_{a}^{+}\beta_{j}-t_{a}^{-}\beta_{j}^{-1}-V)(E_{\rm OBC}-t_{b}^{+}\beta_{j}-t_{b}^{-}\beta_{j}^{-1}+V) ,\\
&\phi^{(j)}_{\rm B}=f_{j}\phi^{(j)}_{\rm A},
\end{align} we have 
\begin{eqnarray}
\left\{ \begin{array}{l}
t_{0}\sum_{j=1}^{4}\beta_{j}f_{j}\phi^{(j)}_{\rm A}+t_{a}^{+}\sum_{j=1}^{4}\beta_{j}^{2}\phi_{\rm A}^{(j)}+V\sum_{j=1}^{4}\beta_{j}\phi^{(j)}_{\rm A}=E_{\rm OBC}\sum_{j=1}^{4}\beta_{j}\phi_{\rm A}^{(j)},   \vspace{5pt}\\
t_{0}\sum_{j=1}^{4}\beta_{j}\phi_{\rm A}^{(j)}+t_{b}^{+}\sum_{j=1}^{4}\beta_{j}^{2}f_{j}\phi^{(j)}_{\rm A}-V\sum_{j=1}^{4}\beta_{j}f_{j}\phi^{(j)}_{\rm A}=E_{\rm OBC}\sum_{j=1}^{4}\beta_{j}f_{j}\phi^{(j)}_{\rm A},   \vspace{5pt}\\
t_{a}^{-}\sum_{j=1}^{4}\beta_{j}^{L-1}\phi_{\rm A}^{(j)}+t_{0}\sum_{j=1}^{4}\beta_{j}^{L}f_{j}\phi^{(j)}_{\rm A}+V\sum_{j=1}^{4}\beta_{j}^{L}\phi_{\rm A}^{(j)}=E_{\rm OBC}\sum_{j=1}^{4}\beta_{j}^{L}\phi_{\rm A}^{(j)}, \vspace{5pt}\\
t_{b}^{-}\sum_{j=1}^{4}\beta_{j}^{L-1}f_{j}\phi^{(j)}_{\rm A}+t_{0}\sum_{j=1}^{4}\beta_{j}^{L}\phi_{\rm A}^{(j)}-V\sum_{j=1}^{4}\beta_{j}^{L}f_{j}\phi^{(j)}_{\rm A}=E_{\rm OBC}\sum_{j=1}^{4}\beta_{j}^{L}f_{j}\phi^{(j)}_{\rm A},
\end{array}\right.
\end{eqnarray}
\begin{eqnarray}
\left\{ \begin{array}{l}
\sum_{j=1}^{4}(E_{\rm OBC}-t_{a}^{+}\beta_{j}-t_{a}^{-}\beta_{j}^{-1}-V)\beta_{j}\phi^{(j)}_{\rm A}+t_{a}^{+}\sum_{j=1}^{4}\beta_{j}^{2}\phi_{\rm A}^{(j)}+V\sum_{j=1}^{4}\beta_{j}\phi^{(j)}_{\rm A}=E_{\rm OBC}\sum_{j=1}^{4}\beta_{j}\phi_{\rm A}^{(j)},   \vspace{5pt}\\
t_{0}\sum_{j=1}^{4}\beta_{j}\phi_{\rm A}^{(j)}+\sum_{j=1}^{4}(t_{b}^{+}\beta_{j}-V-E_{\rm OBC})\beta_{j}f_{j}\phi^{(j)}_{\rm A}=0,   \vspace{5pt}\\
t_{a}^{-}\sum_{j=1}^{4}\beta_{j}^{L-1}\phi_{\rm A}^{(j)}+\sum_{j=1}^{4}(E_{\rm OBC}-t_{a}^{+}\beta_{j}-t_{a}^{-}\beta_{j}^{-1}-V)\beta_{j}^{L}\phi^{(j)}_{\rm A}+V\sum_{j=1}^{4}\beta_{j}^{L}\phi_{\rm A}^{(j)}=E_{\rm OBC}\sum_{j=1}^{4}\beta_{j}^{L}\phi_{\rm A}^{(j)}, \vspace{5pt}\\
\sum_{j=1}^{4}(t_{b}^{-}\beta_{j}^{-1}-V-E_{\rm OBC})\beta_{j}^{L}f_{j}\phi^{(j)}_{\rm A}+t_{0}\sum_{j=1}^{4}\beta_{j}^{L}\phi_{\rm A}^{(j)}=0,
\end{array}\right.
\end{eqnarray}
\begin{eqnarray}
\left\{ \begin{array}{l}
\sum_{j=1}^{4}(-t_{a}^{-}\beta_{j}^{-1})\beta_{j}\phi^{(j)}_{\rm A}=0,   \vspace{5pt}\\
\sum_{j=1}^{4}\beta_{j}\phi_{\rm A}^{(j)}+\sum_{j=1}^{4}\frac{(t_{b}^{+}\beta_{j}-V-E_{\rm OBC})}{(E_{\rm OBC}-t_{b}^{+}\beta_{j}-t_{b}^{-}\beta_{j}^{-1}+V)}\beta_{j}\phi^{(j)}_{\rm A}=0,   \vspace{5pt}\\
\sum_{j=1}^{4}(-t_{a}^{+}\beta_{j})\beta_{j}^{L}\phi^{(j)}_{\rm A}=0, \vspace{5pt}\\
\sum_{j=1}^{4}\frac{(t_{b}^{-}\beta_{j}^{-1}-V-E_{\rm OBC})}{(E_{\rm OBC}-t_{b}^{+}\beta_{j}-t_{b}^{-}\beta_{j}^{-1}+V)}\beta_{j}^{L}\phi^{(j)}_{\rm A}+\sum_{j=1}^{4}\beta_{j}^{L}\phi_{\rm A}^{(j)}=0,
\end{array}\right.
\end{eqnarray}
\begin{eqnarray}
\left\{ \begin{array}{l}
\sum_{j=1}^{4}\phi^{(j)}_{\rm A}=0,   \vspace{5pt}\\
\sum_{j=1}^{4}\frac{1}{(E_{\rm OBC}-t_{b}^{+}\beta_{j}-t_{b}^{-}\beta_{j}^{-1}+V)}\phi^{(j)}_{\rm A}=0,~\sum_{j=1}^{4}\phi^{(j)}_{\rm A}=\sum_{j=1}^{4}(E_{\rm OBC}-t_{a}^{+}\beta_{j}-t_{a}^{-}\beta_{j}^{-1}-V)\phi^{(j)}_{\rm A}=0,   \vspace{5pt}\\
\sum_{j=1}^{4}\beta_{j}^{L+1}\phi^{(j)}_{\rm A}=0, \vspace{5pt}\\
\sum_{j=1}^{4}\frac{1}{(E_{\rm OBC}-t_{b}^{+}\beta_{j}-t_{b}^{-}\beta_{j}^{-1}+V)}\beta_{j}^{L+1}\phi^{(j)}_{\rm A}=0,~\sum_{j=1}^{4}(E_{\rm OBC}-t_{a}^{+}\beta_{j}-t_{a}^{-}\beta_{j}^{-1}-V)\beta_{j}^{L+1}\phi^{(j)}_{\rm A}=0,
\end{array}\right.
\end{eqnarray} where we have used the characteristic dispersion equation
\begin{eqnarray}
\frac{1}{(E_{\rm OBC}-t_{b}^{+}\beta_{j}-t_{b}^{-}\beta_{j}^{-1}+V)}=\frac{(E_{\rm OBC}-t_{a}^{+}\beta_{j}-t_{a}^{-}\beta_{j}^{-1}-V)}{t_{0}^{2}}.
\end{eqnarray}

Imposing the condition that $\phi_{\rm A}^{\left(j\right)}~(j=1,2,3,4)$ do not vanish, we must have the vanishing determinant:
\begin{eqnarray}
\left| \begin{array}{cccc}
1                             & 1                             & 1                             & 1                             \vspace{5pt}\\
X_1                           & X_2                           & X_3                           & X_4                           \vspace{5pt}\\
\beta_1^{L+1}    & \beta_2^{L+1}    & \beta_3^{L+1}    & \beta_4^{L+1}    \vspace{5pt}\\
X_1\beta_1^{L+1} & X_2\beta_2^{L+1} & X_3\beta_3^{L+1} & X_4\beta_4^{L+1}
\end{array}\right|=0,
\label{eqapp2a}
\end{eqnarray}
where $\left|\beta_1\right|\leqslant\left|\beta_2\right|\leqslant\left|\beta_3\right|\leqslant\left|\beta_4\right|$. Here, $X_j~(j=1,2,3,4)$ are defined as
\begin{equation}
X_j\equiv E_{\rm OBC}-t_{a}^{+}\beta_j-t_{a}^{-}\beta_j^{-1}-V,~\left(j=1,2,3,4\right).
\label{eq:Xj}
\end{equation}
Simplifying, we obtain the boundary equation (\ref{eq:boundary_equation0}) from Eq.~(\ref{eqapp2a}):
\begin{eqnarray}\label{eq:boundary_L1}
X_{1,4}X_{2,3}\left[\left(\beta_1\beta_4\right)^{L+1}+\left(\beta_2\beta_3\right)^{L+1}\right] 
-X_{1,3}X_{2,4}\left[\left(\beta_1\beta_3\right)^{L+1}+\left(\beta_2\beta_4\right)^{L+1}\right] 
+X_{1,2}X_{3,4}\left[\left(\beta_1\beta_2\right)^{L+1}+\left(\beta_3\beta_4\right)^{L+1}\right]=0,\nonumber\\
\end{eqnarray}
where $X_{i,j}~(i,j=1,2,3,4)$ are defined as
\begin{equation}
X_{i,j}\equiv X_{i}-X_{j}=t_{a}^{+}(\beta_j-\beta_i)+t_{a}^{-}(\beta_j^{-1}-\beta_i^{-1}),~\left(i,j=1,2,3,4\right).
\end{equation}

\section{Derivation of Eq.~(\ref{eq:boundary_equation1})}\label{Appendix_V}

We start from the characteristic equation of our coupled Hatano-Nelson model with offset:
\begin{eqnarray}
t_{a}^{+}t_{b}^{+}\beta^2 
&+&[-(t_{a}^{+}+t_{b}^{+})E_{\rm OBC} - (t_{a}^{+}-t_{b}^{+})V ]\beta + \left(t_{a}^{+}t_{b}^{-}+t_{a}^{-}t_{b}^{+}+E_{\rm OBC}^2-t_{0}^2-V^2\right) \nonumber\\
&+&[-(t_{a}^{-}+t_{b}^{-})E_{\rm OBC} - (t_{a}^{-}-t_{b}^{-})V ]\beta^{-1}+t_{a}^{-}t_{b}^{-}\beta^{-2}=0.
\label{eq:characteristic_S}
\end{eqnarray}
We consider a perturbative solution up to the second order in $t_0$ by first expanding in terms of the $\beta$s:
\begin{eqnarray}
\left\{ \begin{array}{l}
\beta_1\simeq x_-^{(a)}+y_-^{(a)}t_{0}^2, \vspace{5pt}\\
\beta_2\simeq x_+^{(a)}+y_+^{(a)}t_{0}^2, \vspace{5pt}\\
\beta_3\simeq x_-^{(b)}+y_-^{(b)}t_{0}^2, \vspace{5pt}\\
\beta_4\simeq x_+^{(b)}+y_+^{(b)}t_{0}^2,
\end{array}\right.
\label{eq:characteristic_solutions_S}
\end{eqnarray}
where $t_{a}^{+}>t_{a}^{-}$, $t_{b}^{+}<t_{b}^{-}$, $t_{a}^{+}>t_{b}^{+}$, $V>0$, $\left|\beta_1\right|\leqslant\left|\beta_2\right|\leqslant\left|\beta_3\right|\leqslant\left|\beta_4\right|$, 
\begin{eqnarray}
\left\{ \begin{array}{l}
\displaystyle x_\pm^{(a)}=\frac{1}{2t_{a}^{+}}(E_{\rm OBC}-V\pm \Delta_a), \vspace{5pt}\\
\displaystyle x_\pm^{(b)}=\frac{1}{2t_{b}^{+}}(E_{\rm OBC}+V\pm \Delta_b), \vspace{5pt}\\
\displaystyle y_\pm^{(a)}=\frac{-[E_{\rm OBC}^2-2t_{a}^{+}t_{a}^{-}+V(V\mp\Delta_a)-(2V\mp\Delta_a)E_{\rm OBC}]}{g_{\pm}^{(a)}}, \vspace{5pt}\\
\displaystyle y_\pm^{(b)}=\frac{[E_{\rm OBC}^2-2t_{b}^{+}t_{b}^{-}+V(V\pm\Delta_b)+(2V\pm\Delta_b)E_{\rm OBC}]}{g_{\pm}^{(b)}},
\end{array}\right. 
\end{eqnarray}
and
\begin{eqnarray}
\Delta_a&=&\sqrt{(E_{\rm OBC}-V)^{2} - 4t_{a}^{+}t_{a}^{-} },\\
\Delta_b&=&\sqrt{(E_{\rm OBC}+V)^{2} - 4t_{b}^{+}t_{b}^{-} },\\
g_{\pm}^{(a)}&=&E_{\rm OBC}^{3}(t_{a}^{+}-t_{b}^{+})-4t_{a}^{+}t_{a}^{-}(t_{a}^{+}+t_{b}^{+})V+(t_{a}^{+}+t_{b}^{+})V^{3} \mp 2t_{a}^{+}(t_{a}^{+}t_{b}^{-}-t_{a}^{-}t_{b}^{+})\Delta_{a} \nonumber\\
&&\mp (t_{a}^{+}+t_{b}^{+})\Delta_{a}V^{2} + E_{\rm OBC}^{2}[(3t_{b}^{+}-t_{a}^{+})V\pm(t_{a}^{+}-t_{b}^{+})\Delta_{a}] \nonumber\\ 
&&+ E_{\rm OBC}[-4t_{a}^{+}t_{a}^{-}(t_{a}^{+}-t_{b}^{+}) - (t_{a}^{+}+3t_{b}^{+})V^{2} \pm 2t_{b}^{+}\Delta_{a}V] ,\\
g_{\pm}^{(b)}&=&E_{\rm OBC}^{3}(t_{a}^{+}-t_{b}^{+})-4t_{b}^{+}t_{b}^{-}(t_{a}^{+}+t_{b}^{+})V+(t_{a}^{+}+t_{b}^{+})V^{3} \mp 2t_{b}^{+}(t_{a}^{+}t_{b}^{-}-t_{a}^{-}t_{b}^{+})\Delta_{b} \nonumber\\
&&\pm (t_{a}^{+}+t_{b}^{+})\Delta_{b}V^{2} + E_{\rm OBC}^{2}[(3t_{a}^{+}-t_{b}^{+})V\pm(t_{a}^{+}-t_{b}^{+})\Delta_{b}] \nonumber\\ 
&&+ E_{\rm OBC}[-4t_{b}^{+}t_{b}^{-}(t_{a}^{+}-t_{b}^{+}) + (t_{b}^{+}+3t_{a}^{+})V^{2} \pm 2t_{a}^{+}\Delta_{b}V].
\end{eqnarray}
With Eq.~(\ref{eq:characteristic_solutions_S}) and (\ref{eq:Xj}), we can get
\begin{eqnarray}
\left\{ \begin{array}{l}
\displaystyle X_{1,2}=\frac{(t_{a}^{+}t_{b}^{-}-t_{a}^{-}t_{b}^{+})\Delta_{a}t_0^2}{E_{\rm OBC}^{2}(t_{a}^{-}-t_{b}^{-})(t_{a}^{+}-t_{b}^{+}) + (t_{a}^{+}t_{b}^{-}-t_{a}^{-}t_{b}^{+})^{2} + 2E_{\rm OBC}V(t_{a}^{+}t_{a}^{-}-t_{b}^{+}t_{b}^{-}) + (t_{a}^{-}+t_{b}^{-})(t_{a}^{+}+t_{b}^{+})V^{2} }+{\cal O}(t_0^4), \vspace{5pt}\\
\displaystyle X_{3,4}=\left(\frac{t_{a}^{+}}{t_{b}^{+}}-\frac{t_{a}^{-}}{t_{b}^{-}}\right)\Delta_{b}+{\cal O}\left(t_0^2\right), \vspace{5pt}\\
\displaystyle X_{1,3}=\frac{1}{2}\left[\frac{t_{a}^{+}}{t_{b}^{+}}(E_{\rm OBC}+V-\Delta_{b})-(E_{\rm OBC}-V-\Delta_{a})\right]+\left[\frac{2t_{b}^{+}t_{a}^{-}}{E_{\rm OBC}+V-\Delta_{b}}-\frac{2t_{a}^{+}t_{a}^{-}}{E_{\rm OBC}-V-\Delta_{a}}\right]+{\cal O}(t_0^2), \vspace{5pt}\\
\displaystyle X_{2,4}=\frac{1}{2}\left[\frac{t_{a}^{+}}{t_{b}^{+}}(E_{\rm OBC}+V+\Delta_{b})-(E_{\rm OBC}-V+\Delta_{a})\right]+\left[\frac{2t_{b}^{+}t_{a}^{-}}{E_{\rm OBC}+V+\Delta_{b}}-\frac{2t_{a}^{+}t_{a}^{-}}{E_{\rm OBC}-V+\Delta_{a}}\right]+{\cal O}(t_0^2).
\end{array}\right. 
\label{eqapp5c}
\end{eqnarray}
We can obtain Eq.~(\ref{eq:boundary_equation1}) by substituting Eq.~(\ref{eqapp5c}) into $X_{1,2}X_{3,4}/(X_{1,3}X_{2,4})$ and expanding up to the second order in $t_0$.

\section{OBC spectra for topologically coupled chain model}\label{Appendix_VI}

In this appendix, in order to understand why the topological zero modes appear at $E=0$ in the point gap only at sufficiently large system sizes, we show the OBC energy spectra of the topologically coupled chain model (\ref{eq:Lee_Hr_topo}) with $V=0$ for various system sizes.

\begin{figure}[h]
	\includegraphics[width=0.75\linewidth]{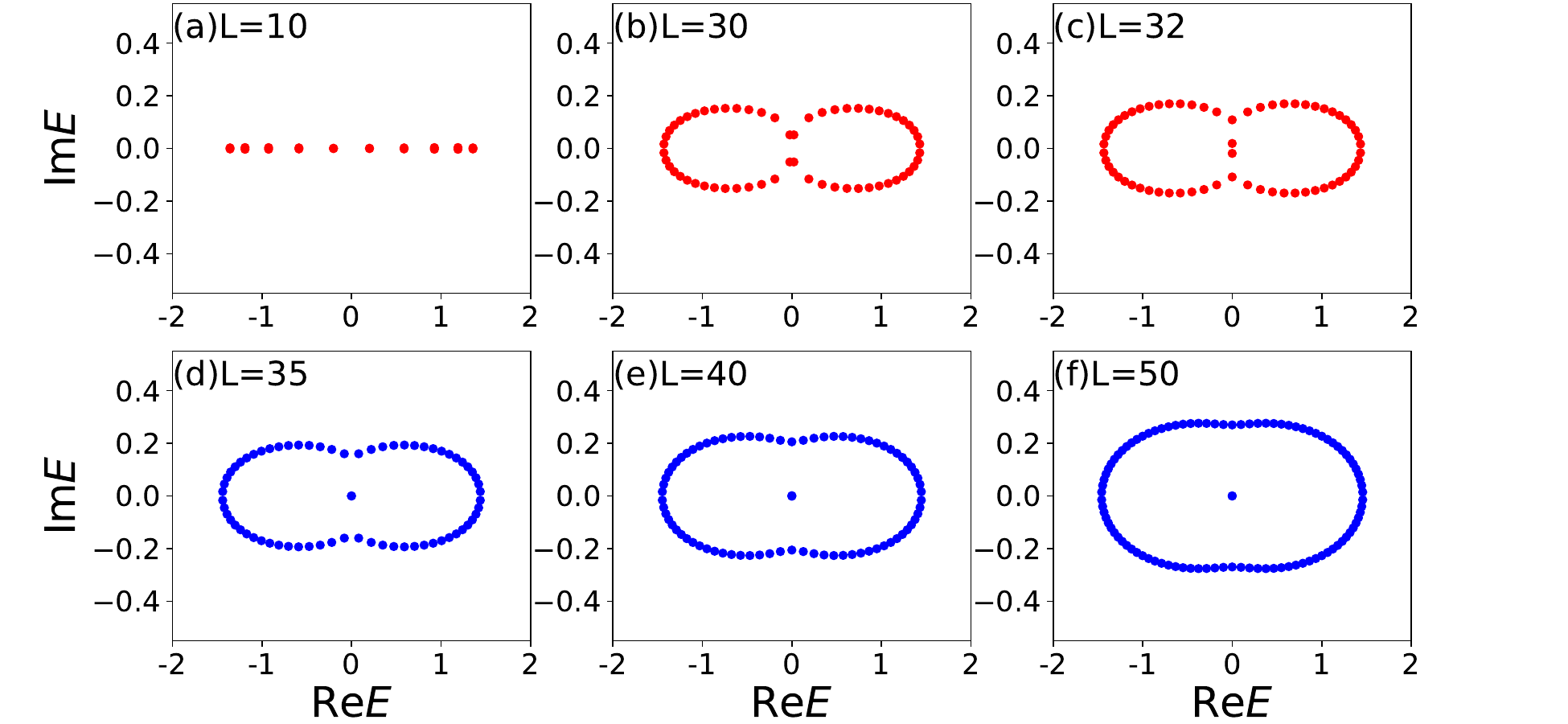}
	\caption{OBC energy spectra of the topologically coupled chain model Hamiltonian (\ref{eq:Lee_Hr_topo}) with $V=0$ at different system sizes (a) $L=10$, (b) $L=30$, (c) $L=32$, (d) $L=35$, (e) $L=40$, (f) $L=50$. Notably, topological zero modes appear at $E=0$ in the point gap only at sufficiently large system sizes of $L=35,40,50$. The other parameters are $\delta_{ab}=0.5\times10^{-3}$, $t_1=0.75$, and $\delta_a=-\delta_b=0.25$.} 
	\label{fig:E_OBC_V0_L_together2_S}
\end{figure}

It is indicated from Fig.~\ref{fig:E_OBC_V0_L_together2_S} that, when we tune the system size $L$ (regarding $L$ as a parameter), the OBC spectrum changes. At a critical $L$, the OBC spectrum's gap closes and after that, topological zero modes appear.


\section{Derivation of Eq.~(\ref{eq:boundary_equation0_topo})}\label{Appendix_VII}

In this appendix, we describe the derivation of Eq.~\eqref{eq:boundary_equation0_topo}, which expresses the OBC constraint of our coupled topological model. Under OBCs, we can write the real-space Schr{\"o}dinger equation ${\cal H}_{\rm t}|\psi\rangle=E_{\rm OBC}|\psi\rangle$ [where ${\cal H}_{\rm t}$ is the Hamiltonian matrix of $H_{\rm t}$ in the basis $(C_{1},C_{2},\cdots,C_{L})^T$], where $|\psi\rangle=\left(\psi_{1,{\rm A}},\psi_{1,{\rm B}},\psi_{2,{\rm A}},\psi_{2,{\rm B}},\dots,\psi_{n,{\rm A}},\psi_{n,{\rm B}},\dots\right)^{\rm T}$, as
\begin{eqnarray}
\left\{ \begin{array}{l}
t_{a}^{-}\psi_{n-1,{\rm A}}+\delta_{ab}\psi_{n-1,{\rm B}}+V\psi_{n,{\rm A}}+t_{a}^{+}\psi_{n+1,{\rm A}}+\delta_{ab}\psi_{n+1,{\rm B}}=E_{\rm OBC}\psi_{n,{\rm A}}, \vspace{3pt}\\
t_{b}^{-}\psi_{n-1,{\rm B}}-\delta_{ab}\psi_{n-1,{\rm A}}-V\psi_{n,{\rm B}}+t_{b}^{+}\psi_{n+1,{\rm B}}-\delta_{ab}\psi_{n+1,{\rm A}}=E_{\rm OBC}\psi_{n,{\rm B}}.
\end{array}\right.\label{eq:OBC_Schrodinger_topo}
\end{eqnarray}
According to the theory of linear difference equations, we can take as an ansatz for the eigenstates the linear combination:
\begin{eqnarray}
\left( \begin{array}{c}
\psi_{n,{\rm A}} \vspace{5pt}\\
\psi_{n,{\rm B}}
\end{array}\right)=\sum_{j=1}^4\beta_j^n\left( \begin{array}{c}
\phi_{\rm A}^{\left(j\right)} \vspace{5pt}\\
\phi_{\rm B}^{\left(j\right)}
\end{array}\right).
\end{eqnarray}
Hence, Eq.~(\ref{eq:OBC_Schrodinger_topo}) can be rewritten as
\begin{eqnarray}
\left( \begin{array}{cc}
t_{a}^{+}\beta+t_{a}^{-}\beta^{-1}+V & \delta_{ab}(\beta+\beta^{-1})                 \vspace{5pt}\\
-\delta_{ab}(\beta+\beta^{-1})                 & t_{b}^{+}\beta+t_{b}^{-}\beta^{-1}-V
\end{array}\right)\left( \begin{array}{c}
\phi_{\rm A} \vspace{5pt}\\
\phi_{\rm B}
\end{array}\right)=E_{\rm OBC}\left( \begin{array}{c}
\phi_{\rm A} \vspace{5pt}\\
\phi_{\rm B}
\end{array}\right). 
\label{eq:OBC_Schrodinger_topo_bulk}
\end{eqnarray}
From the real-space eigenequation in Eq.~(\ref{eq:OBC_Schrodinger_topo}) and the open boundary conditions $\psi_{0,\alpha}=\psi_{L+1,\alpha}=0~(\alpha={\rm A},{\rm B})$, we can get the equations for the eigenstates in real space as
\begin{eqnarray}
\left\{ \begin{array}{l}
\delta_{ab}\psi_{2,{\rm B}}+t_{a}^{+}\psi_{2,{\rm A}}+V\psi_{1,{\rm A}}=E_{\rm OBC}\psi_{1,{\rm A}},   \vspace{5pt}\\
-\delta_{ab}\psi_{2,{\rm A}}+t_{b}^{+}\psi_{2,{\rm B}}-V\psi_{1,{\rm B}}=E_{\rm OBC}\psi_{1,{\rm B}},   \vspace{5pt}\\
t_{a}^{-}\psi_{L-1,{\rm A}}+\delta_{ab}\psi_{L-1,{\rm B}}+V\psi_{L,{\rm A}}=E_{\rm OBC}\psi_{L,{\rm A}}, \vspace{5pt}\\
t_{b}^{-}\psi_{L-1,{\rm B}}-\delta_{ab}\psi_{L-1,{\rm A}}-V\psi_{L,{\rm B}}=E_{\rm OBC}\psi_{L,{\rm B}}.
\end{array}\right.
\label{eqapp1a_topo}
\end{eqnarray}
Now, Eq.~(\ref{eqapp1a_topo}) can be rewritten into coupled equations for the coefficients $\phi_\alpha^{\left(j\right)}~(\alpha={\rm A},{\rm B};~j=1,2,3,4)$ by substituting the general solution $\left( \psi_{n,{\rm A}},  \psi_{n,{\rm B}}\right)^{T}=\sum_{j=1}^4\beta_j^n\left( 
\phi_{\rm A}^{(j)},  \phi_{\rm B}^{(j)}\right)^{T}$ as
\begin{eqnarray}
\left\{ \begin{array}{l}
\delta_{ab}\sum_{j=1}^{4}\beta_{j}^{2}\phi_{\rm B}^{(j)}+t_{a}^{+}\sum_{j=1}^{4}\beta_{j}^{2}\phi_{\rm A}^{(j)}+V\sum_{j=1}^{4}\beta_{j}\phi_{\rm A}^{(j)}=E_{\rm OBC}\sum_{j=1}^{4}\beta_{j}\phi_{\rm A}^{(j)},   \vspace{5pt}\\
-\delta_{ab}\sum_{j=1}^{4}\beta_{j}^{2}\phi_{\rm A}^{(j)}+t_{b}^{+}\sum_{j=1}^{4}\beta_{j}^{2}\phi_{\rm B}^{(j)}-V\sum_{j=1}^{4}\beta_{j}\phi_{\rm B}^{(j)}=E_{\rm OBC}\sum_{j=1}^{4}\beta_{j}\phi_{\rm B}^{(j)},   \vspace{5pt}\\
t_{a}^{-}\sum_{j=1}^{4}\beta_{j}^{L-1}\phi_{\rm A}^{(j)}+\delta_{ab}\sum_{j=1}^{4}\beta_{j}^{L-1}\phi_{\rm B}^{(j)}+V\sum_{j=1}^{4}\beta_{j}^{L}\phi_{\rm A}^{(j)}=E_{\rm OBC}\sum_{j=1}^{4}\beta_{j}^{L}\phi_{\rm A}^{(j)}, \vspace{5pt}\\
t_{b}^{-}\sum_{j=1}^{4}\beta_{j}^{L-1}\phi_{\rm B}^{(j)}-\delta_{ab}\sum_{j=1}^{4}\beta_{j}^{L-1}\phi_{\rm A}^{(j)}-V\sum_{j=1}^{4}\beta_{j}^{L}\phi_{\rm B}^{(j)}=E_{\rm OBC}\sum_{j=1}^{4}\beta_{j}^{L}\phi_{\rm B}^{(j)}.
\end{array}\right.
\label{eqapp1a2_topo}
\end{eqnarray} Furthermore, by using the bulk eigenequation in Eq.~(\ref{eq:OBC_Schrodinger_topo_bulk}):
\begin{eqnarray}
\left\{ \begin{array}{l}
(t_{a}^{+}\beta+t_{a}^{-}\beta^{-1}+V-E_{\rm OBC})\phi_{{\rm A}}+\delta_{ab}(\beta+\beta^{-1})\phi_{{\rm B}}=0,~\frac{\phi_{{\rm B}}}{\phi_{{\rm A}}}=-\frac{(t_{a}^{+}\beta+t_{a}^{-}\beta^{-1}+V-E_{\rm OBC})}{\delta_{ab}(\beta+\beta^{-1})},   \\
-\delta_{ab}(\beta+\beta^{-1})\phi_{{\rm A}}+(t_{b}^{+}\beta+t_{b}^{-}\beta^{-1}-V-E_{\rm OBC})\phi_{{\rm B}}=0,~\frac{\phi_{{\rm B}}}{\phi_{{\rm A}}}=\frac{\delta_{ab}(\beta+\beta^{-1})}{(t_{b}^{+}\beta+t_{b}^{-}\beta^{-1}-V-E_{\rm OBC})},
\end{array}\right.
\label{eqapp1a3_topo}
\end{eqnarray}  i.e.,
\begin{align}\label{eq:phi}
&\frac{\phi^{(j)}_{\rm B}}{\phi_{\rm A}^{(j)}} = \frac{(E_{\rm OBC}-t_{a}^{+}\beta_{j}-t_{a}^{-}\beta_{j}^{-1}-V)}{\delta_{ab}(\beta+\beta^{-1})} = \frac{-\delta_{ab}(\beta+\beta^{-1})}{(E_{\rm OBC}-t_{b}^{+}\beta_{j}-t_{b}^{-}\beta_{j}^{-1}+V)} = f_{j},\\
&-\delta_{ab}^{2}(\beta+\beta^{-1})^{2}=(E_{\rm OBC}-t_{a}^{+}\beta_{j}-t_{a}^{-}\beta_{j}^{-1}-V)(E_{\rm OBC}-t_{b}^{+}\beta_{j}-t_{b}^{-}\beta_{j}^{-1}+V) ,\\
&\phi^{(j)}_{\rm B}=f_{j}\phi^{(j)}_{\rm A}.
\end{align}
The general solution is written as a linear combination:
\begin{align}
\begin{pmatrix}
    \psi_{n,{\rm A}} \\
    \psi_{n,{\rm B}}  
\end{pmatrix} = 
\beta_{1}^{n}\begin{pmatrix}
    \phi_{\rm A}^{(1)} \\
    \phi_{\rm B} ^{(1)} 
\end{pmatrix} + 
\beta_{2}^{n}\begin{pmatrix}
    \phi_{\rm A}^{(2)} \\
    \phi_{\rm B} ^{(2)} 
\end{pmatrix} + 
\beta_{3}^{n}\begin{pmatrix}
    \phi_{\rm A}^{(3)} \\
    \phi_{\rm B} ^{(3)} 
\end{pmatrix} + 
\beta_{4}^{n}\begin{pmatrix}
    \phi_{\rm A}^{(4)} \\
    \phi_{\rm B} ^{(4)} 
\end{pmatrix}
\end{align}
which should satisfy the {open boundary conditions (\ref{eqapp1a2_topo}):
\begin{eqnarray}
\left\{ \begin{array}{l}
\delta_{ab}\sum_{j=1}^{4}\beta_{j}^{2}f_{j}\phi^{(j)}_{\rm A}+t_{a}^{+}\sum_{j=1}^{4}\beta_{j}^{2}\phi_{\rm A}^{(j)}+V\sum_{j=1}^{4}\beta_{j}\phi^{(j)}_{\rm A}=E_{\rm OBC}\sum_{j=1}^{4}\beta_{j}\phi_{\rm A}^{(j)},   \vspace{5pt}\\
-\delta_{ab}\sum_{j=1}^{4}\beta_{j}^{2}\phi_{\rm A}^{(j)}+t_{b}^{+}\sum_{j=1}^{4}\beta_{j}^{2}f_{j}\phi^{(j)}_{\rm A}-V\sum_{j=1}^{4}\beta_{j}f_{j}\phi^{(j)}_{\rm A}=E_{\rm OBC}\sum_{j=1}^{4}\beta_{j}f_{j}\phi^{(j)}_{\rm A},   \vspace{5pt}\\
t_{a}^{-}\sum_{j=1}^{4}\beta_{j}^{L-1}\phi_{\rm A}^{(j)}+\delta_{ab}\sum_{j=1}^{4}\beta_{j}^{L-1}f_{j}\phi^{(j)}_{\rm A}+V\sum_{j=1}^{4}\beta_{j}^{L}\phi_{\rm A}^{(j)}=E_{\rm OBC}\sum_{j=1}^{4}\beta_{j}^{L}\phi_{\rm A}^{(j)}, \vspace{5pt}\\
t_{b}^{-}\sum_{j=1}^{4}\beta_{j}^{L-1}f_{j}\phi^{(j)}_{\rm A}-\delta_{ab}\sum_{j=1}^{4}\beta_{j}^{L-1}\phi_{\rm A}^{(j)}-V\sum_{j=1}^{4}\beta_{j}^{L}f_{j}\phi^{(j)}_{\rm A}=E_{\rm OBC}\sum_{j=1}^{4}\beta_{j}^{L}f_{j}\phi^{(j)}_{\rm A},
\end{array}\right.
\label{eqapp1a4_topo}
\end{eqnarray}
\begin{eqnarray}
\left\{ \begin{array}{l}
\sum_{j=1}^{4}\frac{(E_{\rm OBC}-t_{a}^{+}\beta_{j}-t_{a}^{-}\beta_{j}^{-1}-V)}{\beta_{j}+\beta_{j}^{-1}}\beta_{j}^{2}\phi^{(j)}_{\rm A}=\sum_{j=1}^{4}(E_{\rm OBC}-t_{a}^{+}\beta_{j}-V)\beta_{j}\phi_{\rm A}^{(j)},   \vspace{5pt}\\
-\delta_{ab}\sum_{j=1}^{4}\beta_{j}^{2}\phi_{\rm A}^{(j)}+\sum_{j=1}^{4}(t_{b}^{+}\beta_{j}-V-E_{\rm OBC})\beta_{j}f_{j}\phi^{(j)}_{\rm A}=0,   \vspace{5pt}\\
\sum_{j=1}^{4}\frac{(E_{\rm OBC}-t_{a}^{+}\beta_{j}-t_{a}^{-}\beta_{j}^{-1}-V)}{\beta_{j}+\beta_{j}^{-1}}\beta_{j}^{L-1}\phi^{(j)}_{\rm A}=\sum_{j=1}^{4}(E_{\rm OBC}-t_{a}^{-}\beta_{j}^{-1}-V)\beta_{j}^{L}\phi_{\rm A}^{(j)}, \vspace{5pt}\\
\sum_{j=1}^{4}(t_{b}^{-}\beta_{j}^{-1}-V-E_{\rm OBC})\beta_{j}^{L}f_{j}\phi^{(j)}_{\rm A}-\delta_{ab}\sum_{j=1}^{4}\beta_{j}^{L-1}\phi_{\rm A}^{(j)}=0,
\end{array}\right.
\label{eqapp1a5_topo}
\end{eqnarray}
\begin{eqnarray}
\left\{ \begin{array}{l}
\sum_{j=1}^{4}\left[(E_{\rm OBC}-t_{a}^{+}\beta_{j}-t_{a}^{-}\beta_{j}^{-1}-V)\beta_{j}-(E_{\rm OBC}-t_{a}^{+}\beta_{j}-V)(\beta_{j}+\beta_{j}^{-1}) \right]\beta_{j}\phi^{(j)}_{\rm A}=0,   \vspace{5pt}\\
\sum_{j=1}^{4}\beta_{j}^{2}\phi_{\rm A}^{(j)}+\sum_{j=1}^{4}\frac{(\beta_{j}+\beta_{j}^{-1})(t_{b}^{+}\beta_{j}-V-E_{\rm OBC})}{(E_{\rm OBC}-t_{b}^{+}\beta_{j}-t_{b}^{-}\beta_{j}^{-1}+V)}\beta_{j}\phi^{(j)}_{\rm A}=0,   \vspace{5pt}\\
\sum_{j=1}^{4}\left[(E_{\rm OBC}-t_{a}^{+}\beta_{j}-t_{a}^{-}\beta_{j}^{-1}-V)\beta_{j}^{-1}-(E_{\rm OBC}-t_{a}^{-}\beta_{j}^{-1}-V)(\beta_{j}+\beta_{j}^{-1}) \right]\beta_{j}^{L}\phi^{(j)}_{\rm A}=0, \vspace{5pt}\\
\sum_{j=1}^{4}\frac{(\beta_{j}+\beta_{j}^{-1})(t_{b}^{-}\beta_{j}^{-1}-V-E_{\rm OBC})}{(E_{\rm OBC}-t_{b}^{+}\beta_{j}-t_{b}^{-}\beta_{j}^{-1}+V)}\beta_{j}^{L}\phi^{(j)}_{\rm A}+\sum_{j=1}^{4}\beta_{j}^{L-1}\phi_{\rm A}^{(j)}=0,
\end{array}\right.
\label{eqapp1a6_topo}
\end{eqnarray}
\begin{eqnarray}
\left\{ \begin{array}{l}
\sum_{j=1}^{4}\left[-t_{a}^{-}-(E_{\rm OBC}-t_{a}^{+}\beta_{j}-V)\beta_{j}^{-1} \right]\beta_{j}\phi^{(j)}_{\rm A}=0,   \vspace{5pt}\\
\sum_{j=1}^{4}\left[(E_{\rm OBC}-t_{b}^{+}\beta_{j}-t_{b}^{-}\beta_{j}^{-1}+V)\beta_{j}-(E_{\rm OBC}-t_{b}^{+}\beta_{j}+V)(\beta_{j}+\beta_{j}^{-1})\right]\beta_{j}\phi^{(j)}_{\rm A}=0,   \vspace{5pt}\\
\sum_{j=1}^{4}\left[-t_{a}^{+}-(E_{\rm OBC}-t_{a}^{-}\beta_{j}^{-1}-V)\beta_{j} \right]\beta_{j}^{L}\phi^{(j)}_{\rm A}=0, \vspace{5pt}\\
\sum_{j=1}^{4}\left[(E_{\rm OBC}-t_{b}^{+}\beta_{j}-t_{b}^{-}\beta_{j}^{-1}+V)\beta_{j}^{-1} - (E_{\rm OBC} - t_{b}^{-}\beta_{j}^{-1}+V)(\beta_{j}+\beta_{j}^{-1})\right]\beta_{j}^{L}\phi^{(j)}_{\rm A}=0,
\end{array}\right.
\end{eqnarray}
\begin{eqnarray}
\left\{ \begin{array}{l}
\sum_{j=1}^{4}\left[-t_{a}^{-}-(E_{\rm OBC}-t_{a}^{+}\beta_{j}-V)\beta_{j}^{-1} \right]\beta_{j}\phi^{(j)}_{\rm A}=0,   \vspace{5pt}\\
\sum_{j=1}^{4}\left[-t_{b}^{-}-(E_{\rm OBC}-t_{b}^{+}\beta_{j}+V)\beta_{j}^{-1}\right]\beta_{j}\phi^{(j)}_{\rm A}=0,   \vspace{5pt}\\
\sum_{j=1}^{4}\left[-t_{a}^{+}-(E_{\rm OBC}-t_{a}^{-}\beta_{j}^{-1}-V)\beta_{j} \right]\beta_{j}^{L}\phi^{(j)}_{\rm A}=0, \vspace{5pt}\\
\sum_{j=1}^{4}\left[-t_{b}^{+} - (E_{\rm OBC} - t_{b}^{-}\beta_{j}^{-1}+V)\beta_{j}\right]\beta_{j}^{L}\phi^{(j)}_{\rm A}=0,
\end{array}\right.
\end{eqnarray}
\begin{eqnarray}
\left\{ \begin{array}{l}
\sum_{j=1}^{4}\left[E_{\rm OBC}-(t_{a}^{+}-t_{a}^{-})\beta_{j}-V\right]\phi^{(j)}_{\rm A}=0,   \vspace{5pt}\\
\sum_{j=1}^{4}\left[E_{\rm OBC}-(t_{b}^{+}-t_{b}^{-})\beta_{j}+V\right]\phi^{(j)}_{\rm A}=0,   \vspace{5pt}\\
\sum_{j=1}^{4}\left[E_{\rm OBC}+(t_{a}^{+}-t_{a}^{-})\beta_{j}^{-1}-V \right]\beta_{j}^{L+1}\phi^{(j)}_{\rm A}=0, \vspace{5pt}\\
\sum_{j=1}^{4}\left[E_{\rm OBC} + (t_{b}^{+} - t_{b}^{-})\beta_{j}^{-1}+V\right]\beta_{j}^{L+1}\phi^{(j)}_{\rm A}=0.
\end{array}\right.\label{eqapp1a7_topo}
\end{eqnarray}
Here, we have obtained the coupled equations in terms of only $\phi_{\rm A}^{\left(j\right)}~(j=1,2,3,4)$. 
For $\phi_{\rm A}^{\left(j\right)}~(j=1,2,3,4)$ to have nonzero values, the determinant condition is
\begin{eqnarray}
\left| \begin{array}{cccc}
X_{1}^{(a)} & X_{2}^{(a)} & X_{3}^{(a)} & X_{4}^{(a)} \vspace{5pt}\\
Y_{1}^{(a)} & Y_{2}^{(a)} & Y_{3}^{(a)} & Y_{4}^{(a)} \vspace{5pt}\\
X_{1}^{(b)}\beta_{1}^{L+1} & X_{2}^{(b)}\beta_{2}^{L+1} & X_{3}^{(b)}\beta_{3}^{L+1} & X_{4}^{(b)}\beta_{4}^{L+1} \vspace{5pt}\\
Y_{1}^{(b)}\beta_{1}^{L+1} & Y_{2}^{(b)}\beta_{2}^{L+1} & Y_{3}^{(b)}\beta_{3}^{L+1} & Y_{4}^{(b)}\beta_{4}^{L+1}
\end{array}\right|=0 \label{eqapp2a_topo}
\end{eqnarray}
with $\left|\beta_1\right|\leqslant\left|\beta_2\right|\leqslant\left|\beta_3\right|\leqslant\left|\beta_4\right|$. Here, $X_j$ and $Y_j$ $(j=1,\dots,4)$ are defined as
\begin{eqnarray}
X_{j}^{(a)}&=&E_{\rm OBC}\!-\!(t_{a}^{+}\!-\!t_{a}^{-})\beta_{j}-V=E_{\rm OBC}\!-\!2\delta_{a}\beta_{j}-V,~\left(j=1,\dots,4\right),\\
Y_{j}^{(a)}&=&E_{\rm OBC}\!-\!(t_{b}^{+}\!-\!t_{b}^{-})\beta_{j}+V=E_{\rm OBC}\!-\!2\delta_{b}\beta_{j}+V,~\left(j=1,\dots,4\right),\\
X_{j}^{(b)}&=&E_{\rm OBC} \!+\! (t_{a}^{+} \!-\! t_{a}^{-})\beta_{j}^{-1}-V=E_{\rm OBC} \!+\! 2\delta_{a}\beta_{j}^{-1}-V,~\left(j=1,\dots,4\right),\\
Y_{j}^{(b)}&=&E_{\rm OBC} \!+\! (t_{b}^{+} \!-\! t_{b}^{-})\beta_{j}^{-1}+V=E_{\rm OBC} \!+\! 2\delta_{b}\beta_{j}^{-1}+V,~\left(j=1,\dots,4\right).
\label{eqapp3a_topo}
\end{eqnarray}
Finally, we can obtain the boundary equation (\ref{eq:boundary_equation0_topo}) from Eq.~(\ref{eqapp2a_topo}) as 
\begin{eqnarray}
&&~~\left[Z_{1,4}^{(b)}Z_{2,3}^{(a)}\left(\beta_1\beta_4\right)^{L+1}+Z_{1,4}^{(a)}Z_{2,3}^{(b)}\left(\beta_2\beta_3\right)^{L+1}\right] \nonumber\\
&&-\left[Z_{1,3}^{(b)}Z_{2,4}^{(a)}\left(\beta_1\beta_3\right)^{L+1}+Z_{1,3}^{(a)}Z_{2,4}^{(b)}\left(\beta_2\beta_4\right)^{L+1}\right] \nonumber\\
&&+\left[Z_{1,2}^{(b)}Z_{3,4}^{(a)}\left(\beta_1\beta_2\right)^{L+1}+Z_{1,2}^{(a)}Z_{3,4}^{(b)}\left(\beta_3\beta_4\right)^{L+1}\right]=0,
\end{eqnarray}
where $\beta_j~(j=1,2,3,4)$ satisfy $\left|\beta_1\right|\leqslant\left|\beta_2\right|\leqslant\left|\beta_3\right|\leqslant\left|\beta_4\right|$, and $Z_{i,j}^{(c)}~(i,j=1,2,3,4;~c=a,b)$ are defined as
\begin{eqnarray}
Z_{i,j}^{(c)}&=&X_{i}^{(c)}Y_{j}^{(c)}-X_{j}^{(c)}Y_{i}^{(c)}\\
&=&\begin{cases}
[(t_{b}^{+} - t_{b}^{-})(E_{\rm OBC}-V)-(t_{a}^{+} - t_{a}^{-})(E_{\rm OBC}+V)](\beta_i-\beta_j), & \mbox{$c=a$}\\
[(t_{b}^{+} - t_{b}^{-})(E_{\rm OBC}-V)-(t_{a}^{+} - t_{a}^{-})(E_{\rm OBC}+V)](\beta_j^{-1}-\beta_i^{-1}), & \mbox{$c=b$}
\end{cases},
\end{eqnarray} where $i,j=1,2,3,4;~c=a,b$.

\end{document}